\documentclass[aip,jcp,footinbib,reprint,graphicx]{revtex4-1}
\usepackage{etex}
\usepackage[all]{xy}
\usepackage[T1]{fontenc}
\usepackage[utf8]{inputenc}
\usepackage[english]{babel}
\usepackage[english]{varioref}
\usepackage{amsthm}
\usepackage{amsmath}
\usepackage[version=4]{mhchem}
\usepackage{graphicx} 
\usepackage{subfig}
\usepackage{stmaryrd}
\usepackage{amssymb}
\usepackage{amscd}
\usepackage{tensor}
\usepackage{tabularx}
\newcolumntype{C}[1]{>{\centering\arraybackslash}p{#1}}
\usepackage{mathtools}
\usepackage{amsbsy}
\usepackage{calligra}
\usepackage{dsfont}
\usepackage{bbm}
\usepackage{bbold}
\usepackage{float}
\DeclareMathAlphabet{\mathpzc}{OT1}{pzc}{m}{it}

\usepackage{mathrsfs}

\newcommand{\abs}[1]{\left| #1 \right|} 
\newcommand{\avg}[1]{\left< #1 \right>} 
\newcommand{\gv}[1]{\ensuremath{\mbox{\boldmath$ #1 $}}}
\newcommand{\ket}[1]{\left| #1 \right>} 
\newcommand{\bra}[1]{\left< #1 \right|} 
\newcommand{\braket}[2]{\left< #1 \vphantom{#2} \right|
 \left. #2 \vphantom{#1} \right>} 
\newcommand{\matrixel}[3]{\left< #1 \vphantom{#2#3} \right|
 #2 \left| #3 \vphantom{#1#2} \right>} 

\usepackage{hyperref}

\begin{document}

\title{Semiclassical Quantum Markovian Master Equations. \\Case Study: Continuous Wave Magnetic Resonance of Multispin Systems.} 
\author{J.A. Gyamfi}
\email{jerryman.gyamfi@sns.it}
\affiliation{Scuola Normale Superiore di Pisa, Piazza dei Cavalieri 7, 56126 Pisa, Italy.}

\date{\today}

\begin{abstract}
We propose a method for deriving Lindblad-like master equations when the environment/reservoir is consigned to a classical description. As a proof of concept, we apply the method to continuous wave (CW) magnetic resonance. We make use of a perturbation scheme we have termed \emph{affine commutation perturbation} (ACP). Unlike traditional perturbation methods, ACP has the advantage of incorporating some effects of the perturbation even at its zeroth-order approximation. Indeed, we concentrate here on the zeroth-order, and show how -- even at this lowest order -- the ACP scheme can still yield non-trivial and equally important results.
\par In contradistinction to the purely quantum Markovian master equations in the literature, we explicitly keep the term linear in the system-environment interaction -- at all orders of the perturbation. At the zeroth-order, we show that this results in a dynamics whose map is non-CP (Completely Positive) but approaches asymptotically a CP map as $t \to +\infty$. We also argue that this linear term accounts for the linear response of the system to the presence of the environment -- thus the harbinger for a linear response theory (LRT) within the confines of such (semiclassical) Lindblad-like master equations. The adiabatic process limit of the dynamics is also defined, and considerably explored in the context of CW magnetic resonance. Here, the same linear term emerges as the preeminent link between standard (adiabatic process) LRT (as formulated by Kubo and co.) and Lindblad-like master equations.  And with it, we show how simple stick-plot CW magnetic resonance spectra of multispin systems can be easily generated under certain conditions.
\end{abstract}

\maketitle
 
\section{Introduction}
The overwhelming vast majority of phenomena and processes we study in our labs originate from the interaction between a focus quantum system and its environment. This makes the theory of open quantum systems (TOQS), undeniably, an indispensable one to physicists, chemists and biologists. Of the various approaches to TOQS, the so-called `system-plus-reservoir' ($S+R$) approach has so far been the most successful one
\citep{book:Weiss-2008}. Within the $S+R$ approach, the methodologies and ideas developed over the years based on the Gorini-Kossakowski-Sudarshan-Lindblad (GKSL) equation continue to gain traction due to the essential role this branch of research plays in quantum computation and quantum information theory
\citep{art:GKS-1976,art:Lindblad-1976,art:Gorini-1978,book:Breuer-2007,book:Nielsen-2011, book:Huelga-2012, book:Wilde-2013, book:Kitaev-2002, art:Pascazio-2017}. This is perhaps due to the fact that the GKSL approach successfully offers a way to investigate quantum dynamics and quantum measurement theory with the same mathematical paraphernalia. 
\par Like all $S+R$ approaches
\citep{book:Weiss-2008}, the GKSL approach requires the reservoir/environment to be treated quantum mechanically. Such a requirement naturally impedes its application in those areas where, for all practical purposes, it suffices to treat the environment (or part of it) at the classical level. Cogent examples here include the description of processes like vibronic dynamics, molecular vibrations and electronically nonadiabatic molecular dynamics -- in condensed-phase systems\citep{art:Kapral-1999, art:Yan-1988, art:Richardson-2020}. This has motivated the development of diverse open quantum system models and techniques where the focused system is treated quantum mechanically while the whole environment (or part of it) is described classically. In some of these approaches, the quantum bath correlation functions in the equation of motion for the quantum focused subsystem are substituted with appropriately symmeterized classical correlation functions
\citep{art:Neufeld-2003}. One other approach which has gained traction goes as follows: one begins with the Liouville-von Neumann equation for the fully quantized $S+R$ system, and then performs a partial Wigner transformation over the relevant external degrees of freedom of the environment
\citep{art:Kapral-1999}. This has the advantage of imparting a classical character to the environment's degrees of freedom while maintaining their operator character. One may go further and define an appropriate projection operator for the quantum system's degrees of freedom and then derive a Nakajima-Zwanzig equation from the equation of motion resulting from the partial Wigner transformation described earlier
\citep{art:Mohamad-2005}; this yields an equation of motion for the reduced density matrix $\rho_S(t)$ of the focused system $S$. What seems to be missing in the literature, however, is a similar development within the GKSL approach.

\par We show in this paper that it is possible to derive a quantum master equation for the quantum subsystem $S$ along the lines of the GKSL equation, even when part of the environment (if not the whole) is not explicitly treated quantum mechanically. We illustrate this by developing a quantum theory for continuous wave (CW) magnetic resonance within the GKSL approach.  We shall not take into account any of the molecular non-spin degrees of freedom -- for example, molecular tumbling or rotations\citep{art:Hubbard-1961,art:Zerbetto-2016}, to name a few.

\par Magnetic resonance experiments, both ESR (electron spin resonance) and NMR (nuclear magnetic resonance), provide a very simple and reliable test ground to study and understand open quantum systems and its quantum technological applications
\citep{art:Havel-1997,art:Chuang-1997,art:Warren-1997,art:Jones-1998,art:Price-1999, art:Vlatko-2000, inbook:Glaser-2005, art:Ladd-2010, art:J_Jones-2011, art:Blank-2013, book:Takui-2016, art:Sessoli-2019}. This is one of the reasons why it is only fitting that we bridge the gap between the theory of quantum magnetic resonance as formulated by pioneers like Bloch, Wangsness, Purcell, Pound, Bloembergen, Anderson, Kubo etc during the early decades of research in magnetic resonance
\citep{art:Bloembergen-1948,art:Ramsey_Purcell-1952, art:Wangsness-1953,art:Gutowsky-1953, art:Kubo-Tomita-1954, art:Anderson_JP-1954, art:Corio-1960,book:Corio-1966, book:Andrew-1969, book:Pake-1973, book:Abragam-1983, book:Wertz-1986, book:Ernst-1990, book:Schweiger-2001, book:Cowan-2005} and the theory of open quantum systems (within the GKSL framework)
\citep{art:GKS-1976, art:Lindblad-1976, art:Gorini-1978, book:Breuer-2007, book:Nielsen-2011, book:Huelga-2012, book:Wilde-2013, book:Kitaev-2002, art:Pascazio-2017}. We focus here on CW magnetic resonance -- instead of the pulsed technique -- for historical reasons. In addition, there has been a renewed interest in the CW technique because of advancements made in electronic engineering, which could improve the sensitivity and reduce the cost of its application in research laboratories\citep{art:Newton-2017}.

\subsection{Wavefunction vs density matrix formalism. CP vs non-CP maps.}
Like in many other areas, the history of magnetic resonance shows that the theoretical investigation of the dynamics of the focused quantum system has been developed along the lines of two main formalisms. These are: the wavefunction and the density matrix formalism\citep{book:Blum-1981}.
\par In the wavefunction formalism, the primary object of interest is the transition probability per unit time (or transition rate) between various pairs of spin states. These transition rates are then employed to derive an expression for the spectrum
\citep{art:Solomon-1955}. The derivation of expressions for the transition rate are based on a first-order time-dependent perturbation approximation of the probability amplitudes
\citep{art:Solomon-1955}. This is normally accompanied by the assumption that the system is initially in a given specific normalized pure state
\footnote{See Supplemental Material}. The time-dependent perturbation theory used is that due to Dirac
\citep{art:Dirac-1926}, where one obtains the probability amplitudes by solving a system of differential equations
\cite{Note1}. 
\par The wavefunction formalism has long been known to be inadequate for the full quantum mechanical description of the spin dynamics
\citep{art:Redfield-1957, Note1, book:Abragam-1983}. In the study of magnetic resonance, the density matrix formalism
\citep{book:Blum-1981, book:Pathria-1996} has been widely used to treat relaxation processes in the presence of a random perturbation. The Wangsness-Bloch-Redfield
\citep{art:Redfield-1957,art:Wangsness-1953} theory is the archetypal example. In all such theories, the goal is to find a reasonably approximated expression for the density matrix $\rho_S(t)$ (or $\frac{d}{dt} \rho_S$) of the focused spin system, neglecting, however, key questions regarding the nature of $\rho_S(t)$'s evolution in time. The GKSL approach, on the other hand, pays particular attention to these aspects and tries to provide general requisites on the properties of the map under which $\rho_S(t)$ evolves. The completely positive trace-preserving (CPT)
\citep{art:GKS-1976, art:Lindblad-1976, art:Gorini-1978, book:Breuer-2007, book:Nielsen-2011, book:Huelga-2012, book:Wilde-2013, misc:Lidar-2019, art:Choi-1975, book:Alicki-2007} property is arguably the most celebrated of these. Even more, it has been the common view that the CPT property is a fundamental requisite of any reputable quantum map
\citep{art:Alicki_comment-1995}. This view, however, has been challenged by some authors
\citep{art:Pechukas-1994,art:Shaji-2005}. According to the opposing view, the fundamental requisite to be required of a quantum map is that it preserves Hermiticity, trace and positivity
\citep{art:Shaji-2005, art:Sudarshan-1961}. The unclenching of the view of the CPT requirement as an inescapable one has been met with an increase in research on non-CP maps
\citep{book:McCracken-2014, art:Yu-2000, art:Buzek-2001, art:Carteret-2008, inbook:Shabani_co-2014}. After all, studies
\cite{art:Mazzola-2012, art:R-Rosario-2012, art:Cory-2004} have shown a close connection between non-CP maps and non-Markovian
\citep{art:Breuer-2007, art:Giovannetti-2013, art:Mazzola_Breuer-2010, art:Pomyalov-2005, art:Aspuru-2009} dynamics -- the latter being a hot topic. 
\par All in all, applying the GKSL approach to some problems of considerable interest like multispin magnetic resonance is not an easy sell because the environment (\emph{i.e.} the applied magnetic fields, in the case of magnetic resonance) needs to be fully quantized. In the theory of magnetic resonance, as formulated by the above-mentioned pioneers and others, the so-called Maxwell-Bloch scheme
\citep{art:Jeener-2002} is used. In this scheme, the spin system is quantized while the external magnetic fields are consigned to a classical description (Maxwell equations). Naturally, we could quantize the applied electromagnetic fields and carry out our derivations without any significant conceptual hurdle. Using quantized applied electromagnetic fields in magnetic resonance theory has been done, for example, by Jeener and Henin
\citep{art:Jeener-2002}, and also by Engelke
\citep{art:Engelke-2010}. The results one obtains are in good agreement with those obtained under the Maxwell-Bloch scheme
\citep{art:Jeener-2002,art:Engelke-2010}. A fully quantized electromagnetic field, though, may be necessary under more sophisticated experiments, but for what concerns standard NMR and ESR experiments, it suffices to treat the external fields classically. And this will be our strategy.
\par We propose in this paper a way to still apply the GKSL approach even when the environment is not fully quantized. We also introduce an approximation scheme we have termed \emph{affine commutation perturbation} (ACP). In the case of CW magnetic resonance, we show that the ACP scheme entails non-trivial results even at the zeroth-order approximation. No less important is the fact that the quantum map we get at this order is non-CP due to the term linear in the system-environment interaction that we keep, and which poses as an inhomogeneous term in the master equation (ME). Without it, the map is CP. And it is interesting to observe that while most of the non-CP maps hitherto studied in the literature (with a quantized environment) are a result of initial system-environment correlations
\cite{book:McCracken-2014, art:Buzek-2001, art:Carteret-2008, art:Mazzola-2012, art:R-Rosario-2012, art:Cory-2004, inbook:Shabani_co-2014}, in our case, such a map stems from the presence of this inhomogeneous term in the master equation. As we shall show, this term is also crucial for the correct theoretical description of CW magnetic resonance experiments.  
\par The content below is organized into three main sections: The first, \S \ref{sec:Summary}, summarizes our proposal for treating classical environments within the GKSL formalism. In the second section, \S \ref{sec:QMME2}, we apply the method to study CW magnetic resonance and we give concluding remarks in \S \ref{sec:Conclusion}.
\par More specifically, under section \S \ref{sec:QMME2}, we present the ACP scheme in \S \ref{subsec:ACP}, and derive in \S \ref{subsec:zeroth-order} the Lindblad-like master equation (for a CW magnetic resonance experiment with an arbitrary ensemble of a multispin system as the focus system) at the zeroth-order of the ACP scheme and under the weak-coupling assumption. This is followed by the subsection \S \ref{subsec:Applications} where we apply the Lindblad-like master equation derived in \S \ref{subsec:zeroth-order} to concrete problems like the CW experiment with an ensemble of spin-1/2 particles (\S \ref{subsec:spin-half}) and the generation of simple stick-plot CW resonance spectra (\S \ref{subsec:spectrum}). In the same subsection, \S \ref{subsec:Applications}, we explore the connection to linear response theory in \S \ref{subsec:LRT}. In the last subsection of \S \ref{sec:QMME2}, \S \ref{subsec:higher-orders}, we briefly discuss the higher-order terms of the ACP scheme.

\section{Proposal on how to handle classical environments}\label{sec:Summary}
The steps involved in the usual microscopic derivation of the GKSL equation (where both $S$ and $R$ are considered quantum entities) -- under the assumptions of weak-coupling limit and a separable initial state -- may be summarized as follows: i) start with a Liouville-von Neumann equation for $\rho_{R+S}(t)$ (the density matrix for the isolated $S+R$ bipartite system); ii) transition to the interaction picture; iii) introduce the Markov approximation; iv) trace out the environment degrees of freedom and assume the stability condition to obtain an equation of motion for the reduced density matrix $\rho_S(t)$; and v) introduce the secular approximation. 
\par When $R$ is considered classical, a GKSL-like equation may be obtained in the weak-coupling limit as follows: i) start with a Liouville-von Neumann equation for the reduced density matrix $\rho_S(t)$. The environment's degrees of freedom appear here as (time-dependent) factors in $S$'s Hamiltonian; ii) transform the equation of motion into the interaction picture; iii)  introduce the Markov approximation; and then iv) perform the secular approximation. It is implicitly assumed in step i) that the initial state of $S+R$ is `separable' (\emph{i.e.} no initial correlations between the quantum system and the classical environment). Somewhere between steps ii) and iv), one has to define an appropriate set of Lindblad operators; this may prove difficult to accomplish for some problems or may require some very ingenious choices, but, nonetheless, there is, somehow, a general understanding on how to proceed\citep{book:Breuer-2007}. This is more so when the resonance condition consists of a set of independent conditions which must occur concurrently. CW magnetic resonance is a good example in this regard. 
\section{Semiclassical Quantum Markovian Master Equation Approach to CW Magnetic Resonance}\label{sec:QMME2}
\subsection{Preamble}
In standard CW magnetic resonance experiments, the sample is subjected to an oscillating field $\mathbf{B}_1(t)$ of constant frequency $\omega$ while, simultaneously, a steady magnetic field $\mathbf{B}_o$ (perpendicular to $\mathbf{B}_1(t)$, with $\Vert \mathbf{B}_1(t) \Vert \ll \Vert \mathbf{B}_o \Vert$) is sweepingly applied so as to tune the focus system to resonance.\footnote{Another alternative is to hold the frequency $\mathbf{B}_o$ constant and vary $\omega$, but this scheme is not the experimentally preferred way of doing business.} Keeping strict adherence to this faithful description will land us into what we may call generalized Landau-Zerner\citep{art:Majorana-1932, *art:Landau-1932, *art:Zener-1932} transition problems, obscuring the central effort of the present paper -- which is, to derive a GKSL-like equation for CW magnetic resonance experiments whereby we treat the applied fields ($\mathbf{B}_o$ and $\mathbf{B}_1$) as classical entities. For the sake of argument, we shall not explicitly take into account the sweeping of $\mathbf{B}_o$. Rather, we take the view that for any instance of $\mathbf{B}_o$, the spin system settles very fast to an equilibrium state (solely dependent on $\mathbf{B}_o$) upon its interaction with the latter field, before it begins to adjust to the presence of $\mathbf{B}_1(t)$. Put in other terms, we may view the experiment as a two step process, whereby we first apply $\mathbf{B}_o$ and then $\mathbf{B}_1(t)$. If any justification at all is to be allowed for this simplified view of the CW experiment under discussion, we may invoke the fact that: 1) except at very low temperatures, the scale of resonance energy in magnetic resonance experiments is quite small compared to thermal energy (\emph{high-temperature approximation})
\citep{book:Gamliel_Levanon-1995}, and 2) $\Vert \mathbf{B}_1(t) \Vert \ll \Vert \mathbf{B}_o \Vert$. Naturally, the function of the steady field $\mathbf{B}_o$ is to create the Zeeman effect, while $\mathbf{B}_1(t)$ stimulates transitions between the energy levels resulting from the Zeeman effect. 
\par Bearing in mind the above reinterpretation of the experiment, consider an ensemble of noninteracting molecules in some condensed phase environment. Each member of the ensemble is a multispin system $\mathpzc{A}$ with the isotropic Hamiltonian $H_{spin-spin}$, where:
				\begin{equation}
				\label{eq:H_spin-spin}
				H_{spin-spin} := \sum_{i>j} T_{ij} \gv{S}_i\cdot \gv{S}_j 
				\end{equation}
where $\gv{S}_i \left( \equiv S^x_i \gv{e}_x + S^y_i \gv{e}_y + S^z_i \gv{e}_z\right)$ is the $i-$th element's spin vector operator, and $T_{ij}$ is the coupling constant between spins $i$ and $j$. $S^\alpha_i$ is the spin operator along the axis $\alpha \in \{x,y,z\}$ for spin $i$. Note that all the other degrees of freedom with the exception of the spin degrees in question constitute the environment, so in a fully quantum treatment, the coupling constants will result from the tracing out of the environment's degrees of freedom. A fully quantized condensed phase like the liquid phase -- in which many of these experiments are carried out -- is all but easy to manage. The coupling constants $T_{ij}$ here are, therefore, assumed to have been determined by some other means\citep{book:Kaup-2006, art:Fukui-1999, art:Laszlo-1967, art:NeugebauerJCP-2005, art:Dracinsky-2010}. Surely, these constants incorporate the influence of the environment. And in the liquid phase, one usually speaks of `solvent effects'\cite{art:Laszlo-1967, art:NeugebauerJCP-2005, art:Dracinsky-2010}.
\par If we take $\mathbf{B}_o$ as lying along the $z-$axis (i.e. $\mathbf{B}_o=B_o \gv{e}_z$), and consider the latter as the axis of quantization, then the Hamiltonian of the multispin system acquires a new term (i.e. the Zeeman term):
				\begin{equation}
				\label{eq:def_H_o}
				H_o = H_{spin-spin} + \xi^z B_o
				\end{equation}	
where,
				\begin{equation}
				\label{eq:xi_alpha}
				\xi^\alpha := -\mu^\alpha =-\sum_i \gamma_i S^\alpha_i \ \qquad \alpha \in \{x,y,z\}
				\end{equation}	
where $\mu^\alpha$ indicates the total magnetic moment operator of the multispin system along the axis $\alpha$, and $\gamma_i$ is the gyromagnetic ratio of the $i-$th spin. Suppose we apply the oscillating field $\mathbf{B}_1(t)$ at the instant $t_o$, and assume the spin system had reached its thermal equilibrium state under $\mathbf{B}_o$ prior to $t_o$. Then, the density matrix of the multispin system immediately before the instant $t_o$, $\rho_S(t_o)$, is:
				\begin{equation}
				\label{eq:initial_Boltzmann_state}
				\rho_S(t_o) = \frac{e^{-\beta H_o}}{\mathcal{Z}} \ , \qquad \mathcal{Z}:= \mbox{Tr} \left[e^{-\beta H_o} \right] \ ,
				\end{equation}
where $\beta \equiv \frac{1}{k_B T}$ ($k_B$ is the Boltzmann constant and $T$ is the absolute temperature), and $H_o$ is defined in Eq. \eqref{eq:def_H_o}. The operator $\rho_S$ (which we now, henceforth, simply indicate as $\rho$) -- after the application of $\mathbf{B}_1(t)$ -- must then satisfy the Cauchy initial value problem:
				\begin{equation}
				\label{eq:rho(t)}
				\begin{cases}
				\frac{d}{dt} \rho(t) = -i \left[H_o + V(t) , \rho(t)\right] \ , & \quad (t>t_o)\\
				\rho(t)= \frac{e^{-\beta H_o}}{\mathcal{Z}}\ , & \quad (t=t_o)
				\end{cases}
				\end{equation}
where,
				\begin{equation}
				V(t) := \gv{\xi}\cdot \mathbf{B}_1(t)
				\end{equation}
with the components of $\gv{\xi}$ defined in Eq. \eqref{eq:xi_alpha}. Two observations are due here: First of all, it is implicit in the initial condition on $\rho$, 	Eq. \eqref{eq:rho(t)}, that there is no correlation (quantum or classical) between the spin system and the oscillating field. This is tantamount to the Born approximation. Secondly, the same equation will have us think $\rho(t)$ may evolve by means of a unitary evolution superoperator, but that would contradict the fact that $\rho(t)$ represents the density matrix of an open quantum system. Nevertheless, although the experimental conditions largely justify the series of approximations (like the Markovian) we shall introduce in the course of our discussion, they also have the advantage of leading to a non-unitary evolution of $\rho(t)$. 
\par Taking into account the fact that the oscillating magnetic fields actually used in experiments are not perfectly monochromatic, it is only reasonable that we take $\mathbf{B}_1(t)$ to be a superposition of various independent oscillating fields (for simplicity, all of zero phase and with the same maximum):
				\begin{equation}
				\label{eq:B_1(t)}
				\mathbf{B}_1(t) = \sum_r 2B_1 \cos(\omega_r t) \mathbf{e}_x  
				\end{equation}
where $2B_1$ and $\omega_r$ are the maximum amplitude and frequency of the $r-$th oscillating field, respectively. The frequencies $\omega_r$ are distributed around a central frequency $\omega$, and we assume $\omega \gg \left\vert \delta\omega_r \right\vert$, where $\delta\omega_r \equiv \omega-\omega_r$. $\mathbf{B}_1(t)$ in Eq. \eqref{eq:B_1(t)} is a generalization of the usual $\mathbf{B}_1(t) = 2B_1	\cos(\omega t)\mathbf{e}_x$ used in the literature\citep{book:Corio-1966,book:Andrew-1969,art:Bloembergen-1948}. For the sake of clarity, we choose to decompose $\mathbf{B}_1(t)$ into two cluster of rotating fields in the $x-y$ plane, both with the same intensity but each having a sense of rotation opposite to the other:
				\begin{equation}
				\mathbf{B}_1(t) = \mathbf{B}_{1,+}(t) + \mathbf{B}_{1,-}(t)
				\end{equation}
where,
				\begin{equation}
				\mathbf{B}_{1,\pm}(t) := \sum_r B_1 \left[\cos(\omega_r t) \mathbf{e}_x \pm \sin(\omega_r t) \mathbf{e}_y\right] \ .
				\end{equation}
$\mathbf{B}_{1,\pm}(t)$ rotate in the anticlockwise and clockwise directions, respectively, when observed from the top of the direction parallel to that of $\mathbf{B}_o$. As it is well-known, for a given spin, only one of these may give rise to the resonance phenomenon depending on the sign of its Larmor frequency\cite{book:Abragam-1983}. On similar footing, we may decompose the interaction term $V(t)$ as follows:
				\begin{equation}
				V(t) = V_+(t) + V_-(t)
				\end{equation}
with
				\begin{subequations}
				\begin{align}
				V_\pm(t) & = B_1 \sum_r  \left[\xi^x\cos(\omega_r t)  \pm \xi^y \sin(\omega_r t)  \right] \\
				& = B_1 \sum_r  \  e^{\mp i\omega_r t S^z} \xi^x e^{\pm i\omega_r t S^z}
				\end{align}
				\end{subequations}	
where $S^z:= \sum_i S^z_i$, i.e. the total spin operator along the $z-$axis. According to the sign of the Larmor frequency of the spin, only one of $V_\pm(t)$ contributes significantly to the observed resonance spectra; for example, if the Larmor frequency is positive (thus, negative gyromagnetic ratio), then the observed spectra is primarily due to the interaction term $V_+(t)$, with  negligible contributions from $V_-(t)$. 
\par As noted earlier, in the usual experimental setup, $B_1 \ll B_o$, so we can consider $V(t)$ as a perturbation with respect to $H_o$. We may then take Eq. \eqref{eq:rho(t)} into the interaction picture. The result is:
				\begin{equation}
				\label{eq:varrho(t)}
				\begin{cases}
				\frac{d}{dt} \varrho(t) = -i \left[\mathscr{V}(t) , \varrho(t)\right] \ , & \quad (t>t_o)\\
				\varrho(t_o)= \frac{e^{-\beta H_o}}{\mathcal{Z}}\ , & \quad (t=t_o)
				\end{cases}
				\end{equation}
where,
				\begin{subequations}
				\begin{align}
				\varrho(t) & := e^{itH_o} \rho(t) e^{-itH_o} \label{eq:def_varrho_t}	\\
				\mathscr{V}(t) &:= e^{itH_o} V(t) e^{-itH_o} = 	\mathscr{V}_+(t) + \mathscr{V}_-(t) \ .	\label{eq:V(t)_interaction_pic}
				\end{align}
				\end{subequations}
From Eq. \eqref{eq:varrho(t)}, it follows that:
				\begin{multline}
				\label{eq:varrho(t)_3}
				\frac{d}{dt} \varrho(t) = -i \left[\mathscr{V}(t) , \varrho(t_o)\right] \\
				- \int^{t-t_o}_0 d\tau \left[\mathscr{V}(t) ,\left[\mathscr{V}(t-\tau) , \varrho(t-\tau)\right]\right] \ .
				\end{multline}
We observe from Eq. \eqref{eq:varrho(t)_3} that the evolution of $\varrho(t)$ depends on its history due to the presence of $\varrho(t-\tau)$ on the l.h.s. Let us assume the evolution of $\varrho$ does effectively depend on its history only in the time-frame $0 \leq \tau \leq \tau_B$, where $\tau_B$ is some characteristic time which depends on the interaction between the spin system and the external fields. Supposing the resolution of our experiment does make the time-frame $0 \leq \tau \leq \tau_B$ practically inaccessible to our investigation
\citep{art:Bernardes-2016}, so that measurements on the spin system effectively refer to $t \gg \tau_B$, instances during which the evolution of $\varrho$ does not depend on its history, but only on its present state, then we may substitute $ \varrho(t-\tau)$ in Eq. \eqref{eq:varrho(t)_3} with $\varrho(t)$ (Markov approximation). (When the experimental technique being employed is capable of appropriately resolving certain system-environment correlations
\citep{art:Bernardes-2016}, it may be necessary to account for the non-Markovian
\citep{art:Breuer-2007, art:Giovannetti-2013, art:Mazzola_Breuer-2010, art:Pomyalov-2005, art:Aspuru-2009} property in $\varrho(t)$'s evolution). If, in addition, we set $t_o=0$ and extend the upper limit of the integral over $\tau$ to infinity (this is just an approximation on the integral over $\tau$; and it is particularly justified for steady-state experiments like the one under discussion), the final result is:
				\begin{equation}
				\label{eq:varrho(t)_5}
				\frac{d}{dt} \varrho(t) = -i \left[\mathscr{V}(t) , \varrho(0)\right] - \int^{+\infty}_0 d\tau \left[\mathscr{V}(t) ,\left[\mathscr{V}(t-\tau) , \varrho(t)\right]\right] \ .
				\end{equation}
Note that by taking the limit $t_o \to -\infty$ (i.e. adiabatic approximation\cite{book:Giuliani-2005}) in Eq. \eqref{eq:varrho(t)_3}, we also get Eq. \eqref{eq:varrho(t)_5} since $\varrho(0)=\varrho(-\infty)=\frac{e^{-\beta H_o}}{\mathcal{Z}}$. Though the adiabatic approximation -- which assumes the system, prior to the application of $\mathbf{B}_1(t)$, had been in the equilibrium state for a very, very long time -- leads essentially to the same equation of motion as Eq. \eqref{eq:varrho(t)_5} for $\varrho(t)$, it requires that the lower limit of the integral over $t$ be $t=-\infty$ instead of $t=0$. This is a subtle but important difference as there could be instances whereby this lower limit of $t$ according to the adiabatic approximation leads to infinite expectation values of some observables. In the following, we stick to Eq. \eqref{eq:varrho(t)_5} bearing in mind that the lower limit of $t$ is $t=0$.
\par One other important thing to note from Eq. \eqref{eq:varrho(t)_5} is the presence of the first-order term in $\mathscr{V}(t)$; this is contrary to what is usually done in such microscopic derivations\citep{book:Breuer-2007}. We are keeping the term simply because -- unlike in the usual derivations in the literature whereby this term usually becomes zero (or is assumed to be) upon a trace operation over the environment's quantum degrees of freedom\citep{book:Breuer-2007} -- here, only the spin system is treated at the quantum level from the beginning and there are no quantum degrees of freedom of the environment (the magnetic fields) to trace over. We have thus no reason to neglect the term. As it will turn out later in our discussion, this term is crucial to the quantum theory of magnetic resonance and allows us to derive a number of important results already known in the literature. Most importantly, we shall show that it is the springboard to the development of a linear response theory from the perspective of quantum Markovian master equations. 
\subsection{Affine commutation perturbation}\label{subsec:ACP}
\par Let us now go back to Eq. \eqref{eq:def_H_o} and analyze $H_o$. Under usual experimental conditions, it is often the case that $\left\Vert \xi^z B_o \right\Vert \gg \left\Vert \sum_{i>j} T_{ij} \gv{S}_i\cdot \gv{S}_j \right\Vert$ (Given two operators $X$ and $Y$ acting on the same (finite) Hilbert space, with the statement `$\left\Vert X\right\Vert \gg \left\Vert Y\right\Vert$' we mean: for any pair of nondegenerate eigenkets $\ket{k}$ and $\ket{k'}$ of $X$ such that $X \ket{k}=E_k \ket{k}$, $\abs{\frac{\matrixel{k'}{Y}{k}}{E_{k'} - E_{k}}} \ll 1$). The spin-spin interaction term, $H_{spin-spin}$, may therefore be treated as a perturbation with respect to the Zeeman term, $\xi^z B_o$. Instead of simply treating $H_{spin-spin}$ as a perturbation term with respect to $\xi^z B_o$, we are going to do what we call \emph{affine commutation perturbation} (ACP). In this scheme, the perturbation term is rewritten as a sum of two operators: $A'+B'$, where the operator $A'$ commutes with the leading term, while $B'$ does not. $A'$ is then added to the leading term and their sum is treated as the new leading term, while $B'$ becomes the new perturbation term and the normal perturbation expansion is then carried out. If $A'$ exists and one performs the ACP expansion, the results one obtains -- compared to those from the  standard perturbation expansion -- are more accurate even at low orders.
\par As we intend to perform an ACP, we rewrite $H_o$ as:
				\begin{equation}
				H_o = \mathscr{Z}_o + \mathscr{X}
				\end{equation}
where,
				\begin{subequations}
				\label{eq:def_Z_X}
				\begin{align}
				\mathscr{Z}_o & := B_o \xi^z + \sum_{i>j} T_{ij} S^z_i S^z_j \\
				\mathscr{X} & := \frac{1}{2} \sum_{i>j} T_{ij} \left(S^+_i S^-_j + S^-_i S^+_j \right)
				\end{align}
				\end{subequations}
$(S^\pm_j \equiv S^x_j \pm i S^y_j)$. Note that $\left\Vert \mathscr{Z}_o \right\Vert \gg \left\Vert \mathscr{X} \right\Vert$. Moreover, $\mathscr{Z}_o$ commutes with the total spin operator along the $z-$axis, $S^z$. And more importantly, the eigenvectors of $\mathscr{Z}_o$ are simply the multispin kets in the uncoupled representation. 
\par Using Feynman's operator calculus\citep{art:Feynman-1951}, we now expand all operators in Eq. \eqref{eq:varrho(t)_5} dependent on $\mathscr{X}$ in powers of the latter. Namely, the rotated operators: 1) $\varrho(t)=e^{itH_o} \rho(t) e^{-itH_o} = e^{it(\mathscr{Z}_o + \lambda \mathscr{X})} \rho(t)e^{-it(\mathscr{Z}_o + \lambda \mathscr{X})} $, 2) $\mathscr{V}(t)= e^{it(\mathscr{Z}_o + \lambda \mathscr{X})} V(t)e^{-it(\mathscr{Z}_o + \lambda \mathscr{X})}$ and 3) $\varrho(0)=\frac{e^{-\beta(\mathscr{Z}_o + \lambda \mathscr{X})}}{\mbox{Tr}[e^{-\beta(\mathscr{Z}_o + \lambda \mathscr{X})}]}$ (where the constant $\lambda$, introduced  here for book-keeping purposes, will be set equal to $1$ at the end). For the first two, we get:
				\begin{subequations}
				\begin{align}
				\varrho(t) & = \sum^\infty_{n=0} \ \lambda^n \ \varrho^{(n)}(t) \label{eq:varrho_expansion_X}\\
				\mathscr{V}(t) & = \sum^\infty_{n=0} \ \lambda^n \  \mathscr{V}^{(n)}(t)\label{eq:V_expansion_X}
				\end{align}
				\end{subequations}
where,
				\begin{equation}
				\label{eq:varrho_expansion_X_n}
				\begin{split}
			 & \varrho^{(n)}(t) \\
			 & =   e^{it \mathscr{Z}_o}\left[  \sum^n_{k=0} \sum^{k}_{k'=0} \mathscr{Y}^{(n-k)}(t) \ \rho^{(k-k')}(t) \ \mathscr{Y}^{(k')\dagger}(t) \right] 
			 e^{-it \mathscr{Z}_o}
			 	\end{split}
				\end{equation}
				\begin{equation}
				\label{eq:V_expansion_X_n}
			 \mathscr{V}^{(n)}(t) =   e^{it \mathscr{Z}_o}\left[  \sum^n_{k=0}  \mathscr{Y}^{(n-k)}(t) \ V(t) \ \mathscr{Y}^{(k)\dagger}(t) \right] e^{-it \mathscr{Z}_o}
				\end{equation}
where, for $n\geq 1$, 	
				\begin{equation}
				\mathscr{Y}^{(n)}(t)  \equiv i^n \int^t_{0} ds_1  \cdots \int^{s_{n-1}}_{0} ds_n \ \mathscr{X}(s_1)   \cdots \mathscr{X}(s_n)  
				\end{equation}
while $\mathscr{Y}^{(n)}(t) \equiv \mathbb{I}$ for $n=0$ -- with
				\begin{equation}
				\mathscr{X}(x) :=e^{-ix \mathscr{Z}_o} \mathscr{X} e^{ix \mathscr{Z}_o} \ .
				\end{equation}
Naturally, $\rho(t)$ depends on $\mathscr{X}$. This dependence slightly complicates the expansion of $\varrho(t)$ in powers of $\mathscr{X}$, compared to $\mathscr{V}(t)$. The operator $\rho^{(m)}(t)$ in Eq. \eqref{eq:varrho_expansion_X_n} denotes the $m-$th term coming from the formal expansion of $\rho(t)$ in powers of $\mathscr{X}$. Put differently, $\rho^{(m)}(t)$ is the equivalent of $\varrho^{(m)}(t)$ in the Schr\"odinger picture. In general, the relation between the two is not a simple unitary transformation. As a matter of fact, only $\varrho^{(0)}(t)$ and $\rho^{(0)}(t)$ are related through a unitary transformation. To illustrate this very important point, consider, for example, the cases $n=0$ and $n=1$ from Eq. \eqref{eq:varrho_expansion_X_n}; these yield the following expressions:
				\begin{subequations}
				\begin{align}
				\varrho^{(0)}(t) & = \ e^{it \mathscr{Z}_o} \rho^{(0)}(t) e^{-it \mathscr{Z}_o}  \label{eq:varrho_A^0}\\
				\varrho^{(1)}(t) & = e^{it \mathscr{Z}_o} \rho^{(1)}(t) e^{-it \mathscr{Z}_o} + i \int^t_0 ds \  \left[\varrho^{(0)}(t), \mathscr{X}(s-t) \right] .
				\end{align}
				\end{subequations}
from which we derive that:
				\begin{subequations}
				\begin{align}
				\rho^{(0)}(t) & = \ e^{-it \mathscr{Z}_o} \varrho^{(0)}(t) e^{it \mathscr{Z}_o}  \label{eq:rho_A^0}\\
				\rho^{(1)}(t) & =e^{-it \mathscr{Z}_o} \varrho^{(1)}(t) e^{it \mathscr{Z}_o} - i \int^t_0 ds \  \left[\rho^{(0)}(t), \mathscr{X}(s) \right] .
				\end{align}
				\end{subequations}
Moving on, to fully expand Eq. \eqref{eq:varrho(t)_5} in powers of $\mathscr{X}$, we are only left with the expansion of $\varrho(0)$. Resorting once more to Feynman's operator calculus\citep{art:Feynman-1951}, one can show that, for a fixed $\beta=\frac{1}{k_B T}$:
				\begin{equation}
				\varrho(0) = \sum^{\infty}_{n=0} \lambda^n \varrho^{(n)}(0)
				\end{equation}
where,
				\begin{equation}
				\label{eq:varrho_n=0}
			\varrho^{(n)}(0) \equiv \varrho^{(0)}(0) \sum^{n}_{n'=0}  \zeta_{n'}(i\beta) \ \mathscr{Y}^{(n-n')}(i\beta) 
				\end{equation}
and
				\begin{equation}
				\label{eq:initial_rho_n=0}
				\varrho^{(0)}(0) \equiv \frac{e^{-\beta \mathscr{Z}_o}}{\mbox{Tr}\left[e^{-\beta \mathscr{Z}_o} \right]} \ .
				\end{equation}
The coefficients $\{\zeta_{n}(i\beta)\}$ are the solution to a system of linear equations (of infinite dimension), characterized by a coefficient matrix which is a lower triangular Toeplitz matrix. It can be verified that every $\zeta_n(i\beta)$ is proportional to the determinant of an upper Hessenberg matrix. Indeed, for $n\geq 1$,
			\begin{widetext}
			\begin{align}
			\label{eq:def_zeta_n}
			\zeta_{n}(x) = (-1)^n \ \det \begin{vmatrix}
			\avg{\mathscr{Y}^{(1)}(x)}_o & \avg{\mathscr{Y}^{(2)}(x)}_o & \avg{\mathscr{Y}^{(3)}(x)}_o & \avg{\mathscr{Y}^{(4)}(x)}_o & \ldots & \avg{\mathscr{Y}^{(n-1)}(x)}_o & \avg{\mathscr{Y}^{(n)}(x)}_o \\
			1 & \avg{\mathscr{Y}^{(1)}(x)}_o & \avg{\mathscr{Y}^{(2)}(x)}_o  & \avg{\mathscr{Y}^{(3)}(x)}_o & \ldots & \avg{\mathscr{Y}^{(n-2)}(x)}_o & \avg{\mathscr{Y}^{(n-1)}(x)}_o \\
			0 & 1 & \avg{\mathscr{Y}^{(1)}(x)}_o & \avg{\mathscr{Y}^{(2)}(x)}_o & \ldots & \avg{\mathscr{Y}^{(n-3)}(x)}_o & \avg{\mathscr{Y}^{(n-2)}(x)}_o \\
			0 & 0 & 1 & \avg{\mathscr{Y}^{(1)}(x)}_o & \ldots & \avg{\mathscr{Y}^{(n-4)}(x)}_o & \avg{\mathscr{Y}^{(n-3)}(x)}_o \\
			\vdots & \vdots & \vdots & \vdots & \ldots &\vdots &  \vdots\\ 
			0 & 0 & 0 & 0 & \ldots & 1 & \avg{\mathscr{Y}^{(1)}(x)}_o\\ 
			\end{vmatrix}
			\end{align}
			\end{widetext}
while for $n=0$, $\zeta_n(x)=1$. Moreover,
			\begin{equation}
			\avg{\mathscr{Y}^{(n)}(x)}_o \equiv \mbox{Tr} \left[\varrho^{(0)}(0) \ \mathscr{Y}^{(n)}(x) \right] \ .
			\end{equation}
(On passing, we would like to draw the Reader's attention to an evident connection between the expression for the coefficient $\zeta_n(x)$ as given in Eq. \eqref{eq:def_zeta_n} and the determinant expression for the $n-$th complete Bell polynomial\citep{inbook:Ono-2017}.) The recursive relation for the coefficients $\{\zeta_n(x)\}$, for $n\geq 1$, is as follows:
			\begin{equation}
			\label{eq:def_zeta_n_b}
			\zeta_n(x) = -  \sum^{n-1}_{n'=0} \zeta_{n'}(x) \avg{\mathscr{Y}^{(n-n')}(x)}_o\ .
			\end{equation}
(Compare Eqs. \eqref{eq:def_zeta_n} and \eqref{eq:def_zeta_n_b} with Theorem I of [\onlinecite{art:Milan-2010}].)
With these expansions of the operators $\varrho(t), \mathscr{V}(t)$ and $\varrho(0)$ in $\mathscr{X}$, Eq. \eqref{eq:varrho(t)_5} turns out to be:
				\begin{multline}
				\label{eq:lambda_expansion_of_eq_mo}
				\sum^{\infty}_{n=0} \lambda^n \frac{d}{dt} \varrho^{(n)}(t) = -i \sum^{\infty}_{n=0}\sum^{\infty}_{n'=0} \lambda^{n+n'} \left[\mathscr{V}^{(n)}(t) , \varrho^{(n')}(0)\right] \\
				- \sum^{\infty}_{n=0}\sum^{\infty}_{n'=0} \sum^{\infty}_{n''=0}\int^{+\infty}_0 d\tau \ \lambda^{n+n'+n''}\\
				\times \left[\mathscr{V}^{(n)}(t) ,\left[\mathscr{V}^{(n')}(t-\tau) , \varrho^{(n'')}(t)\right]\right]				
				\end{multline}
Equating terms of the same order in $\lambda$ on both sides of Eq. \eqref{eq:lambda_expansion_of_eq_mo} yields a non-homogeneous system of triangular differential equations for  $\{\varrho^{(n)}(t)\}$, which can be solved step-by-step beginning with the line $n=0$. Indeed, one can easily derive from Eq. \eqref{eq:lambda_expansion_of_eq_mo} that the generic $\varrho^{(n)}(t)$ satisfies the differential equation:
				{\small
				\begin{multline}
				\label{eq:differential_varrho_n}
				\frac{d}{dt} \varrho^{(n)}(t) 
				= -i \sum^n_{k=0} \left[ \mathscr{V}^{(n)}(t) , \varrho^{(n-k)}(0)\right]\\ 
				\vspace{-0.8cm} - \sum^n_{k=0} \sum^k_{k'=0} \int^{+\infty}_0 d\tau 
				\left[\mathscr{V}^{(n-k)}(t) ,\left[\mathscr{V}^{(k-k')}(t-\tau) , \varrho^{(k')}(t)\right]\right] \ .
				\end{multline}
				}
\subsection{The zeroth-order approximation}\label{subsec:zeroth-order}
\par In standard perturbation theory, one has to necessarily go to first-order or beyond in order to see the effects of the perturbation term. This is not the case with ACP, where some effects of the perturbation are already manifest at zeroth-order. We demonstrate this point by showing below that if we simply take $\varrho(t)=\varrho^{(0)}(t) + O(\mathscr{X})$, i.e. the zeroth-order approximation, the results we obtain are in excellent agreement with experiments. But before that, we introduce the so-called Holstein-Primakoff\cite{misc:Gyamfi-2019} (HP) representation of spin states in \S \ref{subsec:HP} -- which we will find very useful in subsequent subsections when dealing with multispin systems. The microscopic derivation of the semiclassical GKSL-like equation at zeroth-order is expounded in \S \ref{subsec:MicroscopicD}.
\subsubsection{The Holstein-Primakoff representation and the index compression map $\eta_o$}\label{subsec:HP}
\par The HP representation is just an alternative way of representing spin states in the uncoupled representation. Its main advantage is that all spin projections on the quantization axis take integral values, independent of the spin quantum number\cite{misc:Gyamfi-2019}. 
\par Let $\mathpzc{A}$ be the multiset of spins composing the focused spin system, i.e. $\mathpzc{A}=\{j_1,j_2, \ldots, j_N\}$, where $j_i$ is the spin quantum number of the $i-$th spin. Then, a generic state of the multispin system in the uncoupled representation is of the form: $\ket{j_1,m_1} \ket{j_2,m_2} \cdots \ket{j_N,m_N}$, where $-j_i \leq m_i \leq j_i$ is the spin magnetic quantum number of the $i-$th spin. Subject to the Holstein-Primakoff (HP) transformation, this uncoupled state undergoes the transformation:
				\begin{multline}
				\label{eq:HP_representation_states}
				\ket{j_1,m_1} \ket{j_2,m_2} \cdots \ket{j_N,m_N} \\
				\mapsto \ket{j_1, n_1}\ket{j_2, n_2} \cdots \ket{j_N, n_N}\\
				\equiv \ket{n_1,n_2, \ldots , n_N}
				\end{multline}
where $n_i := j_i-m_i$. It is clear that the $\{n_i\}$ are nonnegative integers, and $0 \leq n_i \leq 2j_i$. The nonnegative integers are said to represent the occupation numbers of the Holstein-Primakoff bosons\citep{misc:Gyamfi-2019}. We observe from Eq. \eqref{eq:HP_representation_states} that in the HP representation, the multispin state $\ket{j_1,m_1} \ket{j_2,m_2} \cdots \ket{j_N,m_N}$ is simply indicated by a multiset of nonnegative integers in the form $\ket{n_1, n_2, \ldots, n_N}$. For a given state $\ket{n_1, n_2, \ldots, n_N}$, the so-called index compression map $\eta_o$\citep{misc:Gyamfi-2018} maps the string of ordered integers $(n_1,n_2,\ldots,n_N)$ to a unique nonnegative integer $\mathbb{n}$ as follows\citep{misc:Gyamfi-2019}:
				\begin{equation}
				\begin{split}
				\mathbb{n} & = \eta_o(n_1,n_2, \ldots, n_N) = \sum^N_{i=1} \mathpzc{W}_i n_i \\
				\mathpzc{W}_i & := \delta_{N,i} + (1-\delta_{N,i}) \sum^{N-i}_{k=1}d_{i+k}
				\end{split}
				\end{equation}
where $d_i \equiv (2j_i+1)$, i.e. the dimension of the $i-$th spin Hilbert space. In other words, for $i\neq N$, $\mathpzc{W}_i$ is the product of all $d_{i'}$ with $i'>i$; while for $i=N$, $\mathpzc{W}_i=1$. For example, given a spin multiset of three spin$-1/2$, i.e. $\mathpzc{A} = \left\lbrace j_1,j_2,j_3 \right\rbrace=\left\lbrace \frac{1}{2},\frac{1}{2},\frac{1}{2} \right\rbrace$, we see that $0 \leq n_1, n_2, n_3 \leq 1$. A generic multispin state of such a system in the uncoupled representation is $\ket{n_1,n_2,n_3}$. With the index compression map $\eta_o$, we can resort to a notation where the state $\ket{n_1,n_2,n_3}$ can be indicated by a unique nonnegative integer in the form of $\ket{\mathbb{n}}$, where 
				\begin{equation}
				\mathbb{n} = 4n_1 + 2n_2 + n_3 \ .
				\end{equation}
So, for example, $\ket{1,0,1} \overset{\eta_o}{\mapsto} \ket{\mathbb{5}}$. In Table \ref{tab:three_spin_half}, we give the basis kets for $\mathpzc{A}=\{\frac{1}{2},\frac{1}{2}, \frac{1}{2} \}$ according to the three representations discussed above.

\begin{table}[htb!]
	\centering
	\begin{tabular}{| c | c | c |}
	\hline	
	$ \ket{m_1,m_2,m_3}$ & $ \ket{n_1,n_2,n_3}$ & $\ket{\mathbb{n}}$ \\
	\hline
	$ \ket{+\frac{1}{2},+\frac{1}{2},+\frac{1}{2}}$ & $ \ket{0,0,0}$ & $\ket{\mathbb{0}}$ \\
	$ \ket{+\frac{1}{2},+\frac{1}{2},-\frac{1}{2}}$ & $ \ket{0,0,1}$ & $\ket{\mathbb{1}}$ \\
	$ \ket{+\frac{1}{2},-\frac{1}{2},+\frac{1}{2}}$ & $ \ket{0,1,0}$ & $\ket{\mathbb{2}}$ \\
	$ \ket{+\frac{1}{2},+\frac{1}{2},-\frac{1}{2}}$ & $ \ket{0,1,1}$ & $\ket{\mathbb{3}}$ \\
	$ \ket{-\frac{1}{2},+\frac{1}{2},+\frac{1}{2}}$ & $ \ket{1,0,0}$ & $\ket{\mathbb{4}}$ \\	
	$ \ket{-\frac{1}{2},+\frac{1}{2},-\frac{1}{2}}$ & $ \ket{1,0,1}$ & $\ket{\mathbb{5}}$ \\
	$ \ket{-\frac{1}{2},-\frac{1}{2},+\frac{1}{2}}$ & $ \ket{1,1,0}$ & $\ket{\mathbb{6}}$ \\
	$ \ket{-\frac{1}{2},-\frac{1}{2},-\frac{1}{2}}$ & $ \ket{1,1,1}$ & $\ket{\mathbb{7}}$ \\
	\hline
	\end{tabular}
	\caption{Spin Hilbert space basis kets of three qubits according to 1) the usual uncoupled representation ($\ket{m_1,m_2,m_3}$), 2) the HP representation ($\ket{n_1,n_2,n_3}$), and 3) their shorthand notation $\ket{\mathbb{n}}$ according to the index compression map $\eta_0$.}
	\label{tab:three_spin_half}
	\end{table}
\par The map $\eta_o$, therefore, encodes the string $(n_1, n_2, \ldots, n_N)$ into a single integer $\mathbb{n}$. We remark that $\eta_o$ is invertible. Thus, given $\mathbb{n}$, one can easily recover the corresponding integers $(n_1,n_2, \ldots, n_N)$ -- if the spin quantum numbers $j_1, j_2, \ldots, j_N$ are known\citep{misc:Gyamfi-2019}. For a given multiset of spins $\mathpzc{A}=\{j_1,j_2, \ldots, j_N\}$, $\mathbb{n}$'s range is $\mathbb{0} \leq \mathbb{n} \leq (D_\mathcal{H}-1)$ -- where $D_\mathcal{H}$ is the dimension of the multispin Hilbert space: $D_\mathcal{H}=\prod^N_{i=1} d_i$. Naturally, $\braket{\mathbb{n}'}{\mathbb{n}}=\delta_{\mathbb{n}',\mathbb{n}}$.
\par  As remarked earlier, the eigenstates of $\mathscr{Z}_o$ are simply the uncoupled multispin states $\{\ket{j_1,m_1}\ket{j_2,m_2}\cdots \ket{j_N,m_N}\}$, which we have just seen  can be simply represented as $\{\ket{\mathbb{n}}\}$. Thus, by virtue of the HP representation and the index compression map $\eta_o$, we have that:
				\begin{subequations}
				\label{eq:eigenvectors_Z_o}
				\begin{align}
				\mathscr{Z}_o \ket{\mathbb{n}} & = \epsilon_\mathbb{n} \ket{\mathbb{n}} \\
				S^z \ket{\mathbb{n}} & = M_\mathbb{n} \ket{\mathbb{n}}
				\end{align}
				\end{subequations}
where,
				\begin{subequations}
				\begin{align}
				\epsilon_\mathbb{n} & := \matrixel{\mathbb{n}}{\mathscr{Z}_o}{\mathbb{n}}\\
				M_\mathbb{n}& := \matrixel{\mathbb{n}}{S^z}{\mathbb{n}} \ .
				\end{align}
				\end{subequations}
$M_\mathbb{n}$ is the multispin state $\ket{\mathbb{n}}$'s total spin magnetic quantum number along the axis of quantization. 
\par In the following, we are going to assume that \emph{if $\epsilon_\mathbb{n'}-\epsilon_\mathbb{n} = \epsilon_\mathbb{n''} -\epsilon_\mathbb{n}$, then $\mathbb{n'}=\mathbb{n''}$}. That is, we are assuming there are no accidental degeneracies of the multispin states in relation to the Hamiltonian $\mathscr{Z}_o$.
\subsubsection{Derivation of zeroth-order approximation to the semiclassical quantum Markovian master equation}\label{subsec:MicroscopicD}
\par As remarked earlier, with the zeroth-order approximation, we are taking $\varrho(t) = \varrho^{(0)}(t) + O(\mathscr{X})$, and -- according to Eq. \eqref{eq:differential_varrho_n} -- $\varrho^{(0)}(t)$ satisfies the differential equation
				\begin{multline}
				\label{eq:diff_varrho_n=0}
				\frac{d}{dt} \varrho^{(0)}(t)  = -i  \left[ \mathscr{V}^{(0)}(t)  , \varrho^{(0)}(0)\right] \\
				-  \int^{+\infty}_0 d\tau \left[\mathscr{V}^{(0)}(t) ,\left[\mathscr{V}^{(0)}(t-\tau) , \varrho^{(0)}(t)\right]\right] \ .
				\end{multline}
We now present a microscopic derivation of a GKSL-like equation for $\varrho^{(0)}(t)$ starting from Eq. \eqref{eq:diff_varrho_n=0}.
\par We begin our derivation by noting that Eq. \eqref{eq:diff_varrho_n=0} may be rewritten in the following form:
			\begin{multline}
			\frac{d}{dt} \varrho^{(0)}(t)  = -i  \left[ \mathscr{V}^{(0)}(t)  , \varrho^{(0)}(0)\right] \\
			- \int^{+\infty}_0 d\tau \left( \left[\mathscr{V}^{(0)\dagger}(t) ,\mathscr{V}^{(0)}(t-\tau)  \varrho^{(0)}(t)\right] + h.c. \right) \label{eq:diff_varrho_n=0_1}
			\end{multline}
where $h.c.$ denotes the presence of the Hermitian conjugate term.
By virtue of Eqs. \eqref{eq:V(t)_interaction_pic} and \eqref{eq:V_expansion_X_n}, 
			\begin{subequations}
			\label{eq:V_(0)_t_1}
			\begin{align}
			\mathscr{V}^{(0)}(t) & =  \mathscr{V}^{(0)}_+(t) + \mathscr{V}^{(0)}_-(t)\label{eq:V_(0)_t_1_1}\\
			\mathscr{V}^{(0)}_\pm(t)& = B_1 \sum_r \  e^{ it(\mathscr{Z}_o\mp\omega_r S^z)} \xi^x e^{- it(\mathscr{Z}_o\mp\omega_r S^z)} \label{eq:V_(0)_t_1_2}\ .
			\end{align}
			\end{subequations}
Eq. \eqref{eq:V_(0)_t_1_1} allows us to divide the r.h.s. of Eq. \eqref{eq:diff_varrho_n=0_1} into contributions from $\mathscr{V}^{(0)}_\pm(t)$, with cross-terms (\emph{i.e.} terms involving the factors $\mathscr{V}^{(0)}_+ \ , \mathscr{V}^{(0)}_-$ -- simultaneously) coming from the second term. Assuming these cross-terms do not contribute, we may reduce Eq. \eqref{eq:diff_varrho_n=0_1}  to the form:
			\begin{multline}
			\frac{d}{dt} \varrho^{(0)}(t)  = -i  \left[ \mathscr{V}^{(0)}_+(t) + \mathscr{V}^{(0)}_-(t)  , \varrho^{(0)}(0)\right] \\
			- \int^{+\infty}_0 d\tau \left( \left[\mathscr{V}^{(0)\dagger}_+(t) ,\mathscr{V}^{(0)}_+(t-\tau)  \varrho^{(0)}(t)\right] + h.c. \right)\\
			- \int^{+\infty}_0 d\tau \left( \left[\mathscr{V}^{(0)\dagger}_-(t) ,\mathscr{V}^{(0)}_-(t-\tau)  \varrho^{(0)}(t)\right] + h.c. \right) \label{eq:diff_varrho_n=0_1_+}
			\end{multline}	
\par It is most convenient, at this point, to proceed with our derivation by expanding the operator $\xi^x$ in the eigenbasis of $\mathscr{Z}_o$ in the following manner:
			\begin{equation}
			\label{eq:xi^x_sum_n_omega_o}
			\xi^x = \sum_{n,\omega_o} \xi^x(n,\omega_o)
			\end{equation}		
where,
			\begin{equation}
			\label{eq:xi_x_n_omega_o}
			\xi^x(n,\omega_o) := \sum_{\mathbb{n},\mathbb{n}'} \ket{\mathbb{n}}\matrixel{\mathbb{n}}{\xi^x}{\mathbb{n}'}\bra{\mathbb{n}'} \delta_{\omega_o,\epsilon_{\mathbb{n}'}-\epsilon_\mathbb{n}} \delta_{n,M_{\mathbb{n}'}-M_\mathbb{n}}
			\end{equation}			
where $\{\ket{\mathbb{n}}\}$ are the eigenvectors of $\mathscr{Z}_o$, Eq. \eqref{eq:eigenvectors_Z_o}. Note that the $\{n\}$ in Eqs. \eqref{eq:xi^x_sum_n_omega_o} and \eqref{eq:xi_x_n_omega_o} are necessarily integers, and the $\{\omega_o\}$ are the pairwise frequency separation between the eigenvalues of $\mathscr{Z}_o$. Interestingly, one also observes that:
			\begin{subequations}
			\label{eq:commutations_xi}
			\begin{align}
			\left[\mathscr{Z}_o, \xi^x(n,\omega_o) \right]& =-\omega_o \ \xi^x(n,\omega_o)\\
			\left[S^z, \xi^x(n,\omega_o) \right]& =-n \ \xi^x(n,\omega_o)
			\end{align}
			\end{subequations}		
while,
			\begin{subequations}
			\label{eq:commutations_xi_dagger}
			\begin{align}
			\left[\mathscr{Z}_o, \xi^{x\dagger}(n,\omega_o) \right]& =\omega_o \ \xi^{x\dagger}(n,\omega_o)\\
			\left[S^z, \xi^{x\dagger}(n,\omega_o) \right]& =n \ \xi^{x\dagger}(n,\omega_o)
			\end{align}
			\end{subequations}	
Thus, from Eqs. \eqref{eq:commutations_xi} and \eqref{eq:commutations_xi_dagger}, the identity:
			\begin{equation}
			\label{eq:xi_dagger_xi_identity}
			\xi^{x\dagger}(n,\omega_o) = \xi^x(-n,-\omega_o)
			\end{equation}	
readily follows. The same identity could have been proved directly from Eq. \eqref{eq:xi_x_n_omega_o}. The commutation relations in Eqs. \eqref{eq:commutations_xi} and \eqref{eq:xi_dagger_xi_identity} indicate that the operator $\xi^x(n,\omega_o)$ is a generalized ladder operator which, when applied to a generic eigenket $\ket{\mathbb{n}}$ of $\mathscr{Z}_o$, transforms $\ket{\mathbb{n}}$ into a weighted sum of other eigenkets of $\mathscr{Z}_o$, who all share the same eigenvalue $\equiv (\epsilon_\mathbb{n}-\omega_o)$, as well as the same total spin magnetic quantum number $\equiv (M_\mathbb{n}-n)$. Likewise, $\xi^{x\dagger}(n,\omega_o)$ transforms $\ket{\mathbb{n}}$ into a sum of other eigenkets of $\mathscr{Z}_o$, each of which is characterized by the same eigenvalue and total spin magnetic quantum numbers; namely,  $(\epsilon_\mathbb{n}+\omega_o)$ and $(M_\mathbb{n} + n)$, respectively. Indeed, it follows from Eq. \eqref{eq:xi^x_sum_n_omega_o} that:
			\begin{subequations}
			\begin{align}
			\xi^x(n,\omega_o) \ket{\mathbb{n}} & = \sum_{\mathbb{n}'}C_{\mathbb{n}',\mathbb{n}}(n,\omega_o) \ket{\mathbb{n}'}\\
			\xi^{x\dagger}(n,\omega_o)\ket{\mathbb{n}} & = \sum_{\mathbb{n}'}C_{\mathbb{n}',\mathbb{n}}(-n,-\omega_o) \ket{\mathbb{n}'}
			\end{align}
			\end{subequations}	
where,
			\begin{equation}
			C_{\mathbb{n}',\mathbb{n}}(n,\omega_o):= \matrixel{\mathbb{n}'}{\xi^x}{\mathbb{n}}\delta_{\omega_o,\epsilon_\mathbb{n}-\epsilon_{\mathbb{n}'}} \delta_{n,M_\mathbb{n}-M_{\mathbb{n}'}} \ .
			\end{equation}		
\par Going back to Eq. \eqref{eq:V_(0)_t_1_2}, we note that with the introduction of the decomposition of $\xi^x$ according to Eq. \eqref{eq:xi^x_sum_n_omega_o}, $\mathscr{V}^{(0)}_\pm(t)$ simplifies to:
			\begin{equation}
			\label{eq:V_(0)_t_2}
			\mathscr{V}^{(0)}_\pm(t) =  B_1 \sum_r \sum_{n,\omega_o}   e^{-it(\omega_o\mp\omega_r n)} \xi^x(n, \omega_o)\ .			
			\end{equation}		
With this new expression for $\mathscr{V}^{(0)}_\pm(t)$, Eq. \eqref{eq:diff_varrho_n=0_1_+} becomes:
			\begin{widetext}
			\begin{multline}
			\label{eq:diff_varrho_n=0_2}
			\frac{d}{dt} \varrho^{(0)}(t)  = -i B_1 \sum_r \sum_{n,\omega_o}   \left(e^{-it(\omega_o-\omega_r n)} + e^{-it(\omega_o+\omega_r n)} \right) \left[\xi^x(n, \omega_o) , \varrho^{(0)}(0)\right] \\
			-  \left(\sum_{r,r'} \sum_{n,\omega_o}\sum_{n',\omega'_o} e^{it\left[ (\omega_o - \omega'_o) - (\omega_r n - \omega_{r'} n')\right]}\ \Gamma(\omega'_o - n'\omega_{r'}) \left[\xi^{x\dagger}(n,\omega_o) ,\xi^{x}(n',\omega'_o)  \varrho^{(0)}(t)\right] + h.c. \right)\\
			-  \left(\sum_{r,r'} \sum_{n,\omega_o}\sum_{n',\omega'_o} e^{it\left[ (\omega_o - \omega'_o) + (\omega_r n - \omega_{r'} n')\right]}\ \Gamma(\omega'_o + n'\omega_{r'}) \left[\xi^{x\dagger}(n,\omega_o) ,\xi^{x}(n',\omega'_o)  \varrho^{(0)}(t)\right] + h.c. \right)
			\end{multline}
			\end{widetext}	
where,
			\begin{equation}
			\label{eq:Gamma}
			\begin{split}
			\Gamma(\omega'_o\pm\omega_{r'}n')& := B^2_1\int^{+\infty}_0 d\tau \ e^{i\tau(\omega'_o \pm \omega_{r'}n')} \\
			& = B^2_1\left[\pi \delta(\omega'_o \pm n'\omega_{r'}) + i \ \mathcal{P}\left( \frac{1}{\omega'_o \pm n'\omega_{r'}} \right)\right]
			\end{split}
			\end{equation}		
and the operation $\mathcal{P}(\bullet)$ indicates Cauchy's Principal Value. 
\par The time-dependent factors in Eq. \eqref{eq:diff_varrho_n=0_2} are all complex exponential functions. For the first term in Eq. \eqref{eq:diff_varrho_n=0_2}, these factors lead to what we have termed the \emph{linear response Hamiltonian}, $H_{LR}(t)$ (see below). Unlike the time-dependent factors in the first term, those in the second and third terms of Eq. \eqref{eq:diff_varrho_n=0_2} are functions of the differences between the eigenvalues $\{\omega_o\}$, as well as weighted differences between the field frequencies $\omega_r$. These factors can be rapidly oscillating when the frequency function multiplying $t$ is far from zero. The contribution of those rapidly oscillating factors to the evolution of $\varrho^{(0)}(t)$ is negligible compared to those terms where the frequency function multiplying $t$ is in the neighborhood of zero. We may therefore discard those rapidly oscillating terms in the second and third terms of Eq. \eqref{eq:diff_varrho_n=0_2} (\emph{secular approximation}). This leads to the condition: $(\omega_o - \omega'_o) \pm (\omega_r n - \omega_{r'} n')=0$, which is easily satisfied if $\omega_o=\omega'_o$, $\omega_r=\omega_{r'}$ and $n=n'$. We emphasize that this is also the only solution compatible with our assumption (see Eq. \eqref{eq:diff_varrho_n=0_1_+}) that the cross-terms involving $\mathscr{V}^{(0)}_+(t)$ and  $\mathscr{V}^{(0)}_-(t)$ do not contribute at second-order in $B_1^2$ to the equation of motion. If we further assume a continuous distribution of the oscillating field's frequency, we get the final result\cite{Note1}:
			\begin{equation}
			\label{eq:diff_varrho_n=0_++}
			\frac{d}{dt} \varrho^{(0)}(t) =  \mathcal{A}(t)\varrho^{(0)}(0) + \mathcal{L}\varrho^{(0)}(t) 
			\end{equation}
where\cite{Note1}:
			\begin{equation}
			\label{eq:mathcal_A}
			\mathcal{A}(t)\varrho^{(0)}(0)   := -i \left[H_{LR}(t), \varrho^{(0)}(0)\right]	
			\end{equation}
with $H_{LR}(t)$, the linear response Hamiltonian, defined as:
			\begin{equation}
			\label{eq:H_LR+}
			H_{LR}(t) := 2B_1 \Re[\varphi_{f}(t)]\sum_{\omega_o} e^{-it\omega_o}\xi^x(+1, \omega_o)+ h.c.
			\end{equation}
We point out that $\varphi_{f}(t)$ is the characteristic function of $\rho_f(\omega')$ --  where $\rho_f(\omega')$  is the probability density function of the frequencies $\omega'$ in the oscillating field, centered on $\omega$\cite{Note1}. Moreover, the superoperator $\mathcal{L}$ in Eq. \eqref{eq:diff_varrho_n=0_++} is the generator of a quantum dynamical semigroup, and is given as follows:
			\begin{equation}
			\label{eq:diff_varrho_n=0_b}
			\mathcal{L}\varrho^{(0)}(t)  := - i \left[H_{LS}, \varrho^{(0)}(t) \right] + \mathcal{D}\left[\varrho^{(0)}(t) \right] 
			\end{equation}
where the first and second terms in Eq. \eqref{eq:diff_varrho_n=0_b} are the unitary evolution and dissipator terms at zeroth-order (in $\mathscr{X}$), respectively. $H_{LS}$ is the Lamb shift Hamiltonian at zeroth-order, and is given by the expression\cite{Note1}:
			\begin{equation}
			H_{LS} = H_{LS+} + H_{LS-}
			\end{equation}
with
			\begin{equation}
			H_{LS\pm}= \pm \pi B^2_1 \sum_{\omega_o} \rho^\succ_f (\pm \omega_o)  \left[\xi^{x\dagger}(+1,\omega_o), \xi^{x}(+1,\omega_o)\right]\label{eq:Lamb-shift}			
			\end{equation}
where $\rho^\succ_f(\pm\omega_o)$ indicates the Hilbert transform\citep{book:King-2009} of $\rho_f(\omega')$ centered on $\pm \omega_o$:
			\begin{equation}
			\rho^\succ_f(\pm \omega_o) :=  \frac{1}{\pi}\mathcal{P} \int^{+\infty}_{-\infty} d\omega' \ \frac{\rho_f(\omega')}{\pm \omega_o - \omega'}
			\end{equation}
where, again, $\mathcal{P}$ represents the Cauchy principal value operator. We remark that $H_{LS\pm}$ originate from $\mathscr{V}^{(0)}_\pm(t)$, respectively. 
\par Similarly, the expression for the dissipator term, $\mathcal{D}\left[\varrho^{(0)}(t) \right]$, in Eq. \eqref{eq:diff_varrho_n=0_b} may also be decomposed into contributions from $\mathscr{V}^{(0)}_\pm(t)$) as follows\cite{Note1}:
			\begin{equation}
			\mathcal{D}\left[\varrho^{(0)}(t) \right] = \mathcal{D}_+\left[\varrho^{(0)}(t) \right] + \mathcal{D}_-\left[\varrho^{(0)}(t) \right]
			\end{equation}
where
			\begin{widetext}
			\begin{multline}
			\label{eq:dissipator_2}
			\mathcal{D}_\pm \left[\varrho^{(0)}(t) \right] 
			= \sum_{\omega_o} 2\pi B^2_1 \rho_f(\pm \omega_o) \left[\xi^{x}(+1,\omega_o)\varrho^{(0)}(t)\xi^{x\dagger}(+1,\omega_o) - \frac{1}{2}\left\lbrace \xi^{x\dagger}(+1,\omega_o)\xi^{x}(+1,\omega_o) , \varrho^{(0)}(t) \right\rbrace  \right]\\
			+ \sum_{\omega_o} 2\pi B^2_1  \rho_f(\pm \omega_o) \left[\xi^{x\dagger}(+1,\omega_o)\varrho^{(0)}(t)\xi^{x}(+1,\omega_o) - \frac{1}{2}\left\lbrace \xi^{x}(+1,\omega_o)\xi^{x\dagger}(+1,\omega_o) , \varrho^{(0)}(t) \right\rbrace  \right] \ .
			\end{multline}
			\end{widetext}
For $\omega_o >0$, the operator $\xi^x(+1,\omega_o)$ may be described as responsible for emission of a photon at the frequency $\omega_o$; while in the case $\omega_o < 0$, the equivalent absorption process may be attributed to the same operator. It is also important to recognize that for a radiation field with a sharply peaked frequency distribution $\rho_f(\omega')$, either $\mathcal{D}_+$ or $\mathcal{D}_-$ will contribute significantly, at resonance, to the dissipator term -- depending on the sign of $\omega_o$; for example, if $\omega_o > 0$, $\mathcal{D}_+$ will dominate.  In any case, it is quite clear from Eq. \eqref{eq:dissipator_2} that the rate of emission and absorption coincide. This comes at no surprise since we treated the oscillating field as classical. Had we treated the field quantum mechanically, we would have found a contribution to the dissipator term due to spontaneous emission
\citep{book:Breuer-2007}.
\par Perhaps, a number of other noteworthy observations are also due here. First of all, we note that despite the fact that the master equation in Eq. \eqref{eq:diff_varrho_n=0_++} is local in time, the presence of the term $\mathcal{A}(t)\varrho^{(0)}(0)$ deprives the quantum map of the semigroup property. However, we may deduce a very important feature of the superoperator $\mathcal{A}(t)$ due to the presence of the characteristic function $\varphi_{f}(t)$ in the expression for the linear response Hamiltonian $H_{LR+}(t)$, Eq. \eqref{eq:H_LR+}. Indeed, for all practical reasons, $\rho_f$ is a real-valued continuous symmetric function in the shift frequency $\delta \omega \equiv \omega - \omega'$ ($\omega'$ is a random frequency), and whose range coincides with $\mathbb{R}$. (Infact, the most common distributions for $\rho_f$ are Lorentzian, Gaussian and Voight -- which are symmetric in $\delta \omega$.) According to Pòlya's theorem\citep{inproceed:Polya-1949}, we must expect the characteristic function of $\rho_f(\delta \omega)$, $\varphi_{f}(t)$, to satisfy the following properties: i) be a real-valued, symmetric and continuous function defined for all real values of $t$; ii) with maximum at $t=0$ -- specifically, $\varphi_f(0)=1$;  iii) with $\lim_{t\to \infty} \varphi_f(t)=0$ and iv) be convex for $t >0$. The most important property to notice here, for the purpose of our discussion, is property iii). Its implication is that when $t \gg \tau_f$ (where $\tau_f$ is the time scale of relaxation of $\varphi_f(t)$), $H_{LR}(t)$ approaches zero, leading therefore the $\mathcal{A}(t)$ term in Eq. \eqref{eq:diff_varrho_n=0_++} to effectively become negligible and the equation becomes a true quantum Markovian master equation, thus restoring the semigroup property. For example, suppose the distribution $\rho_f(\omega')$ is taken to be a Lorentzian, such that:
			\begin{equation}
			\rho_f(\omega') = \frac{1}{\pi}\frac{\left(\frac{\Delta \nu}{2} \right)^2}{\left(\frac{\Delta \nu}{2} \right)^2 + (\omega -\omega')^2}
			\end{equation}
where $\Delta \nu$ is the distribution's full-width at half maximum (FWHM). Then, the corresponding characteristic function, $\varphi_{f}(t)$, is:
			\begin{equation}
			\varphi_{f}(t) = \int^{+\infty}_{-\infty} d\omega' \ e^{ i\omega't}\rho_f(\omega') = e^{ i\omega t} e^{-\left( \frac{\Delta \nu}{2}\right) \abs{t}}
			\end{equation}
hence, with $\tau_f^{-1} = \frac{\Delta \nu}{2}$. 
\par Moreover, we note that the superoperator $\mathcal{A}(t)$ does not operate on $\varrho^{(0)}(t)$, for $t>0$. This means the ME in Eq. \eqref{eq:diff_varrho_n=0_++} may therefore be regarded as a time-dependent Markovian\citep{art:Mazzola_Breuer-2010, art:Dann-2018} one, whereby the time-dependence is not to be found in the rate constants present in the quantum dynamical semigroup's generator $\mathcal{L}$, but in the inhomogeneous term present in the ME which is independent of $\varrho^{(0)}(t>0)$.
\par Formally solving Eq \eqref{eq:diff_varrho_n=0_++} for $\varrho^{(0)}(t)$, we find that:
			\begin{equation}
			\varrho^{(0)}(t) = \Lambda(t)\varrho^{(0)}(0)
			\end{equation}		
with 	
			\begin{equation}
			\label{eq:Lambda_t_general}
			\Lambda(t) := e^{\mathcal{L} t} + \int^t_0 dt'\  e^{\mathcal{L}(t-t')} \mathcal{A}(t') \ .
			\end{equation}	
Surely, with $\mathcal{A}(t)=0$, the map $\Lambda(t)$ is clearly CPT. However, $\Lambda(t)$ --  as given in Eq. \eqref{eq:Lambda_t_general} -- is in general non-CP
\cite{Note1} (though it remains trace-preserving). It is interesting to note that $\Lambda(t)$ replicates the same structure of linear non-CP maps present in the literature\citep{art:Buzek-2001, art:Carteret-2008} (where both the focus system and the environment are quantized). Namely, it is the sum of two terms: the first term is a CPT map, while the second (or inhomogeneous) term is a more complicated map which is traceless\citep{art:Buzek-2001, art:Carteret-2008, art:Cory-2004}. Furthermore, $\Lambda(t)$ may also be written as the difference between two CP maps (see [\onlinecite[\S II.C]{Note1}]) -- which again proves $\Lambda(t)$ to be non-CP\cite{art:Yu-2000, inbook:Shabani_co-2014}. It is important to notice that, here in the zeroth-order, the theory restricts the domain of the map $\Lambda(t)$, $\{ \varrho^{(0)}(0) \}$ ,  to $\varrho^{(0)}(0)=\frac{e^{-\beta \mathscr{Z}_o}}{\mbox{Tr}\left[e^{-\beta \mathscr{Z}_o} \right]}$, Eq. \eqref{eq:initial_rho_n=0} -- which is a Boltzmann state. Failing to do so may lead to unphysical results (see [\onlinecite[\S II.E]{Note1}]). Moreover, there are strong indications that $\Lambda(t)$ is positive on its domain. We provide a heuristic argument in favor of this proposition in [\onlinecite[\S II.D]{Note1}]. A more careful analysis of the positivity of $\Lambda(t)$ needs to be done.
\par As we shall see below, despite the fact that it breaks the CPT property of $\Lambda(t)$ in Eq. \eqref{eq:Lambda_t_general}, the presence of $\mathcal{A}(t)$ puts the predictions of the theory impressively in line with experimental results; on the other hand, the predictions become quite incompatible with experimental results if we choose to neglect the inhomegeneous term. 
\subsection{Applications}\label{subsec:Applications}
\par In this section, we apply the basic results of the preceding section to some specific problems in CW magnetic resonance. In \S \ref{subsec:spin-half}, we study the CW experiment with an ensemble of spin-$1/2$ particles. There, we also draw a connection to linear response theory which is further expounded on in \S \ref{subsec:LRT}. In \S \ref{subsec:spectrum} we show how simple stick-plot ESR spectra can be obtained from the theory, and illustrate the method with some specific molecules.
\subsubsection{CW experiment with an ensemble of spin-$1/2$ particles}\label{subsec:spin-half}
\par To illustrate the application of the above equations and concepts, it may help to consider an ensemble of particles of spin-$1/2$. In this case, there won't be any need of ACP, given that we do not have the spin-spin coupling term, $H_{spin-spin}$, Eq. \eqref{eq:H_spin-spin}, to begin with. Indeed, all the above equations, from Eq. \eqref{eq:def_H_o} to Eq. \eqref{eq:Lambda_t_general}, apply here -- only that we just need to set $H_{spin-spin} \to 0$, which also translates into setting $\mathscr{X} \to 0$ and $\mathscr{Z}_o \to B_o \xi^z$, Eq. \eqref{eq:def_Z_X}.  With these transformations, we note that the non-homogeneous system of triangular differential equations in Eq. \eqref{eq:differential_varrho_n} reduces to a single differential equation, namely Eq. \eqref{eq:diff_varrho_n=0}. Actually, the overall result of putting $H_{spin-spin} \to 0$ is that $\varrho(t)$ coincides now exactly with $\varrho^{(0)}(t)$, and Eq. \eqref{eq:diff_varrho_n=0_++} becomes:
			\begin{equation}
			\label{eq:diff_varrho_n=0_++_spin-1/2}
			\frac{d}{dt} \varrho(t) = \mathcal{A}(t)\varrho(0)  + \mathcal{L}\varrho(t)
			\end{equation}
which is now, by default, the relative exact equation of motion for $\varrho(t)$ -- \emph{i.e.} within the very limitations of the approximations and assumptions which led to Eq. \eqref{eq:diff_varrho_n=0_++}. 
\par All that is left now is to determine the operators $\{\xi^x(+1,\omega_o)\}$ and substitute them into Eqs. \eqref{eq:mathcal_A}, \eqref{eq:Lamb-shift} and \eqref{eq:dissipator_2}. To make the connection with some known results in the literature more intelligible, we shall make use of the Pauli matrices:
			\begin{equation}
			\sigma_1 = \begin{pmatrix}
			0 & 1\\
			1 & 0
			\end{pmatrix} , \quad \sigma_2 = \begin{pmatrix}
			0 & -i\\
			i & 0
			\end{pmatrix}, \quad \sigma_3 = \begin{pmatrix}
			1 & 0 \\
			0 & -1
			\end{pmatrix} \ .
			\end{equation}
In this regard, $\mathscr{Z}_o$, as stated in the previous paragraph, becomes:
			\begin{equation}
			\mathscr{Z}_o = -\frac{\gamma B_o}{2} \sigma_3
			\end{equation}
with $\gamma$ being the gyromagnetic ratio of the spin-$1/2$, and $B_o$ the magnitude of the steady magnetic field. The eigenkets of $\mathscr{Z}_o$ are: $\ket{+\frac{1}{2}}$ and $\ket{-\frac{1}{2}}$, which according to the HP representation and the index compression map $\eta_o$, may also be denoted as $\ket{\mathbb{0}}$ and $\ket{\mathbb{1}}$, respectively. With just two eigenkets, it is easily observed that the possible frequency difference $\omega_o$ between these eigenkets are: $\omega_o \in \{0, \pm \gamma B_o\}$.   
\par Furthermore, it is clear from the definition given in Eq.\eqref{eq:xi_x_n_omega_o} that:
			\begin{equation}
			\label{eq:xi_x_Pauli}
			\xi^x(+1,\omega_o) = -\frac{\gamma}{2}\sigma_- , \quad \xi^x(-1,-\omega_o) =-\frac{\gamma}{2}\sigma_+
			\end{equation}
when $\omega_o = -\gamma B_o$, i.e. the Larmor frequency -- with
			\begin{equation}
			\label{eq:sigma_pm}
			\sigma_\pm = \frac{1}{2}\left(\sigma_1 \pm i \sigma_2 \right) \ . 
			\end{equation}
However, when $\omega_o=0$, $\xi^x(n,\omega_o)=0$ for all possible values of $n$. We may, therefore, in the following, intend $\omega_o$ as the Larmor frequency without loss of generality. Interestingly, given Eq. \eqref{eq:xi_x_Pauli}, it readily follows from Eqs. \eqref{eq:commutations_xi} and \eqref{eq:commutations_xi_dagger} that:
			\begin{equation}
			\left[\mathscr{Z}_o,\sigma_\pm \right]=\pm \omega_o\ \sigma_\pm
			\end{equation}	
and the initial density matrix $\varrho(0)$ is
			\begin{equation}
			\varrho(0) = \frac{e^{-\beta \mathscr{Z}_o}}{\mbox{Tr}[e^{-\beta \mathscr{Z}_o}]}	
			\end{equation}					 
\par With the identity of the operators $\{\xi^x(n,\omega_o)\}$ in our possession, thanks to Eq. \eqref{eq:xi_x_Pauli}, we now determine $H_{LR}(t)$, $H_{LS}$ and $\mathcal{D}_\pm\left[\bullet \right]$  through Eqs. \eqref{eq:H_LR+}, \eqref{eq:Lamb-shift} and \eqref{eq:dissipator_2}, respectively. Namely:
			\begin{equation}
			\label{eq:H_LR+-1/2}
			H_{LR}(t)  = \Re[\varphi_{f}(t)]\omega_1  e^{-it\omega_o} \sigma_- + h.c.
			\end{equation}

			\begin{equation}
			\label{eq:H_LS_def}
			\begin{split}
			H_{LS} & = \pi \left(\frac{\omega_1}{2} \right)^2 \left[\rho^\succ_f(\omega_o) - \rho^\succ_f(-\omega_o)\right] \left[ \sigma_-, \sigma_+\right]\\
			& = - \pi \left(\frac{\omega_1}{2} \right)^2 \left[\rho^\succ_f(\omega_o) - \rho^\succ_f(-\omega_o)\right] \ \sigma_3
			\end{split}
			\end{equation}
with $\omega_1 := -\gamma B_1$. Regarding the dissipator term, it easily follows from Eqs. \eqref{eq:dissipator_2} and \eqref{eq:xi_x_Pauli} that:
			\begin{multline}
			\label{eq:dissipator_3}
			\mathcal{D}_\pm\left[\varrho(t) \right] \\
			=   2\pi\left(\frac{ \omega_1}{2}\right)^2 \rho_f(\pm \omega_o) \left[\sigma_-\varrho(t)\sigma_+ - \frac{1}{2}\left\lbrace\sigma_+\sigma_-, \varrho(t) \right\rbrace  \right]\\
			+2\pi\left(\frac{ \omega_1}{2}\right)^2 \rho_f(\pm \omega_o) \left[\sigma_+\varrho(t)\sigma_- - \frac{1}{2}\left\lbrace\sigma_-\sigma_+, \varrho(t) \right\rbrace  \right]
			\end{multline}
This dissipator component is reminiscent of its analogue in the quantum optical master equation (see Eq. (3.219) of [\onlinecite{book:Breuer-2007}]). The only difference between the two are the rate constants. In the quantum optical master equation\citep{book:Breuer-2007}, the radiation field is treated quantum mechanically, thus allowing to account for the rate of spontaneous emission, besides the usual stimulated rates of emission and absorption. In Eqs. \eqref{eq:dissipator_2} and \eqref{eq:dissipator_3}, there is no trace of rate of spontaneous emission simply because the oscillating field is treated classically.
\par Regarding the rates of stimulated emission and absorption, it is worth observing that when $\omega_o >0$ (i.e. $\gamma < 0$), the first term in Eq. \eqref{eq:dissipator_3} describes the stimulated emission process  $\ket{-\frac{1}{2}} \leftarrow \ket{+\frac{1}{2}}$ at the rate $\Gamma_{-\frac{1}{2}, +\frac{1}{2}}$, while the second term of the same equation describes the stimulated absorption process  $\ket{+\frac{1}{2}} \leftarrow \ket{-\frac{1}{2}}$ at the rate $\Gamma_{+\frac{1}{2} , -\frac{1}{2}}$. The rate of both processes coincides:
			\begin{subequations}
			\begin{align}
			\Gamma_{-\frac{1}{2}, +\frac{1}{2}}  = \Gamma_{+\frac{1}{2} , -\frac{1}{2}} & = 2\pi\left(\frac{ \omega_1}{2}\right)^2 \left[ \rho_f(\omega_o) +  \rho_f(-\omega_o)\right]	\nonumber \\
			& \equiv \Gamma(\omega_o) \ .
			\end{align}
			\end{subequations}
If we write 
			\begin{equation}
			\label{eq:varrho_qubit}
			\varrho(t) = \frac{1}{2}\left[ \mathbb{I} + 2\avg{\sigma_-(t)}'\sigma_+ +  2\avg{\sigma_+(t)}'\sigma_- + \avg{\sigma_3(t)}' \sigma_3\right]
			\end{equation}
where $\avg{\sigma_\pm(t)}'\equiv \mbox{Tr}[\varrho(t)\sigma_\pm]$ and $\avg{\sigma_3(t)}'\equiv \mbox{Tr}[\varrho(t)\sigma_3]$, (with $\avg{x}'$ we mean the expectation value of $x$ in the interaction picture; the unprimed $\avg{x}$, until otherwise stated, will be the analogous expectation value in the Schr\"odinger picture) then Eqs. \eqref{eq:varrho_qubit} and \eqref{eq:diff_varrho_n=0_++_spin-1/2} lead to the differential equations
			\begin{subequations}
			\label{eq:qubit_diff_eqs_in_int_pic}
			\begin{align}
			\frac{d}{dt}\avg{\sigma_3(t)}' & = - 2\Gamma(\omega_o) \avg{\sigma_3(t)}' \\
			\frac{d}{dt}\avg{\sigma_+(t)}' & = i\omega_1\tanh(\beta \omega_o /2) \Re[\varphi_{f}(t)]  e^{-it\omega_o} \nonumber \\
			& - \left[ \Gamma(\omega_o) + i \varpi(\omega_o)  \right] \avg{\sigma_+(t)}'
			\end{align}
			\end{subequations}
with $\varpi(\omega_o)$ given by the expression:
			\begin{equation}
			\frac{\varpi(\omega_o)}{2} \equiv \pi   \left(\frac{\omega_1}{2} \right)^2 \left[ \rho^\succ_f(\omega_o) - \rho^\succ_f(-\omega_o)\right] \equiv  \omega_{LS}(\omega_o)
			\end{equation}
where $\omega_{LS}$ is the `Lamb shift rate' and originates from $H_{LS}$, Eq. \eqref{eq:H_LS_def}. 
In the following we shall simply write $\Gamma$ and $\varpi$, but their dependence on $\omega_o$ must be kept in mind. Naturally, $\avg{\sigma_-(t)}'=\avg{\sigma_+(t)}'^*$. 
\par Bearing in mind that in the Schr\"odinger picture $\avg{\sigma_\pm(t)}' \mapsto \avg{\sigma_\pm(t)} e^{\mp i \omega_o t}$, we deduce that the stationary solutions for $\avg{\sigma_3(t)}$ and $\avg{\sigma_+(t)}$ -- namely, $\avg{\sigma_3(t)}_s$ and $\avg{\sigma_+(t)}_s$ -- in the Schr\"odinger picture are:
			\begin{subequations}
			\begin{align}
			\avg{\sigma_3(t)}_s & =  0 \\
			\avg{\sigma_\pm (t)}_s &  = -\frac{\omega_1 \tanh(\beta \omega_o/2)}{[\omega_o-\varpi]\pm i\Gamma} \Re[\varphi(t)]
			\end{align}
			\end{subequations}
Since $\Re[\varphi(t)]$ tends to zero as $t \to +\infty$ according to Pòlya's theorem, it follows that $\avg{\sigma_\pm(t)}$ also tend to zero as $t \to +\infty$. Putting this observation together with Eq. \eqref{eq:varrho_qubit}, we readily come to the conclusion that the equilibrium state, $\varrho_{eq}= \rho_{eq}$, of the qubit ensemble in our CW experiment is the corresponding maximally mixed state, i.e. $\rho_{eq}=\frac{1}{2}\mathbb{I}$. This also means that the associated quantum map contracts the Bloch sphere to a point, namely the center. Indeed, solving the differential equations in Eq. \eqref{eq:qubit_diff_eqs_in_int_pic} in the Schr\"odinger picture, we find:
			\begin{equation}
			\label{eq:sigma_3_t}
			\avg{\sigma_3(t)} = -\tanh(\beta \omega_o/2)\ e^{-2\Gamma t}
			\end{equation}
			\begin{equation}
			\avg{\sigma_\pm(t)} = \Re \avg{\sigma_+(t)} \pm i \Im \avg{\sigma_+(t)}
			\end{equation}
where
			\begin{multline}
			\Re \avg{\sigma_+(t)} = e^{-\Gamma t}\omega_1 \tanh(\beta \omega_o/2) \\
						\times	 \int^t_0 dt'\ e^{\Gamma t'}\sin[(\varpi-\omega_o)(t-t')]\ \Re[\varphi_f(t')]  
			\end{multline}
			\begin{multline}
			\Im \avg{\sigma_+(t)} = e^{-\Gamma t}\omega_1 \tanh(\beta \omega_o/2) \\
					\times	\int^t_0 dt'\ e^{\Gamma t'}\cos[(\varpi-\omega_o)(t-t')]\ \Re[\varphi_f(t')]  \ .
			\end{multline}
It is thus evident that $\avg{\sigma_\pm(t)}$ inexorably approaches zero as $t \to +\infty$. From Eq. \eqref{eq:sigma_pm}, we also find that
			\begin{equation}
			\label{eq:sigma_1_and_2_t}
			\avg{\sigma_1(t)} = 2 \Re \avg{\sigma_+(t)} \ \quad \ \text{and} \ \quad \avg{\sigma_2(t)} = 2 \Im \avg{\sigma_+(t)} \ .
			\end{equation}
\par To illustrate how the present model of semiclassical quantum Markovian master equation, Eq. \eqref{eq:diff_varrho_n=0_++_spin-1/2}, also entails known results in LRT (linear response theory), we determine the dynamical structure factor\citep{book:Giuliani-2005} of $\xi^x(+1,\omega_o)$. Since $\xi^x(+1,\omega_o)$ is proportional to $\sigma_-$, Eq. \eqref{eq:xi_x_Pauli}, we can equally concentrate below on the dynamic structure factor of $\sigma_-$, $S_{\sigma_-\sigma_+}(\omega')$:
			\begin{equation}
			\label{eq:sigma_pm_dynamic_structure_factor}
			S_{\sigma_-\sigma_+}(\omega') = \frac{1}{2\pi} \int^{+\infty}_{-\infty} dt \ e^{i\omega' t} \avg{\sigma_-(t)\sigma_+}
			\end{equation}
where the correlation $\avg{\sigma_-(t)\sigma_+} \equiv \mbox{Tr}[\sigma_-(t)\sigma_+ \rho(0)]$ is evaluated in the Heisenberg picture. We therefore need to determine how our quantum map evolves $\sigma_-$ in the Heisenberg picture.
\par Given that any qubit operator $X$ can be written as
			\begin{equation}
			X = \frac{1}{2} \left[c_0(0)\mathbb{I} + c_1(0) \sigma_1 + c_2(0) \sigma_2 + c_3(0)\sigma_3 \right]
			\end{equation}
with 	
			\begin{equation}
			c_i(0) = \mbox{Tr}[X\sigma_i] \ \qquad \ i=\{0,1,2,3\}
			\end{equation}
(where $\sigma_0 \equiv \mathbb{I}$), it naturally follows that the Heisenberg picture evolution of $X$, $X(t)$, must be of the form
			\begin{equation}
			X(t) = \frac{1}{2} \left[c_0(t)\mathbb{I} + c_1(t) \sigma_1 + c_2(t) \sigma_2 + c_3(t)\sigma_3 \right]
			\end{equation}
where the coefficients $c_i(t)$ are to be determined through the condition
			\begin{equation}
			\mbox{Tr}[\rho(t) X] = \mbox{Tr}[\rho(0)X(t)] \ .
			\end{equation}
After some algebra, one finds that with $\rho(0)=\frac{e^{-\beta \mathscr{Z}_o}}{\mbox{Tr}[e^{-\beta \mathscr{Z}_o}]}=\frac{1}{2}[\mathbb{I}+\avg{\sigma_3(0)}\sigma_3]$, where $\avg{\sigma_3(0)}= -\tanh(\beta \omega_o/2)$,
			\begin{equation}
			\begin{pmatrix}c_0(t) \\ c_1(t) \\ c_2(t) \\ c_3(t) \end{pmatrix} = \begin{pmatrix} 
			\kappa_0(t) & \kappa_1(t) & \kappa_2(t) & \kappa_3(t) \\
			\kappa_1(t) & \kappa_0(t) & -i\kappa_3(t) & i\kappa_2(t)\\ 		
			\kappa_2(t) & i\kappa_3(t) & \kappa_0(t) & -i\kappa_1(t)\\	
			\kappa_3(t) & -i\kappa_2(t) & i\kappa_1(t) & \kappa_0(t)\\
\end{pmatrix} 
			\begin{pmatrix}c'_0(t) \\ c'_1(t) \\ c'_2(t) \\ c'_3(t) \end{pmatrix}
			\end{equation}
where
			\begin{equation}
			\begin{pmatrix}c'_0(t) \\ c'_1(t) \\ c'_2(t) \\ c'_3(t) \end{pmatrix} = \begin{pmatrix} 
			1 & 0 & 0 & 0 \\
			0 & \cos(\omega_o t) & \sin(\omega_o t) & 0\\ 		
			0 & -\sin(\omega_o t) & \cos(\omega_o t) & 0\\	
			0 & 0 & 0 & 1\\
\end{pmatrix} 
			\begin{pmatrix}c_0(0) \\ c_1(0) \\ c_2(0) \\ c_3(0) \end{pmatrix}
			\end{equation}
and
			\begin{subequations}
			\begin{align}
			\kappa_0(t) & \equiv \frac{1 -  \avg{\sigma_3(0)}'\avg{\sigma_3(t)}'}{1 - \avg{\sigma_3(0)}'^2} \\
			\kappa_1(t) & \equiv i \left[ \frac{ \avg{\sigma_2(t)}' \avg{\sigma_3(0)}' -i\avg{\sigma_1(t)}'   }{1 - \avg{\sigma_3(0)}'^2} \right]\\
			\kappa_2(t) & \equiv -i \left[\frac{\avg{\sigma_1(t)}'\avg{\sigma_3(0) }' + i\avg{\sigma_2(t)}'   }{1 - \avg{\sigma_3(0)}'^2} \right]\\
			\kappa_3(t) & \equiv \frac{\avg{\sigma_3(t)}'  -\avg{\sigma_3(0)}' }{1 - \avg{\sigma_3(0)}'^2}
			\end{align}
			\end{subequations}
and where, as remarked earlier -- Eq. \eqref{eq:varrho_qubit} -- $\avg{\sigma_i(t)}'$ is the equivalent of $\avg{\sigma_i(t)}$ in the interaction picture. Namely,
			\begin{equation}
			\begin{pmatrix} \avg{\sigma_1(t)}' \\ \avg{\sigma_2(t)}' \\ \avg{\sigma_3(t)}' \end{pmatrix} = \begin{pmatrix} 
			 \cos(\omega_o t) & \sin(\omega_o t) & 0\\ 		
			 -\sin(\omega_o t) & \cos(\omega_o t) & 0\\	
			 0 & 0 & 1\\
\end{pmatrix} 
			\begin{pmatrix} \avg{\sigma_1(t)} \\ \avg{\sigma_2(t)} \\ \avg{\sigma_3(t)} \end{pmatrix}
			\end{equation}			
where $\avg{\sigma_3(t)}$ is given by Eq. \eqref{eq:sigma_3_t} and, $\avg{\sigma_1(t)}$ and $\avg{\sigma_2(t)}$ are defined in Eq. \eqref{eq:sigma_1_and_2_t}.
\par It follows then that with the initial density matrix $\rho(0)=\frac{e^{-\beta \mathscr{Z}_o}}{\mbox{Tr}[e^{-\beta \mathscr{Z}_o}]}$, given a qubit operator $X$, its corresponding dynamical structure factor $S_{XX^\dagger}(\omega')= \frac{1}{2\pi} \int^{+\infty}_{-\infty} dt \ e^{i\omega' t} \avg{X(t)X^\dagger}$ in terms of the coefficients $c_i(0), c_i(t)$ is:
			\begin{multline}
			\label{eq:dynamic_structure_factor_general}
			S_{XX^\dagger}(\omega')  =   \frac{1}{4} \bigg[ \bigg(\mathscr{C}_{3,0}(\omega') +\mathscr{C}_{0,3}(\omega')\bigg) \avg{\sigma_3(0)} + \mathscr{C}_{0,0}(\omega') \\
			+   \mathscr{C}_{3,3}(\omega') 
			 +4\frac{e^{\beta \omega_o /2} }{\mbox{Tr}[e^{-\beta \mathscr{Z}_o}]}\mathscr{C}_{+,+}(\omega') + 4\frac{e^{-\beta \omega_o /2} }{\mbox{Tr}[e^{-\beta \mathscr{Z}_o}]}\mathscr{C}_{-,-}(\omega')\bigg]
			\end{multline}
where
			\begin{equation}
			\mathscr{C}_{\nu,\mu}(\omega') = \frac{1}{2\pi} \int^{+\infty}_{-\infty} dt \ e^{i\omega' t} c_\nu(t)c^{*}_\mu(0) 
			\end{equation}
with
			\begin{equation}
			c_\pm(t)= \frac{1}{2}\left[c_1(t) \pm ic_2(t) \right] \ \qquad \ (t \geq 0) \ .
			\end{equation}
Going back to $S_{\sigma_-\sigma_+}(t)$, Eq. \eqref{eq:sigma_pm_dynamic_structure_factor}, we may apply Eq. \eqref{eq:dynamic_structure_factor_general} by setting $X=\sigma_-$. This reduces Eq. \eqref{eq:dynamic_structure_factor_general} to
			\begin{equation}
			\label{eq:S_-_+}
			\begin{split}
			 & S_{\sigma_-\sigma_+}(\omega')  
			  =  \frac{e^{\beta \omega_o /2} }{\mbox{Tr}[e^{-\beta \mathscr{Z}_o}]}\mathscr{C}_{+,+}(\omega') \\
			& = \frac{e^{\beta \omega_o /2} }{\mbox{Tr}[e^{-\beta \mathscr{Z}_o}]}\frac{1}{2\pi}\int^{+\infty}_{-\infty}dt \ e^{i(\omega'-\omega_o) t}\frac{1 - \avg{\sigma_3(t)}}{1 - \avg{\sigma_3(0)}}\\
			&= \frac{1 }{2}\bigg[  \delta(\omega'-\omega_o)  
			+  \tanh(\beta \omega_o/2) \cdot \frac{1}{\pi} \frac{2\Gamma}{(2\Gamma)^2 + (\omega'-\omega_o)^2} \bigg] \ .
			\end{split}
			\end{equation}
This tells us that the spectrum of the time-dependent fluctuations of $\sigma_-$ has a Lorentzian profile centered on the Larmor frequency $\omega_o$ and its HWHM (half width at half maximum) is twice the transition rate between the two level system. The temperature dependence of the profile is embodied in the factor $\tanh(\beta \omega_o/2)$.
\par Conversely, if we put $X=\sigma_+$, it turns out from Eq. \eqref{eq:dynamic_structure_factor_general} that
			\begin{equation}
			\label{eq:S_+_-}
			\begin{split}
			 & S_{\sigma_+\sigma_-}(\omega')  =  \frac{e^{-\beta \omega_o /2} }{\mbox{Tr}[e^{-\beta \mathscr{Z}_o}]}\mathscr{C}_{-,-}(\omega') \\
			& = \frac{e^{-\beta \omega_o /2} }{\mbox{Tr}[e^{-\beta \mathscr{Z}_o}]}\frac{1}{2\pi}\int^{+\infty}_{-\infty}dt \ e^{i(\omega'+\omega_o) t}\frac{1 + \avg{\sigma_3(t)}}{1 + \avg{\sigma_3(0)}}\\
			&= \frac{1 }{2}\bigg[  \delta(\omega'+\omega_o)  
			- \tanh(\beta \omega_o/2) \cdot \frac{1}{\pi} \frac{2\Gamma}{(2\Gamma)^2 + (\omega'+\omega_o)^2} \bigg] \ .
			\end{split}
			\end{equation}
It is evident that for $\omega_o >0$, $S_{\sigma_-\sigma_+}(\omega')$ is the stimulated absorption spectrum and $S_{\sigma_+\sigma_-}(\omega')$ is the stimulated emission spectrum\citep{book:Giuliani-2005}. At any rate, the two spectra are related through the relation
			\begin{multline}
			\label{eq:fluctuations_relation}
			S_{\sigma_-\sigma_+}(\omega')  - S_{\sigma_+ \sigma_-}(-\omega') \\
			= \tanh(\beta \omega_o/2) \cdot \frac{1}{\pi} \frac{2\Gamma}{(2\Gamma)^2 + (\omega'-\omega_o)^2}   \ .
			\end{multline}
This relation differs from the one given in LRT (see, for example, Eq. (3.73) of  [\onlinecite{book:Giuliani-2005}]). To recover the LRT limit from Eq. \eqref{eq:fluctuations_relation}, we need to remember that the whole edifice of LRT rests on the adiabatic process assumption, where it is assumed the interaction between the system and the bath is weak enough so as not to change appreciably the occupation probabilities of the initial state  of the system, and that the system has remained in its equilibrium state in  the far past prior to its encounter with the bath. This is equivalent to taking the limit $\mathcal{L} \to 0$  in our master equation, Eq. \eqref{eq:diff_varrho_n=0_++_spin-1/2}, and having $t_o \text{(the initial time)} \to -\infty$. In other terms, we can get the adiabatic process limit of a given dynamic structure factor obtained from the full solution of the quantum Markovian ME by taking the limits $\Gamma \to 0$ and $\omega_{LS}\to 0$. Now, if we introduce
			\begin{subequations}
			\begin{align}
			S^{ad}_{\sigma_-\sigma_+}(\omega') & \equiv \lim_{\omega_{LS},\Gamma \to 0} S_{\sigma_-\sigma_+}(\omega')  \\
			 S^{ad}_{\sigma_+\sigma_-}(\omega') & \equiv \lim_{\omega_{LS},\Gamma \to 0} S_{\sigma_+\sigma_-}(\omega')
			\end{align}
			\end{subequations}
-- where $S^{ad}_{\sigma_-\sigma_+}(\omega')$ and $S^{ad}_{\sigma_+\sigma_-}(\omega')$ indicate the adiabatic process limits of $S_{\sigma_-\sigma_+}(\omega')$ and $S_{\sigma_+\sigma_-}(\omega')$, respectively -- then we readily derive from Eqs. \eqref{eq:S_-_+} and \eqref{eq:S_+_-} that
			\begin{equation}
			\label{eq:fluctuation_dissipation_1}
			S^{ad}_{\sigma_+ \sigma_-}(-\omega') = e^{-\beta \omega_o} S^{ad}_{\sigma_-\sigma_+}(\omega')
			\end{equation}
which is the relation between the two spectra according to LRT\citep{book:Giuliani-2005}. 
\par It is certainly worth noting that without the linear response Hamiltonian $H_{LR}(t)$, Eq. \eqref{eq:H_LR+-1/2}, we would have had $\avg{\sigma_\pm(t)}=0 \ \forall t\geq 0$ since $\avg{\sigma_\pm(0)}=0$ (see Eq. \eqref{eq:qubit_diff_eqs_in_int_pic}). As a consequence, the dynamic structure factors $S_{\sigma_\mp\sigma_\pm}(\omega')$, for example, would result to be always zero, meaning we do not observe any absorption or emission spectrum -- but that would have been contrary to experimental observations. Given that $H_{LR}(t)$ stems from the superoperator $\mathcal{A}(t)$, Eq. \eqref{eq:mathcal_A}, the observation just made reinforces the assertion that $\mathcal{A}(t)$ is central to the theory and cannot be simply -- generally speaking -- put to zero (or ignored) in Eq. \eqref{eq:diff_varrho_n=0_++_spin-1/2} in order to make the quantum map $\Lambda(t)$, Eq. \eqref{eq:Lambda_t_general}, CPT.
\subsubsection{More on the Linear Response Theory connection}\label{subsec:LRT}
\par Since the seminal work of Davies and Spohn\citep{art:Davies_Spohn-1978}, there have been a number of works
\citep{art:Chetrite-2012, art:Zanardi-2012, art:Mehboudi-2018} aimed at revisiting at least some aspects of Kubo's LRT\citep{book:Kubo-2012,book:Giuliani-2005,inbook:Gumbs-2011,book:VanVliet-2008,art:Marini-2008} from the perspective of quantum (non-)Markovian master equations. Much of these efforts have been concentrated on deriving fluctuation-dissipation theorems. We showed in the last subsection how the theory and formalism we are developing entail some of the key results in LRT. In this subsection, we shall try to extend some of the results obtained above to the general case of an arbitrary multispin system. We also show how the celebrated LRT fluctuation-dissipation theorem
\citep{book:Giuliani-2005} is easily derived as a limit case.
\par We begin with the following formal solution to Eq. \eqref{eq:diff_varrho_n=0_++}:
			\begin{equation}
			\varrho^{(0)}(t) - \varrho^{(0)}(0)=  \int^t_0 dt' \ \mathcal{A}(t')\varrho^{(0)}(0) + \int^t_0 dt' \ \mathcal{L}\varrho^{(0)}(t')    
			\end{equation}
which in the Schr\"odinger picture (see Eq. \eqref{eq:varrho_A^0}) becomes:
			\begin{multline}
			\label{eq:rho_t_LRT}
			\rho^{(0)}(t) - \rho^{(0)}(0)\\
			= -i \int^t_0 dt' \ \left[e^{-it\mathscr{Z}_o}H_{LR}(t')e^{it\mathscr{Z}_o}, \rho^{(0)}(0)\right]  \\
			+ \int^t_0 dt' \ \mathcal{L}\left[e^{-i(t-t')\mathscr{Z}_o}\rho^{(0)}(t')e^{i(t-t')\mathscr{Z}_o}\right] 
			\end{multline}
Thus, for any given operator $X$ of the multispin system, we derive from Eq. \eqref{eq:rho_t_LRT} that:
			\begin{multline}
			\label{eq:X_LRT}
			\avg{X^{(0)}(t)} - \avg{X^{(0)}(0)}\\
			= \int^t_0 dt' \ \mbox{Tr}\left(X\mathcal{L}\left[e^{-i(t-t')\mathscr{Z}_o}\rho^{(0)}(t')e^{i(t-t')\mathscr{Z}_o}\right] \right)\\
			-i\int^t_0 dt'\ \avg{\left[X,e^{-it\mathscr{Z}_o}H_{LR}(t')e^{it\mathscr{Z}_o} \right]}_o \\
			\end{multline}
where $\avg{X^{(0)}(t)}  \equiv \mbox{Tr}\left[X \rho^{(0)}(t)\right]$ and $\avg{F}_o \equiv \mbox{Tr}\left[ F \rho^{(0)}(0) \right]$. Naturally, $\avg{X^{(0)}(t)} $ is the expectation value of the observable $X$ at zeroth-order in $\mathscr{X}$. We now show that if we take the limit $\mathcal{L}\to 0$ (or if the Lindblad term is negligible with respect to the linear response term), we get a richer version of linear response theory\citep{book:Kubo-2012,book:Giuliani-2005,inbook:Gumbs-2011,book:VanVliet-2008,art:Marini-2008} from Eq. \eqref{eq:X_LRT}. We may term this the `LRT limit' of Eq. \eqref{eq:X_LRT}. Indeed, with $\mathcal{L}\to 0$, Eq. \eqref{eq:X_LRT} reduces to:
			\begin{multline}
			\label{eq:L_to_zero}
			\avg{X^{(0)}(t)} - \avg{X^{(0)}(0)}\\
			= -i\int^t_0 dt'\ \avg{\left[X,e^{-it\mathscr{Z}_o}H_{LR}(t')e^{it\mathscr{Z}_o} \right]}_o \ .
			\end{multline}
In any case, after some simple rearrangements, it can be shown that:
			\begin{multline}
			\label{eq:X_LRT_2}
			-i\int^t_0 dt'\ \avg{\left[X,e^{-it\mathscr{Z}_o}H_{LR}(t')e^{it\mathscr{Z}_o} \right]}_o \\
			 = 2B_1\int^{+\infty}_{-\infty}d\omega' \rho_f(\omega')
			 \sum_{\omega_o}  \frac{e^{it\omega_o}}{2} \big[ \chi_{\omega_o,\infty}(\omega') 
			+  \chi_{\omega_o,t}(\omega') \big]\\ + c.c.
			\end{multline}
with
			\begin{subequations}
			\begin{align}
			\chi_{\omega_o,\infty}(\omega') & \equiv \chi_{+,\omega_o,\infty}(\omega') + \chi_{-,\omega_o,\infty}(\omega')\\
			\chi_{\omega_o,t}(\omega') & \equiv \chi_{+,\omega_o,t}(\omega') + \chi_{-,\omega_o,t}(\omega')
			\end{align}
			\end{subequations}
where $\chi_{\pm,\omega_o,\infty}(\omega')$ and $\chi_{\pm,\omega_o,t}(\omega')$ together define the \emph{linear} frequency-dependent response function of the spin system's observable $X$ to the perturbation defined by the coupling between the rotating fields, $\mathbf{B}_{1,\pm}(t)$, and $\gv{\xi}^x$. In particular, $\chi_{\pm,\omega_o,\infty}(\omega')$, which define the steady-state limit of the linear response, are given by the expressions:
			\begin{equation}
			\label{eq:chi_+_infinito_LRT}
			\chi_{\pm,\omega_o,\infty}(\omega'):=\lim_{\eta \to 0^+} \frac{\avg{\left[X,\xi^x(+1, \omega_o)\right]}_o}{(\pm \omega'-\omega_o)+i\eta}  
			\end{equation}
while for $\chi_{\pm,\omega_o,t}(\omega')$, the transient elements of the response function, we have:
			\begin{equation}
			\chi_{\pm ,\omega_o,t}(\omega'):= -\lim_{\eta \to 0^+}   \frac{\avg{\left[X,\xi^x(+1, \omega_o)\right]}_o}{(\pm \omega'-\omega_o)+i\eta}e^{\left[i(\pm \omega'-\omega_o)-\eta\right] t}
			\end{equation}
where, 
			\begin{equation}
			\avg{\left[X,\xi^x(+1, \omega_o)\right]}_o := \mbox{Tr}\left(\left[X,\xi^x(+1, \omega_o)\right]\rho^{(0)}(0)\right)\ .
			\end{equation}
It is interesting to observe that $\chi_{\pm,\omega_o,\infty}(\omega')$, Eq. \eqref{eq:chi_+_infinito_LRT},  are precisely the usual frequency response functions one would define for the pair of operators $X$ and $\xi^x(+1, \omega_o)$ in LRT under the so-called Lehmann representation\citep{book:Giuliani-2005}. In LRT, one obtains $\chi_{\pm,\omega_o,\infty}(\omega')$ under the assumption of an adiabatic process\citep{book:Kubo-2012, book:Giuliani-2005, inbook:Gumbs-2011, book:VanVliet-2008, art:Marini-2008}, where, as remarked earlier, can be seen as taking the limits $\mathcal{L}\to 0$, $t_o \to -\infty$. It is crucial to note here that while the first limit alone retains the transient components of the response functions, the introduction of the second limit dumps these. For steady-state experiments like CW magnetic resonance, taking the limit $t_o \to -\infty$ is acceptable since it practically translates into the limit $t \to +\infty$, i.e. the steady-state limit; but for transient experiments like pulsed NMR and ESR, these transient response functions play a crucial role in the theory. Interestingly, the integral $\int^{+\infty}_{-\infty}d\omega' \rho_f(\omega') \chi_{\omega_o,t}(\omega')$ is an exponentially decaying oscillatory function, with decay rate $\tau_f$. So, for $t\gg \tau_f$, the transient component of the response function becomes negligible and Eq. \eqref{eq:X_LRT_2} reduces to the form:
			\begin{multline}
			\label{eq:X_LRT_3}
			-i\int^t_0 dt'\ \avg{\left[X,e^{-it\mathscr{Z}_o}H_{LR}(t')e^{it\mathscr{Z}_o} \right]}_o \\
			= 2B_1\int^{+\infty}_{-\infty}d\omega' \rho_f(\omega')\sum_{\omega_o} \left[ \frac{e^{it\omega_o}}{2}  \chi_{\omega_o,\infty}(\omega') +  c.c.\right] \\
			= 2B_1\int^{+\infty}_{-\infty}d\omega' \rho_f(\omega')\sum_{\omega_o} \bigg[ \cos(\omega_o t)  \chi^{'}_{\omega_o,\infty}(\omega') \\
			+  \sin(\omega_o t)  \chi^{''}_{\omega_o,\infty}(\omega') \bigg] 		
			\end{multline}
where:
			\begin{subequations}
			\label{eq:chi_prime_d-prime}
			\begin{align}
			\chi^{'}_{\omega_o,\infty}(\omega') & := \Re \chi_{\omega_o,\infty}(\omega') \nonumber \\
			& = \Re \chi_{+,\omega_o,\infty}(\omega') + \Re \chi_{-,\omega_o,\infty}(\omega')\\
			-\chi^{''}_{\omega_o,\infty}(\omega') & := \Im \chi_{\omega_o,\infty}(\omega') \nonumber\\
			& =\Im \chi_{+,\omega_o,\infty}(\omega') + \Im \chi_{-,\omega_o,\infty}(\omega')
			\end{align}
			\end{subequations}
and
			\begin{multline}
			\Re \chi_{\pm,\omega_o,\infty}(\omega') = \left[\mathcal{P}\left(\frac{\Re \avg{\left[X^\dagger(+1, \omega_o),\xi^x(+1, \omega_o)\right]}_o}{\pm \omega'-\omega_o} \right) \right. \\
			 + \pi \delta(\pm \omega'-\omega_o)\Im \avg{\left[X^\dagger(+1, \omega_o),\xi^x(+1, \omega_o)\right]}_o \Bigg]
			\end{multline}
			\begin{multline}
			\Im \chi_{\pm ,\omega_o,\infty}(\omega') = \left[\mathcal{P}\left(\frac{\Im \avg{\left[X^\dagger(+1, \omega_o),\xi^x(+1, \omega_o)\right]}_o}{\pm \omega'-\omega_o} \right) \right. \\
			 - \pi \delta(\pm \omega'-\omega_o)\Re \avg{\left[X^\dagger(+1, \omega_o),\xi^x(+1, \omega_o)\right]}_o \Bigg]
			\end{multline}
In these last two equations, we have made use of the fact that $\avg{\left[X,\xi^x(+1, \omega_o)\right]}_o = \avg{\left[X^\dagger(+1,\omega_o),\xi^x(+1, \omega_o)\right]}_o$. (The expression for $X(+1,\omega_o)$ follows from Eq. \eqref{eq:xi_x_n_omega_o}.) This identity clearly indicates that the linear response functions $\chi_{\pm,\omega_o,\infty}(\omega')$ and $\chi_{\pm,\omega_o,t}(\omega')$ become identically zero if $X$ is not proportional to the $q=\pm 1$ component of a spherical tensor of rank $k\geq 1$ like $\xi^x(+1, \omega_o)$. For, example, if $X=\xi^z$, which is the zeroth-component of a rank $k=1$ tensor,    $\xi^z(+1, \omega_o)=0$, therefore,  $\avg{\left[\xi^z,\xi^x(+1, \omega_o)\right]}_o=0$ as a consequence. 
\par Moreover, one can easily show that $\Re \chi_{\omega_o,\infty}(\omega') $ is the Hilbert transform of $\Im \chi_{\omega_o,\infty}(\omega') $ (which also means the latter is the Hilbert transform of the former multiplied by $(-1)$), as one would expect from the Kramers-Kr\"onig dispersion relation\citep{book:Giuliani-2005,inbook:Gumbs-2011}.
\par It is worth noting that the LRT limit of Eq. \eqref{eq:X_LRT} is always real, independent of whether $X$ is real Hermitian or not. This is clear from Eq. \eqref{eq:X_LRT_2}. In the limit case whereby $X$ is real Hermitian, $\avg{\left[X^\dagger(+1,\omega_o),\xi^x(+1, \omega_o)\right]}_o$ is also real and we get:
			\begin{equation}
			\label{eq:Re_chi_real}
			\Re \chi_{\pm,\omega_o,\infty}(\omega') = \mathcal{P}\frac{ \avg{\left[X^\dagger(+1, \omega_o),\xi^x(+1, \omega_o)\right]}_o}{\pm \omega'-\omega_o} 
			\end{equation}
and 
			\begin{multline}
			\label{eq:Im_chi_real}
			\Im \chi_{\pm ,\omega_o,\infty}(\omega')\\
			 = - \pi \delta(\pm \omega'-\omega_o) \avg{\left[X^\dagger(+1, \omega_o),\xi^x(+1, \omega_o)\right]}_o \ .
			\end{multline}
These expressions coincide with those from LRT\citep{book:Giuliani-2005}. Many of the results known in LRT can also be derived from the above relations but care must be taken when comparing these relations. Most importantly, one must note that the spin operators here, i.e. $\xi^x(+1, \omega_o)$, which get coupled to the relevant part of the external field are not Hermitian.
\par For example, in the qubit problem discussed in the previous subsection, we saw that $\xi^x(+1,\omega_o)$ is given by Eq. \eqref{eq:xi_x_Pauli}. Then, for the response of $\mu^x$ (the magnetic moment operator of the qubit system along the direction $x$) to the coupling of $\xi^x(+1,\omega_o)$ to $\mathbf{B}_1(t)$, we simply put $X = \mu^x=-\xi^x$ (see Eq. \eqref{eq:xi_alpha}) . And since in this case $\avg{\left[X^\dagger(+1, \omega_o),\xi^x(+1, \omega_o)\right]}_o = -(\gamma/2)^2\avg{\left[\sigma_+,\sigma_-\right]}_o$  is real, Eqs. \eqref{eq:Re_chi_real} and \eqref{eq:Im_chi_real} also hold. Now, if -- in order to keep tradition with the notations in use in LRT\cite{book:Giuliani-2005} -- we write $ \chi_{\omega_o,\infty}(\omega') \equiv (\gamma/2)^2 \chi_{\sigma_- \sigma_+}(\omega')$, we see that for positive $\omega_o$, it follows from Eq. \eqref{eq:Im_chi_real} that
			\begin{equation}
			\label{eq:Im_chi_sigma_-_+_qubit}
			\begin{split}
			\Im \chi_{\sigma_-\sigma_+}(\omega') &  = \pi \delta(\omega'-\omega_o) \avg{\left[\sigma_+,\sigma_-\right]}_o\\
			& = -\pi \delta(\omega'-\omega_o) \tanh(\beta \omega_o/2) \ .
			\end{split}
			\end{equation}
It is interesting to observe that if we now take the adiabatic process limit of Eq. \eqref{eq:fluctuations_relation}, and then make use of Eqs. \eqref{eq:fluctuation_dissipation_1} and \eqref{eq:Im_chi_sigma_-_+_qubit}, we end up with
			\begin{equation}
			 \Im \chi_{\sigma_-\sigma_+}(\omega') = -\pi \left( 1  - e^{-\beta \omega_o} \right)S^{ad}_{\sigma_-\sigma_+}(\omega')
			\end{equation}
which is the celebrated LRT fluctuation-dissipation theorem\citep{book:Giuliani-2005}.
\subsubsection{Theoretical zeroth-order spectrum in the adiabatic process limit}\label{subsec:spectrum}
\par At this stage, it should be evident to the Reader that, when it comes to theoretical spectra, the Lindblad superoperator $\mathcal{L}$ in our ME, Eq. \eqref{eq:diff_varrho_n=0_++}, has the role of primarily allowing for a finite width of the resonance lines. This is quite evident, for example, from the qubit dynamic structure factors we derived in Eqs. \eqref{eq:S_-_+} and \eqref{eq:S_+_-}. On the other hand, by taking the adiabatic process limit, we shrink the finite-width resonance lines to Dirac-delta-like ones. This tells us that if we are only interested in determining the position and intensity of the resonance lines, then we just have to consider the adiabatic process limit of our ME in Eq. \eqref{eq:diff_varrho_n=0_++}. In the following, our object of concern will be the position and intensity of the resonance lines so we consider the adiabatic process limit of Eq. \eqref{eq:diff_varrho_n=0_++}.
\par In CW experiments, the signal detected is the induced voltage $\mathscr{E}$ in the receiver coil caused by the time variation of the magnetic flux therein due to the relaxation of the sample's spin polarization vector. It is known that	\citep{book:Corio-1966, book:Eaton-2010}:
				\begin{equation}
				\label{eq:induced_voltage}
				\mathscr{E} \propto \frac{\mathscr{P}}{B_1}
				\end{equation}				 
where $\mathscr{P}$ is the power absorbed per unit volume of the sample. In the steady-state limit, 
				\begin{equation}
				\label{eq:power}
				\mathscr{P}=\lim_{t\to +\infty}\frac{1}{t}\int^t_0 dt' \ \frac{dE(t')}{dt'}
				\end{equation}
where $\frac{dE(t)}{dt}$ is the rate at which the spin system absorbs energy from the oscillating field, per unit volume of sample. With $\mathbf{B}_{1}(t)$ given by Eq. \eqref{eq:B_1(t)}, we have that,
				\begin{multline}
				\label{eq:rate_of_absorption}
				\frac{dE(t)}{dt} = \mathbf{B}_{1}(t) \cdot \frac{d\avg{\mathbf{M}(t)}}{dt} \\ 					=2\sum_{\omega_{r}}B_1 \cos(\omega_r t) \frac{d\avg{M^{(0)}_x(t)}}{dt}
				+ O(\mathscr{X})
				\end{multline}
where $\avg{M^{(0)}_x(t)}$ is the zeroth-order approximation of the expectation value of the operator $M_x$ ($\equiv (N/V)\mu^x = -(N/V)\xi^x$; $(N/V)$ is the number of particles per unit volume), i.e. the magnetization operator along the $x-$ axis.
Hence, it follows from Eq. \eqref{eq:power} that:
				\begin{equation}
				\mathscr{P}  = \lim_{t\to +\infty} \frac{1}{t} \int^t_0 dt'\  2  \sum_{\omega_{r}} B_1 \cos(\omega_r t') \frac{d\avg{M^{(0)}_x(t')}}{dt'}
				\end{equation}
or alternatively,
				\begin{multline}
				\label{eq:E_bar_2}
				\mathscr{P} =   4  B^2_1 \sum_{\omega_{r'}} \sum_{\omega_r}  \sum_{\omega_o} \omega_o   \\
				 \times \lim_{t\to +\infty} \frac{1}{t} \int^t_o dt' \bigg[- \cos(\omega_{r'}t')\sin(\omega_o t') \chi'_{\omega_o,\infty}(\omega_r) \\
				 + \cos(\omega_{r'}t')\cos(\omega_o t') \chi''_{\omega_o,\infty}(\omega_r)\bigg] \ .
				\end{multline}
Here, we have used the steady-state limit of $\avg{M_x^{(0)}(t)} $, derived directly from Eqs. \eqref{eq:L_to_zero} and \eqref{eq:X_LRT_3}:
			\begin{multline}
			\label{eq:M_x_expectation_steady_limit}
			\avg{M_x^{(0)}(t)}  = 2B_1\int^{+\infty}_{-\infty}d\omega' \rho_f(\omega')\\
			\times \sum_{\omega_o} \left[ \cos(\omega_o t)  \chi^{'}_{\omega_o,\infty}(\omega') 
			+   \sin(\omega_o t)  \chi^{''}_{\omega_o,\infty}(\omega') \right] 
			\end{multline}
where, as usual, the relations in Eq. \eqref{eq:chi_prime_d-prime} hold, with
			\begin{equation}
			\chi'_{\pm,\omega_o,\infty}(\omega') = \mathcal{P}\frac{ \avg{\left[M_x^\dagger(+1, \omega_o),\xi^x(+1, \omega_o)\right]}_o}{\pm \omega'-\omega_o} 
			\end{equation}
			\begin{equation}
			\label{eq:chi''_for_res}
			\chi''_{\pm,\omega_o,\infty}(\omega') = \pi \delta(\pm \omega'-\omega_o) \avg{\left[M_x^\dagger(+1, \omega_o),\xi^x(+1, \omega_o)\right]}_o
			\end{equation}
and
			\begin{multline}
			\label{eq:avg_M_x_P_o}
			\avg{\left[M_x^\dagger(+1, \omega_o),\xi^x(+1, \omega_o)\right]}_o \\
			= \left(\frac{N}{V} \right) \sum_{\mathbb{n},\mathbb{n}'} \ \left|\matrixel{\mathbb{n}}{\xi^x(+1, \omega_o) }{\mathbb{n}'}\right|^2 \left(P^{(0)}_\mathbb{n} - P^{(0)}_{\mathbb{n}'} \right) 
			\end{multline}
-- where
			\begin{equation}
			P^{(0)}_\mathbb{n} \equiv \matrixel{\mathbb{n}}{\rho^{(0)}(0)}{\mathbb{n}} =  \frac{e^{-\beta \epsilon_{\mathbb{n}}}}{\mbox{Tr}\left[e^{-\beta \mathscr{Z}_o} \right]} \ .
			\end{equation}
The fact that we are working in the steady-state limit is confirmed by the absence of the transient magnetic susceptibilities $ \chi'_{\omega_o,t},  \chi''_{\omega_o,t}$ in Eq. \eqref{eq:E_bar_2}. In obtaining Eq. \eqref{eq:E_bar_2}, we have made use of Eq. \eqref{eq:M_x_expectation_steady_limit} and the transformation $\int^{+\infty}_{-\infty} d\omega' \rho_f(\omega') \mapsto \sum_{\omega_r}$. Upon going through with the integration in Eq. \eqref{eq:E_bar_2}, we see that the terms proportional to $\chi'_{\omega_o,\infty}$ either vanish or may be neglected altogether for all practical purposes. Thus,
				\begin{multline}
				\label{eq:E_bar_3}
				\mathscr{P} =  2  B^2_1 \sum_{\omega_{r'}} \sum_{\omega_r}  \sum_{\omega_o} \omega_o 
				 \lim_{t \to +\infty} \bigg(   \mbox{sinc}\left[(\omega_o+\omega_{r'})t \right] 
				 \\ + \mbox{sinc}\left[(\omega_o-\omega_{r'})t \right] \bigg)\chi''_{\omega_o,\infty}(\omega_r)\ .
				\end{multline}
But the fact that $\omega_{r'}$ is always positive, together with the presence of the Dirac delta function in the definition of $\chi''_{\pm,\omega_o,\infty}$, Eq. \eqref{eq:chi''_for_res}, also makes the term proportional to $ \mbox{sinc}\left[(\omega_o\pm \omega_{r'})t \right]$ negligible with respect to the $ \mbox{sinc}\left[(\omega_o\mp \omega_{r'})t \right]$ term for positive and negative $\omega_o$, respectively, reducing, therefore, Eq. \eqref{eq:E_bar_3} to:
				\begin{multline}
				\label{eq:Power_4}
				\frac{\mathscr{P}}{2 B^2_1}  =    \sum_{\omega_r}  \sum_{\omega_o} \omega_o \ \lim_{t \to +\infty} \bigg( \mbox{sinc}\left[(\omega_o-\omega_{r})t \right] \chi''_{+,\omega_o,\infty}(\omega_r) \\
				 +  \mbox{sinc}\left[(\omega_o+\omega_{r})t \right] \chi''_{-,\omega_o,\infty}(\omega_r)\bigg) \\
				  =  \sum_{\omega_o} \omega_o  \int d\omega' \rho_f(\omega') \  \lim_{t \to +\infty} \bigg( \mbox{sinc}\left[(\omega_o-\omega')t \right] \chi''_{+,\omega_o,\infty}(\omega') \\
				  + \mbox{sinc}\left[(\omega_o+\omega')t \right] \chi''_{-,\omega_o,\infty}(\omega')\bigg) \\
				  = \frac{\mathscr{P}_+}{2B_1^2} + \frac{\mathscr{P}_-}{2B_1^2}
				 \end{multline}
where
				\begin{equation}
				\mathscr{P}_\pm = 2 \pi B^2_1 \sum_{\omega_o} \omega_o \ \rho_f(\pm \omega_o)   \avg{\left[M_x^\dagger(+1, \omega_o),\xi^x(+1, \omega_o)\right]}_o \ .
				\end{equation}
Making use of the relation in Eq. \eqref{eq:avg_M_x_P_o}, we may rewrite Eq. \eqref{eq:Power_4} as:
				\begin{equation}
				\label{eq:Power_5}
				\mathscr{P} = \left(\frac{N}{V} \right) \sum_{\omega_o}\omega_o \sum_{\mathbb{n},\mathbb{n}'} \left(P^{(0)}_\mathbb{n} - P^{(0)}_{\mathbb{n}'} \right)\Gamma_{\mathbb{n},\mathbb{n}'}(\omega_o)
				\end{equation}
where
				\begin{subequations}
				\begin{align}
				\Gamma_{\mathbb{n},\mathbb{n}'}(\omega_o) & = \Gamma^+_{\mathbb{n},\mathbb{n}'}(\omega_o) + \Gamma^-_{\mathbb{n},\mathbb{n}'}(\omega_o) \label{eq:Transition_rate_a}\\
				\Gamma^\pm_{\mathbb{n},\mathbb{n}'}(\omega_o) & := 2\pi B_1^2 \rho_f(\pm \omega_o) \left\vert \matrixel{\mathbb{n}}{\xi^x(+1,\omega_o)}{\mathbb{n}'}\right\vert^2 \ . \label{eq:Transition_rate}
				\end{align}
				\end{subequations}
$\Gamma^\pm_{\mathbb{n},\mathbb{n}'}(\omega_o)$ is the transition rate between the states $\ket{\mathbb{n}}$ and $\ket{\mathbb{n'}}$ at the frequency $\pm \omega_o$, respectively -- with $\omega_o=\epsilon_{\mathbb{n}'} - \epsilon_\mathbb{n}$.
The expression for $\Gamma_{\mathbb{n},\mathbb{n}'}(\omega_o)$ in Eq. \eqref{eq:Transition_rate_a} can be easily derived from Eq. \eqref{eq:diff_varrho_n=0_++} if one expands $ \dot{\varrho}^{(0)}_{\mathbb{n},\mathbb{n}}(t)\equiv \matrixel{\mathbb{n}}{d\varrho^{(0)}(t)/dt}{\mathbb{n}}$, and compares the result with the general expression for the Pauli master equation\citep{book:Breuer-2007,book:Alicki-2007}. 
\par If the applied oscillating field has a frequency distribution $\rho_f(\omega')$ sharply peaked at $\omega$, and  $\omega=\pm \omega'_o$, where $\omega'_o$ is one of the allowed transition frequencies of the spin system, we see that only the frequency $\omega'_o$ in the summation $\sum_{\omega_o}$, Eq. \eqref{eq:Power_5}, survives. In this case, one of $\Gamma^\pm_{\mathbb{n},\mathbb{n}'}(\omega'_o)$ dominates the other in the sum in Eq. \eqref{eq:Transition_rate_a}. For example, $\Gamma^+_{\mathbb{n},\mathbb{n}'}(\omega'_o) \gg \Gamma^-_{\mathbb{n},\mathbb{n}'}(\omega'_o)$ if $\omega'_o$ is positive.
\par If we apply the high temperature approximation\citep{book:Gamliel_Levanon-1995}, i.e. $\rho^{(0)}(0) \approx D^{-1}_S \left(\mathbb{I} -\beta \mathscr{Z}_o \right)$, with $\beta= \frac{1}{k_B T}$ ($D_S$ is the dimension of the multispin Hilbert space) in Eq. \eqref{eq:Power_5}, and introduce the obtained zeroth-order approximation for $\mathscr{P}$ (i.e. Eq. \eqref{eq:Power_5}) into Eq. \eqref{eq:induced_voltage}, we get:
				\begin{equation}
				\mathscr{E}  \propto \sum_{\omega_o}\Omega(\omega_o) \cdot \frac{1}{D_S} \sum_{\mathbb{n},{\mathbb{n}'}}  \left\vert \matrixel{\mathbb{n}}{\xi^x(+1,\omega_o)}{\mathbb{n}'}\right\vert^2 \label{eq:rate_of_absorption_3c}
				\end{equation}
				
				\begin{equation}
				\Omega(\omega_o) := \left( \frac{N}{V}\right)2\pi  B_1 \omega^2_o   \beta \left[\rho_f(\omega_o) + \rho_f(-\omega_o)\right]\ .
				\end{equation}
\par For fixed $\Omega(\omega_o)$, we note from Eq. \eqref{eq:rate_of_absorption_3c} that the intensity of the resonance signal at $\omega=\pm \omega_o$, $\texttt{Int}(\omega_o)$, is:
				\begin{equation}
				\label{eq:intensity_formula}
				\texttt{Int}(\omega_o)\propto \frac{1}{D_S}\sum_{\mathbb{n},\mathbb{n}'}  \left\vert \matrixel{\mathbb{n}}{\xi^x(+1,\omega_o)}{\mathbb{n}'}\right\vert^2 \ .
				\end{equation}
This means at zeroth-order, all pair of states $\{\ket{\mathbb{n}},\ket{\mathbb{n}'}\}$ such that $\epsilon_{\mathbb{n}'} - \epsilon_{\mathbb{n}}=\omega_o$ and $M_{\mathbb{n}'} - M_{\mathbb{n}}=+1$, contribute to $\texttt{Int}(\omega_o)$. The operator $\xi^x$ being the sum of single spin operators, Eq. \eqref{eq:xi_alpha}, we also note that another implication of the observation just made is the following: for a specific choice of $\ket{\mathbb{n}},\ket{\mathbb{n}'}$, the nonzero value of $\matrixel{\mathbb{n}}{\xi^x(+1,\omega_o)}{\mathbb{n}'}$ can only be interpreted as a transition involving a single spin of the multiset $\mathpzc{A}=\{j_1,j_2,\ldots,j_N\}$ -- at a given time. Let's call this spin the "resonance spin". We therefore find from Eqs. \eqref{eq:eigenvectors_Z_o} and \eqref{eq:def_Z_X} that:
				\begin{equation}
				\label{eq:epsilon_difference}
				\epsilon_{\mathbb{n}'} - \epsilon_\mathbb{n} = -\gamma_i B_o + \sum_{k \neq i} T_{ik} m_{z,k} 
				\end{equation}
where the resonance spin is assumed to be the $i-$th element of the multiset $\mathpzc{A}=\{j_1,j_2,\ldots,j_N\}$, and $m_{z,k}= \matrixel{\mathbb{n}}{S^z_k}{\mathbb{n}}=\matrixel{\mathbb{n}'}{S^z_k}{\mathbb{n}'}$ is the magnetic quantum number of the $k-$th spin according to the multispin states $\ket{\mathbb{n}}$ and $\ket{\mathbb{n}'}$.
\par To proceed with our discussion, it is much helpful to reconsider the multiset of spins $\mathpzc{A}=\{j_1,j_2,\ldots,j_N\}$ in terms of equivalent spins. By "equivalent" spins we mean a submultiset $\mathpzc{A}'$ of $\mathpzc{A}$ whose elements cannot be distinguished from each other on the basis of their coupling tensors with other spins and external fields\cite{misc:Gyamfi-2019}. Say the resonance spin $i$ belongs to the group of equivalent spins labeled $\alpha$. In terms of equivalent spins, we may rewrite Eq. \eqref{eq:epsilon_difference} as:
				\begin{equation}
				\label{eq:epsilon_difference_b}
				\epsilon_{\mathbb{n}'} - \epsilon_\mathbb{n} = -\gamma_\alpha B_o + \sum_{\alpha' \neq \alpha } T_{\alpha \alpha'} M_{z,\alpha'} \ .
				\end{equation}
$M_{z,\alpha'}$ is the total spin magnetic quantum number of the $\alpha'-$th group of equivalent spins according to the multispin states $\ket{\mathbb{n}}$ and $\ket{\mathbb{n}'}$. In the HP representation, we may express $M_{z,\alpha'}$ as\citep{misc:Gyamfi-2019,art:Gyamfi-2018}:
				\begin{equation}
				M_{z,\alpha'} = J_{\alpha'} - n_{\alpha'}
				\end{equation}
where $J_{\alpha'}=j_{\alpha'}N_{\alpha'}$ is the total spin of the $\alpha'-$th group of equivalent spins ($N_{\alpha'}$ is the cardinality of the group and $j_{\alpha'}$ is the spin quantum number of each member of the group, assumed to be identical for all). The integer $n_{\alpha'}$ is the total number of HP bosons distributed among the $N_{\alpha'}$ spins of the $\alpha'-$th group\cite{art:Gyamfi-2018}.  
\par Suppose in our CW experiment, the frequency $\omega$ and the amplitude $B_1$ of the rotating field are fixed, with $\omega=\omega_o$, while the steady magnetic field $B_o$ is slowly tuned to resonance. If the frequency gap  $\epsilon_{\mathbb{n}'} - \epsilon_\mathbb{n} $ in Eq. \eqref{eq:epsilon_difference_b} coincides with $\omega_o$, then we derive from the latter that the resonance condition in terms of the amplitude of the steady field is (see also [\onlinecite{book:Wertz-1986}]):
				\begin{equation}
				\label{eq:B_o_resonance}
				B_o = B_\alpha(\omega_o) + \sum_{\alpha' \neq \alpha} \lambda_{\alpha \alpha'} n_{\alpha'}
				\end{equation}
where,
				\begin{subequations}
				\begin{align}
				B_\alpha(\omega_o) & := -\frac{\omega_o}{\gamma_\alpha} - \sum_{\alpha' \neq \alpha} \lambda_{\alpha \alpha'} J_{\alpha'} \label{eq:B_alpha}\\
				\lambda_{\alpha \alpha'} & := -\frac{T_{\alpha \alpha'}}{\gamma_\alpha} \ .
				\end{align}
				\end{subequations}
The absolute values $\vert \lambda_{\alpha,\alpha'} \vert $ are the so-called \emph{splitting constants} in magnetic resonance. $B_\alpha(\omega_o)$ is the position of the resonance line originating from the transition event (involving obviously the resonance spin) whereby all the spins of the other groups are with their maximum spin projection along the quantization axis (i.e. $n_{\alpha'}=0 \ \forall \alpha'$). As we can see from Eq. \eqref{eq:B_alpha}, for a fixed frequency $\omega_o$, $B_\alpha(\omega_o)$ is constant. Below, we shall use $B_\alpha(\omega_o)$ as the reference for the other resonance lines, i.e. we shall be considering $\Delta B \equiv [B_o- B_\alpha(\omega_o)]$.
\par We readily infer from Eq. \eqref{eq:B_o_resonance} that, in this weak coupling limit under consideration, the resonance position $B_o$ depends on the total HP bosons' occupation numbers $\{n_{\alpha'}\} $. The intensity of the detected magnetic resonance signal -- in reference to the resonance spin group $\alpha$ -- is proportional to the degeneracy $C_{\alpha,\{n_{\alpha'}\}}$ of the collection $\{n_{\alpha'}\}$. Indeed, if $c_{n_{\alpha'}}$ is the degeneracy of HP boson's total occupation number $n_{\alpha'}$ for the $\alpha'-$th group of equivalent spins (i.e. $c_{n_{\alpha'}}$ is the number of different distinct ways of distributing a total of $n_{\alpha'}$ HP bosons between the $N_{\alpha'}$ spins of the $\alpha'-$th group; or, in other words, $c_{n_{\alpha'}}$ is the number of distinct ways of configuring the spins of the $\alpha'-$th group so as to obtain a total spin magnetic quantum number of $M_{z,\alpha'}=(J_{\alpha'}- n_{\alpha'})$), then it readily follows that:
				\begin{equation}
				\label{eq:C_alpha}
				C_{\alpha,\{n_{\alpha'}\}} = \prod_{\alpha'} c_{n_{\alpha'}}
				\end{equation}
since distinct groups of equivalent spins are independent of each other. It is easy to prove that the generating function for the integers $C_{\alpha,\{n_{\alpha'}\}}$ is the polynomial $P_\alpha(\gv{x})$\citep{art:Gyamfi-2018,misc:Gyamfi-2019}:
				\begin{equation}
				\label{eq:generating_poly}
				\begin{split}
				P_\alpha(\gv{x}) & := \prod_{\alpha' \neq \alpha} \left(1+x_{\alpha'} + \ldots + x^{2j_{\alpha'}}_{\alpha'}\right)^{N_{\alpha'}}\\
				& = \sum_{\{n_{\alpha'}\}} C_{\alpha,\{n_{\alpha'}\}} \prod_{\alpha'} x^{n_{\alpha'}}_{\alpha'} \ .
				\end{split}
				\end{equation}
Having determined $C_{\alpha,\{n_{\alpha'}\}}$, we may now go back to Eq. \eqref{eq:rate_of_absorption_3c}. It is now clear that for fixed $\omega=\pm \omega_o$, if the transition frequency for the $\alpha-$th group of equivalent spins happen to coincide with $\omega_o$, then the induced voltage in the receiver coil is:
				\begin{equation}
				\label{eq:rate_of_absorption_4}
				\mathscr{E} \propto \Omega(\omega_o)\frac{\gamma^2_\alpha N_\alpha}{D_{\alpha, S}}  \binom{2j_\alpha +2}{3} C_{\alpha,\{n_{\alpha'}\}} \ .
				\end{equation}
Consequently,
				\begin{equation}
				\label{eq:intensity_2}
				\mathtt{Int}(B_o) \propto \frac{\gamma^2_\alpha N_\alpha}{D_{\alpha, S}} \binom{2j_\alpha +2}{3} C_{\alpha,\{n_{\alpha'}\}}
				\end{equation}
where $D_{\alpha, S}$ is the dimension of the spin Hilbert subspace comprising the resonance group $\alpha$ and all the other equivalent groups with which it effectively interacts with (i.e. those with $T_{\alpha\alpha'} \neq 0$). In this case, the index $\alpha'$ in Eq. \eqref{eq:generating_poly} (and in Eqs. \eqref{eq:epsilon_difference_b}-\eqref{eq:C_alpha}) may then be re-interpreted as running over only those groups with $T_{\alpha\alpha'} \neq 0$.
From the last two equations above, we note the dependence of the signal intensity on the quantum spin number $j_\alpha$ of the resonance group through the $2j_\alpha-$th tetrahedral number, i.e. $\binom{2j_\alpha +2}{3}$. With all other parameters and conditions held constant, these equations inform us that the higher the spin quantum number of the resonance group, the higher the intensity of the signal. We also note that for a fixed resonance group $\alpha$, the integers $\{C_{\alpha,\{n_{\alpha'}\}}\}$ are effectively the relative intensities of the resonance signals.
\par In our derivation of Eq. \eqref{eq:rate_of_absorption_4}, we have assumed the resonance spin group $\alpha$ is present in all the initial $N$ chemical species of the ensemble. This is not always the case. Eqs. \eqref{eq:rate_of_absorption_4} and \eqref{eq:intensity_2} may therefore be multiplied by the fraction $f_\alpha$ of the initial $N$ which contains the resonance group $\alpha$. In NMR, for example, if the resonance group in the sample has not undergone any alteration of its isotopic concentration $f_\alpha$ becomes the natural abundance of the group.
\par Moreover, if there are more than one resonance spin groups who satisfy the resonance conditions, it is clear that Eqs. \eqref{eq:rate_of_absorption_4} and \eqref{eq:intensity_2} must be summed over such groups since the operator $\xi^x(n,\omega_o)$ is the sum of single spin operators, Eqs. \eqref{eq:xi_alpha} and \eqref{eq:xi^x_sum_n_omega_o}. 
\par
\begin{figure}[h!]
	\centering
	\includegraphics[scale=0.35]{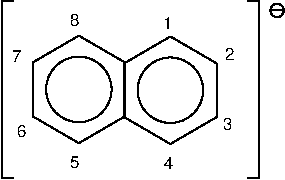}
	\caption{Naphthalene anion}
	\label{fig:naphthalene_chemfig}
	\end{figure}

\begin{figure*}
	\centering
	\includegraphics[trim=2cm 0cm 0cm 0cm, scale=0.6]{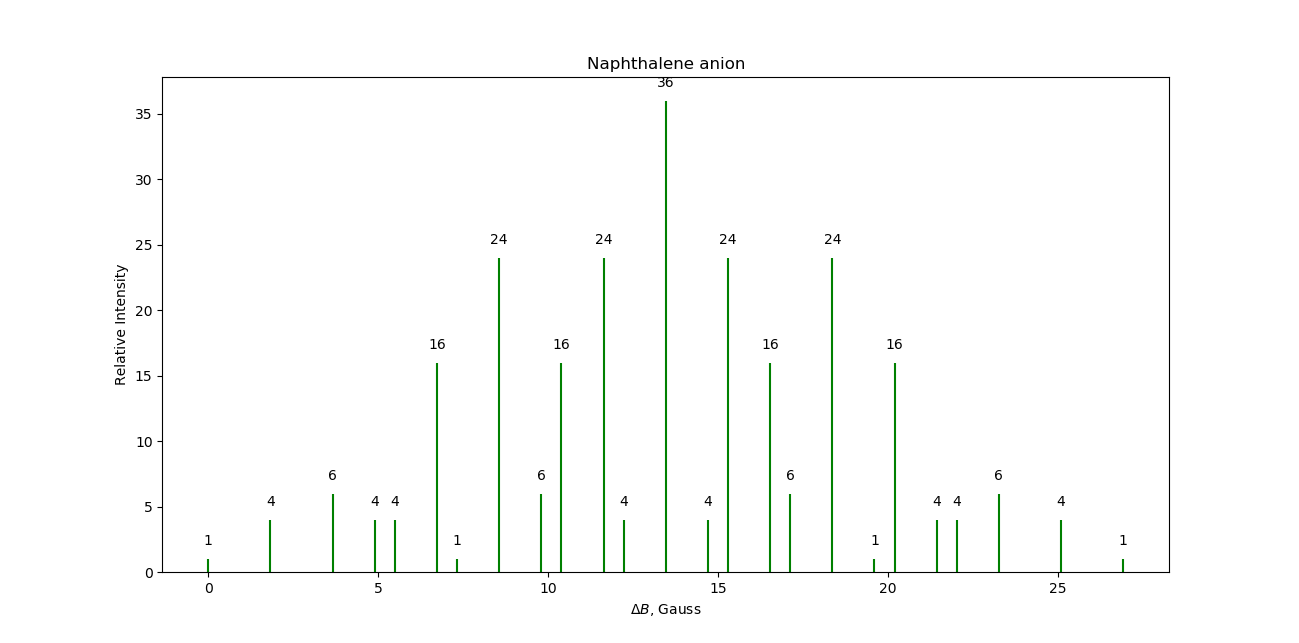} 
	\caption{Theoretical ESR stick-plot spectrum of naphthalene anion (counterion: \ce{K+}). Parameters used were taken from [\onlinecite{book:Wertz-1986}]. Each spectral line is labeled by its relative intensity.}
	\label{fig:Naphthalene}
	\end{figure*}

\begin{figure}[h!]
	\centering
	\includegraphics[scale=0.35]{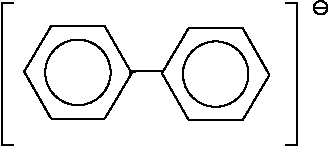}
	\caption{Biphenyl anion}
	\label{fig:biphenyl_chemfig}
	\end{figure}

\begin{figure*}
	\centering
	\includegraphics[trim=2cm 0cm 0cm 0cm, scale=0.6]{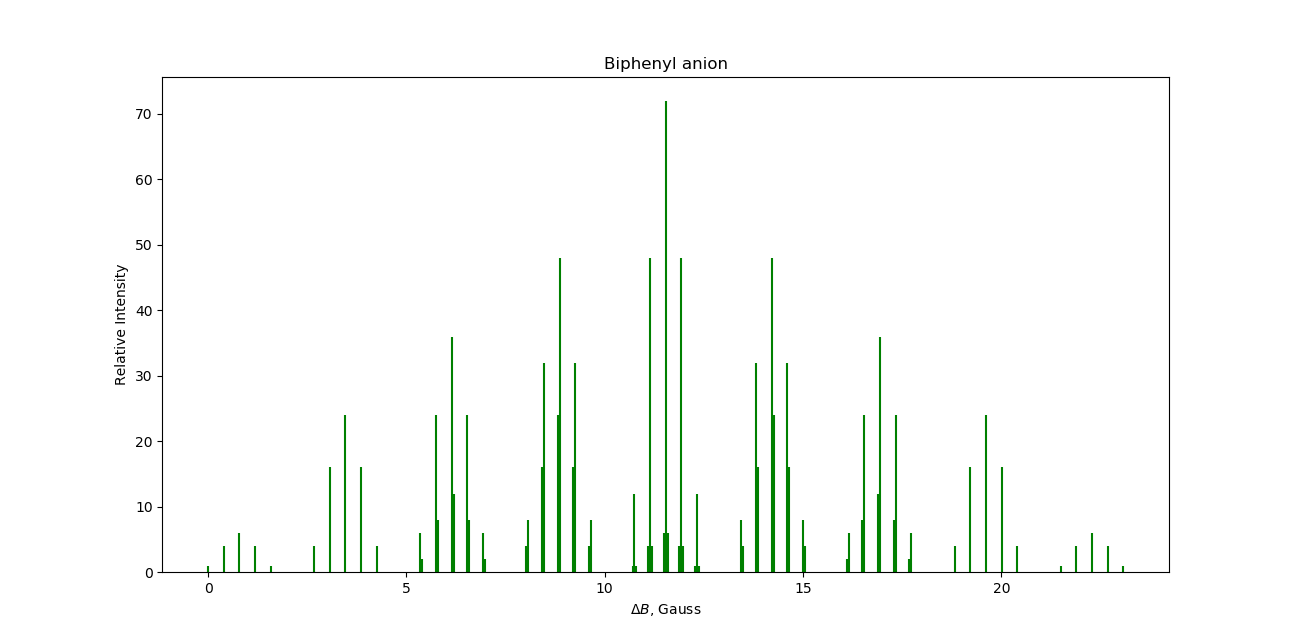} 
	\caption{High-field theoretical ESR stick-plot absorption spectrum of biphenyl anion (counterion: \ce{K+}). Parameters were taken from [\onlinecite{book:Wertz-1986}].}
	\label{fig:biphenyl}
	\end{figure*}

 Eqs. \eqref{eq:B_o_resonance} and \eqref{eq:intensity_2}, together determine the resonance spectrum of the spin system at zeroth-order in $\mathscr{X}$ according to ACP, in the adiabatic process limit. While the former gives the resonance steady field $B_o$ for a given configuration of the spins in terms of $\{n_{\alpha'}\}$, the latter equation gives the intensity of the resonance signal. And the properties of the spectrum -- which one can easily conclude from these two equations -- are in agreement with those reported by Gutowsky, McCall and Slichter\citep{art:Gutowsky-1951,art:Gutowsky-1953}. But more importantly, we must remark that the polynomial $P_\alpha(\gv{x})$, Eq. \eqref{eq:generating_poly}, is the \emph{generating function} for the resonance spectrum. Once we construct $P_\alpha(\gv{x})$ and are in possession of the value of parameters like the constants $\lambda_{\alpha,\alpha'}$, $\gamma_\alpha$ and $\omega_o$, we can easily generate the stick-plot spectrum. Each term of $P_\alpha(\gv{x})$ represents a resonance line: for a given term, the coefficient indicates the relative intensity of the corresponding resonance line, while the exponents of the variables determine -- by means of Eq. \eqref{eq:B_o_resonance} -- the position of the resonance line. The advancement in computer algebra makes the computational implementation of this protocol easy to achieve. We illustrate these points by considering specific examples from ESR, namely the absorption spectrum of naphthalene, anthracene and biphenyl anions. The parameters are taken from [\onlinecite{book:Wertz-1986}] and the plots were generated from a simple Python code which implemented Eqs. \eqref{eq:B_o_resonance} and \eqref{eq:generating_poly} (and an extensive use of the SymPy\citep{art:Sympy} library was made).
\paragraph{Naphthalene anion}
The naphthalene anion, Fig. \ref{fig:naphthalene_chemfig}, has two groups of equivalent nuclei: the first group comprises the hydrogen nuclei in the positions $1,4,5,8$, and those in the positions $2,3,6,7$ form the second group. The splitting constants for the two groups are $\lambda_{e,1}=4.90\ $G and $\lambda_{e,2}=1.83\ $G (counterion is \ce{K+})\citep{book:Wertz-1986}. 
From Eq. \eqref{eq:generating_poly}, we conclude that the generating function for this anion's ESR spectra at zeroth-order is:
		\begin{equation}
		\begin{split}
		P_e(\gv{x})& =(1+x_1)^4(1+x_2)^4\\
		& = x_2^4 x_1^4 + 4 x_2^3 x_1^4 + 6 x_2^2 x_1^4 
		 + 4 x_2 x_1^4 + x_1^4 + 4 x_2^4 x_1^3 \\
		& + 16 x_2^3 x_1^3 + 24 x_2^2 x_1^3 + 16 x_2 x_1^3  
		 + 4 x_1^3 + 6 x_2^4 x_1^2 \\
		& + 24 x_2^3 x_1^2 + 36 x_2^2 x_1^2 + 24 x_2 x_1^2 + 6 x_1^2 + 4 x_2^4 x_1 \\
		& + 16 x_2^3 x_1 + 24 x_2^2 x_1 + 16 x_2 x_1 + 4 x_1 + x_2^4 + 4 x_2^3 \\
		& + 6 x_2^2 + 4 x_2 + 1 \ .	
		\end{split}	
		\end{equation}
Let the exponents of $x_1$ and $x_2$ count the total number of HP bosons held by the first and second group of equivalent nuclei, respectively. As remarked above, every term in the polynomial $P_e(\gv{x})$ represents a resonance line: The coefficient of a given term indicates the relative intensity of the corresponding resonance line and the exponents of the variables of the term determine the position of the resonance line by means of Eq. \eqref{eq:B_o_resonance}. If we take the term $(24 x_2^2 x_1^3)$, for example, the relative intensity of the resonance line it represents is $24$, the number of HP bosons specifying the configuration of group 1 $(x_1)$, and group 2 $(x_2)$ are $3$ and $2$, respectively. So, from Eq. \eqref{eq:B_o_resonance}, we determine that the corresponding resonance line falls at $\Delta B =  3\lambda_{e,1} + 2 \lambda_{e,2} = 18.36 \ $G from the reference position $B_\alpha(\omega_o)$ (which may be set equal to zero for convenience). We show the stick-plot ESR spectrum of the naphthalene anion computed this way in Fig. \ref{fig:Naphthalene}. The experimental\citep{book:Wertz-1986} positions and relative intensities of the spectral lines are in very good agreement with the simple theoretical spectrum in  Fig. \ref{fig:Naphthalene}. 
\paragraph{Biphenyl anion}
The biphenyl anion, Fig. \ref{fig:biphenyl_chemfig}, has three groups of equivalent protons: two of which are of cardinality 4, and the last of cardinality 2. Let $\lambda_{e,1}, \lambda_{e,2}$ be the splitting constants of the first and second groups of equivalent protons (of size 4), and $\lambda_{e,3}$ the splitting constant of the set of equivalent protons of size 2. From the literature\citep{book:Wertz-1986}, we have:  $\lambda_{e,1}=2.675 \ $G, $\lambda_{e,2}=0.394 \ $G and $\lambda_{e,3}=5.387 \ $G.
Once again, we see from Eq. \eqref{eq:generating_poly} that the generating function for biphenyl anion's ESR spectrum is:
			\begin{equation}
			\begin{split}
			\label{eq:gen_func_biphenyl}
			P_e(\gv{x})& =(1+x_1)^4(1+x_2)^4 (1+x_3)^2\\
			& = x_2^4 x_1^4 + 4 x_2^3 x_1^4 + 6 x_2^2 x_1^4 + x_2^4 x_3^2 x_1^4 + 4 x_2^3 x_3^2 x_1^4 \\
			& + 6 x_2^2 x_3^2 x_1^4 + 4 x_2 x_3^2 x_1^4 + x_3^2 x_1^4 + 4 x_2 x_1^4 + 2 x_2^4 x_3 x_1^4 \\
			& + 8 x_2^3 x_3 x_1^4 + 12 x_2^2 x_3 x_1^4 + 8 x_2 x_3 x_1^4 + 2 x_3 x_1^4 + x_1^4  \\
			& + 4 x_2^4 x_1^3 + 16 x_2^3 x_1^3 + 24 x_2^2 x_1^3 + 4 x_2^4 x_3^2 x_1^3 + 16 x_2^3 x_3^2 x_1^3  \\			
			& + 24 x_2^2 x_3^2 x_1^3 + 16 x_2 x_3^2 x_1^3 + 4 x_3^2 x_1^3 + 16 x_2 x_1^3   \\
			& + 8 x_2^4 x_3 x_1^3 + 32 x_2^3 x_3 x_1^3 + 48 x_2^2 x_3 x_1^3  + [\ldots] + 1
			\end{split}
			\end{equation}
For each term, the exponent of $x_i$ corresponds to the number of HP bosons held by the $i-$th group of equivalent protons. The ESR spectrum can be generated from this polynomial as explained above. For example, if we take the term $(4 x_2^3 x_1^4)$, the relative intensity of the corresponding spectral line is $4$, and it falls at a distance of $\Delta B=4\lambda_{e,1}+3\lambda_{e,2}= 11.882 \ $G from $B_\alpha(\omega_o)$, the reference position. The generating function in Eq. \eqref{eq:gen_func_biphenyl} has $75$ terms, which is also the number of spectral lines to expect experimentally. We show in Fig. \ref{fig:biphenyl} the stick-plot ESR spectrum for the biphenyl anion generated from $P_e(\gv{x})$, Eq. \eqref{eq:gen_func_biphenyl}.
	
\begin{figure}[h!]
	\centering
	\includegraphics[scale=0.35]{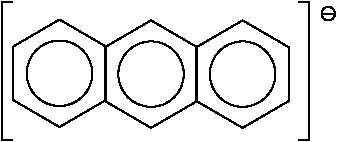}
	\caption{Anthracene anion}
	\label{fig:anthracene_chemfig}
	\end{figure}

\begin{figure*}
	\centering
	\includegraphics[trim=2cm 0cm 0cm 0cm, scale=0.6]{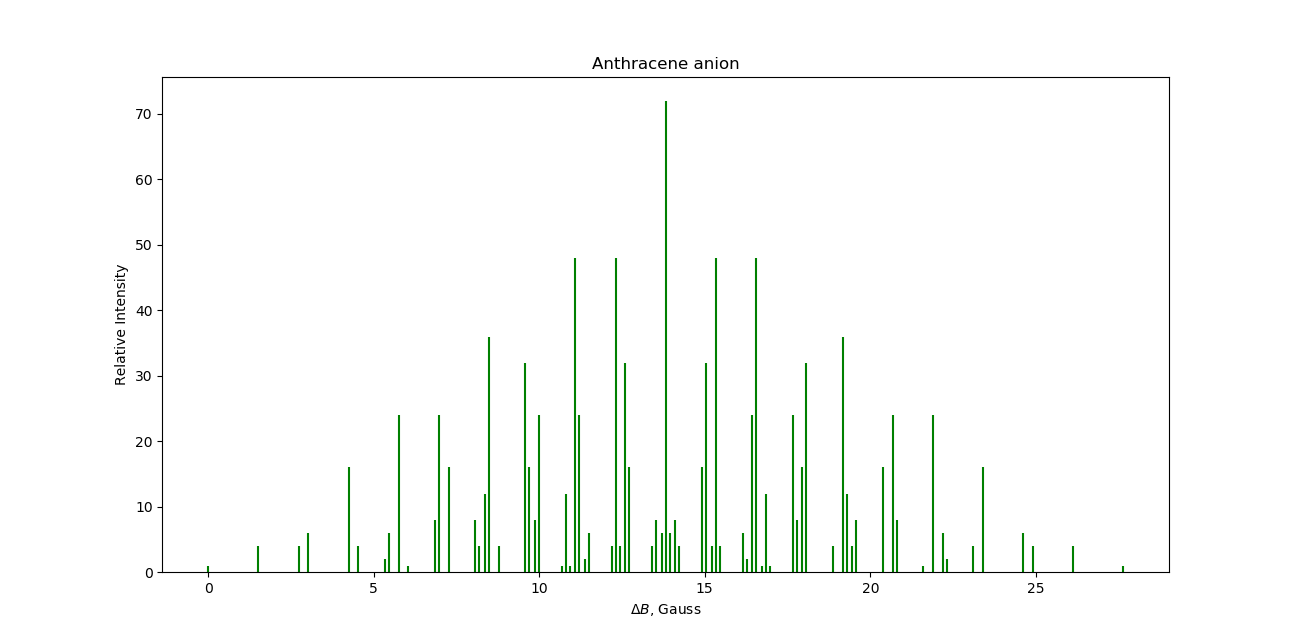} 
	\caption{High-field theoretical ESR stick-plot absorption spectrum of anthracene anion (counterion: \ce{K+}). Parameters were taken from [\onlinecite{book:Wertz-1986}].}
	\label{fig:anthracene}
	\end{figure*}

\paragraph{Anthracene anion}
Like the biphenyl anion, the anthracene anion -- Fig. \ref{fig:anthracene_chemfig} -- has three groups of equivalent protons and of the same dimension as those of the biphenyl anion. 
Let $\lambda_{e,1}$ and $\lambda_{e,2}$ be the splitting constants of the first and second groups of equivalent spins of dimension 4, and $\lambda_{e,3}$ the splitting constant of the group of dimension 2. Their values are taken from the literature\citep{book:Wertz-1986} to be: $\lambda_{e,1}=2.73 \ $G, $\lambda_{e,2}=1.51 \ $G and $\lambda_{e,3}=5.34\ $G. Given that the biphenyl and anthracene anions present the same groups of equivalent spins (cardinality-wise), their high-field spectra also share the same generating function. As a matter of fact, using the generating function in \eqref{eq:gen_func_biphenyl}, we can determine the stick-plot spectrum of the anthracene anion. The result is reported in Fig. \ref{fig:anthracene}. 
Both spectra in Figs. \ref{fig:Naphthalene} and \ref{fig:anthracene} have the same number of resonance lines, and the distribution of the relative intensities of these lines is the same in both spectra. The only difference between the two is the \emph{position} of the resonance lines.
\subsection{Higher-order terms}\label{subsec:higher-orders}
\par We have concerned ourselves so far with the zeroth-order term of the ACP scheme, Eq. \eqref{eq:differential_varrho_n}. If, in deriving the master equation for the higher order terms (i.e. $n \geq 1$) of $\varrho(t)$, we apply the same techniques and reasoning which led to Eq. \eqref{eq:diff_varrho_n=0_++}\citep{Note1}, then, one can see that, in general:
			\begin{equation}
			\label{eq:diff_varrho_m_t}
			\frac{d}{dt}\varrho^{(n)}(t) = \mathcal{L}\varrho^{(n)}(t)  + \mathcal{A}(t)\varrho^{(n)}(0)  + \mathcal{G}^{(n)}(t)
			\end{equation}
where the initial condition, $\varrho^{(n)}(0)$, is given by Eq. \eqref{eq:varrho_n=0}, and where
			\begin{equation}
			\label{eq:def_G_m_t}
			\mathcal{G}^{(n)}(t) := 
			\sum^{n}_{l=1} \mathcal{A}^{(l)}(t)\varrho^{(n-l)}(0)  + \sum^{n}_{l=1} \mathcal{L}^{(l)}\varrho^{(n-l)}(t)
			\end{equation}
for $n \geq 1$ (while for $n=0$, $\mathcal{G}^{(n)}(t) =0$). 
The superoperators $\mathcal{A}(t)$ and $\mathcal{L}$ in Eq. \eqref{eq:diff_varrho_m_t} are still given by Eqs. \eqref{eq:mathcal_A} and \eqref{eq:diff_varrho_n=0_b}, respectively. We note that the master equation for the higher order term $\varrho^{(n\geq 1)}(t)$, Eq. \eqref{eq:diff_varrho_m_t}, is just the same as that for $\varrho^{(0)}(t)$, Eq. \eqref{eq:diff_varrho_n=0_++}, except for the presence of the time-dependent operator $\mathcal{G}^{(n)}(t)$ in the former. Evidently, $\mathcal{G}^{(n)}(t)$ depends on the lower order corrections to $\varrho(0)$ and $\varrho(t)$, i.e. $\varrho^{(n')}(0)$ and $\varrho^{(n')}(t)$ with $n'< n$. We also remark that the superoperators $\mathcal{A}^{(l)}(t)$ and $\mathcal{L}^{(l)}$ in Eq. \eqref{eq:def_G_m_t} differ from $\mathcal{A}(t)$ and $\mathcal{L}$ in Eq. \eqref{eq:diff_varrho_m_t}, and may be seen as some form of higher-order corrections to the latter two, respectively. In the Supplemental Material\cite{Note1} we briefly discuss the first-order correction $\varrho^{(1)}(t)$. 

The formal solution to Eq. \eqref{eq:diff_varrho_m_t} is:
			\begin{multline}
			\label{eq:varrho_n_st}
			\varrho^{(n)}(t) = \left(e^{\mathcal{L}t} + \int^t_0 dt'\  e^{\mathcal{L}(t-t')} \mathcal{A}(t')\right) \varrho^{(n)}(0)\\
			+ \int^t_0 dt' \ e^{\mathcal{L}(t-t')} \mathcal{G}^{(n)}(t') \ .
			\end{multline}
The quantum map involved here is also non-CP. In fact, we see again in Eq. \eqref{eq:varrho_n_st} the same structure observed for non-CP maps\citep{art:Buzek-2001, art:Carteret-2008, art:Cory-2004}: the first term, $e^{\mathcal{L}t}\varrho^{(n)}(0)$, involves a CPT map, while $\int^t_0 dt'\  e^{\mathcal{L}(t-t')} \mathcal{A}(t') \varrho^{(n)}(0) + \int^t_0 dt' \ e^{\mathcal{L}(t-t')} \mathcal{G}^{(n)}(t')$ constitute the traceless inhomogeneous term. It must be mentioned, however, that $e^{\mathcal{L}t}\varrho^{(n)}(0)$ for $n\geq 1$ is also traceless.
\section{Concluding Remarks}\label{sec:Conclusion}
We have shown that it is possible to extend the GKSL formalism to those problems where one wants to treat the environment at a classical level. A quantum theory for CW magnetic resonance was developed in this paper to illustrate the approach. And in the development of the theory, we introduced the affine commutation perturbation (ACP) scheme, which makes it possible to account for some effects of the perturbation even at the zeroth-order approximation. Indeed, we were able to derive the CW magnetic spectra of multispin systems at the zeroth-order of the ACP scheme and computed the ESR spectra for a number of radicals -- which are in good agreement with the experimental spectra. It must be emphasized that the generating function method for computing theoretical spectra expounded in \S \ref{subsec:spectrum} predicates on the weak-coupling assumption, and on the condition that $\left\Vert \mathscr{Z}_o \right\Vert \gg \left\Vert \mathscr{X} \right\Vert$ (in the sense explained at the beginning of \S \ref{subsec:ACP}). We also mention that quadrupolar effects can also be accounted for by adding the corresponding isotropic term to $H_o$ in Eq. \eqref{eq:def_H_o}.
\par We have focused here on quantum Markovian master equations, but the approach can easily be extended to non-Markovian ones as well. More importantly, in discussing the dynamics at the zeroth-order, we have argued and illustrated the importance of the term linear in the system-environment interaction, $\mathcal{A}(t)\varrho^{(0)}(0)$, Eq. \eqref{eq:diff_varrho_n=0_++}. We have shown that this term, which is usually discarded (when not identically zero) in standard microscopic derivations of quantum Markovian master equations, actually leads to a linear response theory (LRT) within the GKSL formalism (\S \ref{subsec:LRT}). With it, we were able to derive some known results in standard LRT\cite{book:Giuliani-2005} as limit cases. Despite its vital importance, we also observed that the presence of this linear term breaks the CPT property of the zeroth-order quantum map $\Lambda(t)$, Eq. \eqref{eq:Lambda_t_general}, turning it into a non-CP map. These observations further exacerbate the debate on whether the CPT requirement is truly a fundamental requisite of all quantum maps. The maps associated with higher-order corrections, Eq. \eqref{eq:varrho_n_st}, are also non-CP, and present a similar structure like that observed in the literature for non-CP maps (whereby both the focus system and the environment are treated quantum mechanically). These standard (linear) non-CP maps treated in the literature\citep{book:McCracken-2014, art:Yu-2000, art:Buzek-2001, art:Carteret-2008, inbook:Shabani_co-2014} often arise when there are initial correlations between the focus system and its environment. In our case, there are no such initial correlations. Moreover, as it often happens with non-CP maps\citep{art:Shaji-2005, art:Carteret-2008}, the maps involved in our case cannot take any arbitrary initial density matrix as the input state, but have well-defined (positivity) domains set by the theory. For the zeroth-order map $\Lambda(t)$, Eq. \eqref{eq:Lambda_t_general}, for example\citep{Note1}, its domain is $\varrho^{(0)}(0)=\frac{e^{-\beta \mathscr{Z}_o}}{\mbox{Tr}\left[e^{-\beta \mathscr{Z}_o} \right]}$, Eq. \eqref{eq:initial_rho_n=0}.
\section*{Supplemental Material}
See Supplemental Material for a more detailed discussion on the limitations of the application of the wavefunction formalism in the quantum theory of magnetic resonance, the derivation of Eq. \eqref{eq:diff_varrho_n=0_++}, a detailed analysis of the map $\Lambda(t)$ and further discussion on the higher-order corrections to $\varrho^{(0)}(t)$ under the ACP scheme. 
\section*{Acknowledgments}
The author would like to express his profound gratitude to Prof. Vittorio Giovannetti and Prof. Davide Rossini for their guidance and for always taking the time to discuss and share ideas with him. The author would also like to thank Prof. Antonino Polimeno and the Theoretical Chemistry Group at the University of Padua for their warm invitation -- an occasion on which part of the results elaborated here was presented to the group. Support from Prof. Henrik Koch is also acknowledged.
\bibliography{ms_biblio}

\end{document}


\title{Supplemental Material for\\
``Semiclassical Quantum Markovian Master Equations. Case study: Continuous Wave Magnetic Resonance of Multispin Systems''}
\author{J.A. Gyamfi}
\email{jerryman.gyamfi@sns.it}
\affiliation{Scuola Normale Superiore di Pisa, Piazza dei Cavalieri 7, 56126 Pisa, Italy.}

\date{\today}

\begin{abstract}
Some limitations of the application of the wavefunction formalism in the quantum theory of magnetic resonance are discussed in more details (\S \ref{subsec:wave}). Equation (52) of the text is derived in \S \ref{subsec:Derivation}. The expression for $\Lambda(t)$, Eq. (64), is derived in \S \ref{subsec:Derivation_Lambda}.  This is followed by a detailed analysis of the same map and its properties in \S \ref{subsec:L1}, \S \ref{subsec:L2}, \S \ref{subsec:L3}. Finally, in \S \ref{subsec:higher-orders}, the higher-order corrections to $\varrho^{(0)}(t)$ under the ACP scheme are briefly discussed.
\end{abstract}

\maketitle

\section{Introduction}
\subsection{Some observations on the wavefunction formalism and Dirac's time-dependent perturbation theory, as commonly applied in traditional quantum magnetic resonance theory.}\label{subsec:wave}
\par We discuss here, in more details, the limitations of the wavefunction formalism, especially in relation to magnetic resonance. We also point out the assumptions on which the often used transition rate equations derived from the formalism rest. (Unlike the rest of the material, the constant $\hbar$ is not set equal to $1$ in this section.)
\par Consider a spin system with its spin Hilbert space $\mathcal{H}_S$, and whose spin Hamiltonian $\mathscr{H}(t)$ is given by the sum
				\begin{equation}
				\label{eq:Hamiltonian_generic}
				\mathscr{H}(t) = \mathscr{H}_o + \mathscr{H}'(t) \ .
				\end{equation}
$\mathscr{H}_o$ in Eq. \eqref{eq:Hamiltonian_generic} is the time-independent component of $\mathscr{H}(t)$. We denote the eigenvectors of $\mathscr{H}_o$ as $\{\ket{k}\}$, where
				\begin{equation}
				\mathscr{H}_o \ket{k} = E_k \ket{k} \ , \quad \ \mbox{and } \quad \braket{k}{k'} = \delta_{k,k'} \ .
				\end{equation}
$\mathscr{H}'(t)$ in Eq. \eqref{eq:Hamiltonian_generic}, on the other hand, is the time-dependent part of $\mathscr{H}(t)$. 
\par According to the time-dependent Schr\"{o}dinger equation, the equation of motion for a generic normalized spin ket $\ket{\psi(t)} \in \mathcal{H}_S$ satisfies the first-order linear differential equation
				\begin{equation}
				\label{eq:Schroedinger_time_dependent}
				i \hbar \ \frac{d}{dt} \ket{\psi(t)} = \mathscr{H}(t) \ket{\psi(t)} \ .
				\end{equation}
Let us assume that at the initial time $t_o$, the corresponding normalized spin ket is $\ket{\psi(t_o)}$ (which is supposed to be known). 
\par If $\mathscr{H}'(t)$ can be considered as the perturbation term -- meaning: for any pair of nondegenerate eigenkets $\ket{k}$ and $\ket{k'}$ of $\mathscr{H}_o$, $\abs{\frac{\matrixel{k'}{\mathscr{H}'(t)}{k}}{E_{k'} - E_{k}}} \ll 1$ -- then, we may solve Eq. \eqref{eq:Schroedinger_time_dependent} for $\ket{\psi(t)}$ through a perturbation expansion. To ensure convergence of this expansion, we may transit to the interaction picture by introducing the following transformation:
				\begin{equation}
				\label{eq:psi_t_Schroedinger}
				\ket{\psi(t)} = \mathscr{U}_o(t,t_o) \ket{\phi(t)}  
				\end{equation}
with,
				\begin{equation}
				\mathscr{U}_o(t,t_o):= e^{-i(t-t_o) \mathscr{H}_o/\hbar}
				\end{equation}
where $\ket{\phi(t)}$ is the spin ket in the interaction picture. Then, in light of this transformation, it follows from Eqs. \eqref{eq:Hamiltonian_generic} and \eqref{eq:Schroedinger_time_dependent} that
				\begin{equation}
				i\hbar \ \frac{d}{dt} \ket{\phi(t)} =  \mathscr{V}_{t_o}(t) \ket{\phi(t)} 
	\ \qquad \
				\mathscr{V}_{t_o}(t):=\mathscr{U}^\dagger_o(t,t_o) \mathscr{H}'(t) \mathscr{U}_o(t,t_o)
				\end{equation}
from which we derive that
				\begin{equation}
				\label{eq:phi(t)}
				\ket{\phi(t)} = \mathscr{U}_I(t,t_o) \ket{\phi(t_o)}
				\end{equation}
where,
				\begin{multline}
				\label{eq:U_I}
				\mathscr{U}_I(t,t_o) := \mathbb{I} + \frac{1}{i\hbar} \int^t_{t_o} dt' \  \mathscr{V}_{t_o}(t') + \left(\frac{1}{i\hbar}\right)^2 \int^t_{t_o} dt' \int^{t'}_{t_o} dt'' \  \mathscr{V}_{t_o}(t')\mathscr{V}_{t_o}(t'') \\
				+  \left(\frac{1}{i\hbar}\right)^3 \int^t_{t_o} dt' \int^{t'}_{t_o} dt'' \int^{t''}_{t_o} dt''' \  \mathscr{V}_{t_o}(t')\mathscr{V}_{t_o}(t'')\mathscr{V}_{t_o}(t''') + \ldots
				\end{multline}			
where $\mathbb{I}$ is the identity operator on $\mathcal{H}_S$. Since the normalized eigenkets $\{\ket{k}\}$ constitute an orthonormal basis for the Hilbert space $\mathcal{H}_S$, we may expand $\ket{\phi(t)}$ in this basis,
				\begin{equation}
				\label{eq:phi_t_Interaction}
				\ket{\phi(t)} = \sum_k \ket{k}\braket{k}{\phi(t)} = \sum_k a_k(t) \ket{k} 
				\end{equation}
where,
				\begin{equation}
				\label{eq:precursor_a_i(t)}
				a_k(t) : = \braket{k}{\phi(t)} \ .
				\end{equation}	
It then definitely follows from Eqs. \eqref{eq:phi(t)} and \eqref{eq:precursor_a_i(t)} that
				\begin{equation}
				\label{eq:a_i(t)}
				a_k(t) = \matrixel{k}{\mathscr{U}_I(t,t_o)}{\phi(t_o)} \ .
				\end{equation}			
Naturally,
				\begin{equation}
				\label{eq:phi(t_o)}
				\ket{\phi(t_o)} = \ket{\psi(t_o)} = \sum_k a_k(t_o) \ket{k} \ .
				\end{equation}	
Since the initial ket $\ket{\psi(t_o)}$ is normalized, it follows that
				\begin{equation}
				\label{eq:a_i(t)_normalization_cond}
				\sum_k \abs{a_k(t)}^2 = 1 \  \qquad (t \geq t_o) \ .
				\end{equation}
Given that the initial ket $\ket{\psi(t_o)}$ is supposed to be known, the coefficients $\{a_k(t_o)\}$ are also known. If we make use of Eq. \eqref{eq:phi(t_o)},  \eqref{eq:a_i(t)} becomes
				\begin{equation}
				\label{eq:a_i(t)_2}
				a_k(t) = \sum_{k'} \matrixel{k}{\mathscr{U}_I(t,t_o)}{k'} a_{k'}(t_o)\ .
				\end{equation}	
From Eqs. \eqref{eq:a_i(t)} and \eqref{eq:a_i(t)_2}, we see that we have succeeded in expressing the coefficients $\{a_k(t)\}$ in function of only known quantities. More importantly, these expressions for $a_k$ are also exact. According to Eq. \eqref{eq:a_i(t)}, we may interpret $a_k(t)$ as the transition amplitude from the initial state $\ket{\phi(t_o)}$ to the eigenstate $\ket{k}$, by means of the evolution operator $\mathscr{U}_I(t,t_o)$. In particular, we note from Eq. \eqref{eq:a_i(t)_2} that $a_k(t)$ is a weighted sum of all the initial probability amplitudes $\{a_{k'}(t_o)\}$; the weighting factor here is the transition amplitude from the generic eigenstate $\ket{k'}$ to $\ket{k}$ by means of $\mathscr{U}_I(t,t_o)$.	
\par Going back to the Schr\"{o}dinger picture, it follows from Eqs. \eqref{eq:psi_t_Schroedinger} and \eqref{eq:phi_t_Interaction} that
				\begin{equation}
				\ket{\psi(t)} = \sum_k a_k(t) e^{-i(t-t_o)E_k/\hbar} \ket{k} \ .
				\end{equation}	
\subsubsection{The first-order approximation and the zero-temperature limit}			
The derivation carried out above is exact. In general, however, it is hardly possible to exactly evaluate Eq. \eqref{eq:a_i(t)_2} for the coefficient $a_k(t)$ without any approximations. In practical computations, the approximations are introduced at the level of the evolution operator $\mathscr{U}_I(t,t_o)$, Eq. \eqref{eq:U_I}. We consider in this subsection and the next expressions for the probabilities $\lvert a_k(t)\rvert^2$, since these are most often of practical interest. 
\par If we choose to approximate $a_k(t)a^*_{l}(t)$ up to first-order in $\mathscr{H}'(t)$, then from Eqs. \eqref{eq:a_i(t)_2} and \eqref{eq:U_I} we have
				\begin{multline}
				\label{eq:a_i(t)_first-order}
				a_k(t) a^*_l(t)  = a_k(t_o)a^*_l(t_o)-\frac{1}{i\hbar}\sum_{k'} \int^t_{t_o} dt_1 \ e^{-i (t_1-t_o) \omega_{l,k'}} \matrixel{k'}{\mathscr{H}'(t_1)}{l} a_{k}(t_o) a^*_{k'}(t_o)\\
				 + \frac{1}{i\hbar}\sum_{k'}  \int^t_{t_o} dt_1 \ e^{-i (t_1-t_o) \omega_{k',k}} \matrixel{k}{\mathscr{H}'(t_1)}{k'}   a_{k'}(t_o) a^*_{l}(t_o) 
				\end{multline}
($\omega_{m,m'} \equiv (E_m-E_{m'}) / \hbar$), from which we derive that
				\begin{equation}
				\label{eq:a_i(t)_modulus_first-order}
				\left\vert a_k(t) \right\vert^2 = \left\vert a_k(t_o) \right\vert^2 +  \frac{2}{\hbar} \Im \left[\sum_{k'}  \int^t_{t_o} dt_1 \ e^{-i (t_1-t_o) \omega_{k',k}} \matrixel{k}{\mathscr{H}'(t_1)}{k'}   a_{k'}(t_o) a^*_{k}(t_o) \right] \ . 
				\end{equation}
Note that these  first-order approximation expressions are generally valid for any initial ket $\ket{\psi(t_o)}$. However, it is common practice in the literature to assume $\ket{\psi(t_o)}$ is precisely one of the eigenkets $\{\ket{k}\}$ of $\mathscr{H}_o$, say $\ket{k_o}$\citep{book:Pake-1973, book:Corio-1966, book:Atkins-2011}. Accordingly, the coefficients $\{a_k(t_o)\}$ are such that
				\begin{equation}
				\label{eq:zero-T_Limit}
				a_k(t_o) = \delta_{k,k_o} \ ,  \qquad \forall \ k.
				\end{equation}
Without loss of generality, we may refer to this assumption as the "zero-temperature limit"\citep{book:Giuliani-2005}. The zero-temperature limit assumption, Eq. \eqref{eq:zero-T_Limit}, is obviously a great simplification even within the wavefunction formalism. In standard magnetic resonance experiments conducted at $T>0 K$, the initial state of the probed spin system is hardly a pure state\citep{book:Blum-1981}, let alone one whose expansion coefficients in $\{\ket{k_o}\}$ satisfy Eq. \eqref{eq:zero-T_Limit}. 
\par At any rate, in the zero-temperature limit, Eq. \eqref{eq:a_i(t)_modulus_first-order} simplifies to:
				\begin{equation}
				\label{eq:a_i(t)_modulus_first-order_ZERO-TEMP}
				\left\vert a_k(t) \right\vert^2 = \delta_{k,k_o} +  \frac{2}{\hbar} \Im \left[  \int^t_{t_o} dt_1 \ e^{-i (t_1-t_o) \omega_{k_o,k}} \matrixel{k}{\mathscr{H}'(t_1)}{k_o}   \delta_{k_o,k} \right] = \delta_{k,k_o} \ .
				\end{equation}
Thus, nothing interesting happens at the first-order approximation in the zero-temperature limit. Nevertheless, the normalization condition, Eq. \eqref{eq:a_i(t)_normalization_cond}, is satisfied. 

\subsubsection{The second-order approximation and the zero-temperature limit}
From Eqs. \eqref{eq:a_i(t)_2} and \eqref{eq:U_I}, we see that the expression for $\left\vert a_k(t) \right\vert^2$ approximated to second-order in $\mathscr{H}'(t)$ yields
				\begin{multline}
				\label{eq:a_i(t)_modulus_second-order}
				\left\vert a_k(t) \right\vert^2 = \left\vert a_k(t_o) \right\vert^2 +  \frac{2}{\hbar} \Im \left[\sum_{k'}  \int^t_{t_o} dt_1  \matrixel{k}{\mathscr{V}_{t_o}(t_1)}{k'}   a_{k'}(t_o) a^*_{k}(t_o) \right] \\
				- 2\left(\frac{1}{\hbar} \right)^2 \Re \left[\sum_{k'} \int^t_{t_o}dt_1 \int^{t_1}_{t_o}dt_2  \matrixel{k}{\mathscr{V}_{t_o}(t_1) \mathscr{V}_{t_o}(t_2)}{k'}  a_{k'}(t_o) a^*_{k}(t_o) \right]\\
				+  \left(\frac{1}{\hbar} \right)^2 \left\vert \sum_{k'}  \int^t_{t_o}dt_1  \matrixel{k}{\mathscr{V}_{t_o}(t_1)}{k'}a_{k'}(t_o) \right\vert^2  \ .
				\end{multline} 
In the zero-temperature limit, Eq. \eqref{eq:a_i(t)_modulus_second-order} simplifies to:
				\begin{multline}
				\label{eq:a_i(t)_modulus_second-order_ZERO-TEMP_0}
				\left\vert a_k(t) \right\vert^2 = \delta_{k_o,k} - 2\left(\frac{1}{\hbar} \right)^2 \Re \left[ \int^t_{t_o}dt_1 \int^{t_1}_{t_o}dt_2  \matrixel{k_o}{\mathscr{V}_{t_o}(t_1) \mathscr{V}_{t_o}(t_2)}{k_o}  \right] \delta_{k,k_o}\\
				+ \left(\frac{1}{\hbar} \right)^2 \left\vert   \int^t_{t_o}dt_1 \ e^{-i (t_1-t_o)\omega_{k_o,k}} \matrixel{k}{\mathscr{H}'(t_1)}{k_o} \right\vert^2 
				\end{multline}	
-- from which follows that $\sum_k \left\vert a_k(t) \right\vert^2=1$. Thus, the normalization condition stated in Eq. \eqref{eq:a_i(t)_normalization_cond} is also satisfied at this order of approximation. This is actually the case for all orders of approximation.
\par It is easily derived from Eq. \eqref{eq:a_i(t)_modulus_second-order_ZERO-TEMP_0} that for $k \neq k_o$,
				\begin{equation}
				\label{eq:a_i(t)_modulus_second-order_ZERO-TEMP}
				\left\vert a_k(t) \right\vert^2 =  \left(\frac{1}{\hbar} \right)^2 \left\vert   \int^t_{t_o}dt_1 \ e^{-i (t_1-t_o)\omega_{k_o,k}} \matrixel{k}{\mathscr{H}'(t_1)}{k_o} \right\vert^2 \ .
				\end{equation}
This is the ubiquitous transition probability equation in the wavefunction formalism
\citep{book:Atkins-2011}, which is applied in many problems -- including magnetic resonance
\citep{book:Pake-1973, book:Corio-1966}. In fact, theoretical derivations in quantum magnetic resonance studies based on the wavefunction formalism are usually carried out: 1) with the product $a_k(t)a^*_l(t)$ approximated to second-order in $\mathscr{H}'(t)$, and 2) assuming the zero-temperature limit condition. Together, these two have been extensively applied both in the theory of electron spin resonance (ESR)\citep{book:Pake-1973} and nuclear magnetic resonance (NMR)\citep{book:Corio-1966}. As limiting as the zero-temperature assumption is, its employment has played an invaluable role in our understanding of the magnetic resonance phenomenon in the framework of quantum mechanics. Certainly, the reduction in mathematical complexity one achieves with it has been a decisive factor in its widespread applications. 
For example, the starting point of Solomon's derivation of expressions for the transition rates between the states in a system of two spins in his 1955 seminal paper\citep{art:Solomon-1955} -- which has had an enormous impact on the field -- is actually the expression in Eq. \eqref{eq:a_i(t)_modulus_second-order_ZERO-TEMP}. Consequently, the expressions for the relaxation times $T_1$ and $T_2$ in [\onlinecite{art:Solomon-1955}] should be used  bearing in mind the limitations of the zero-temperature limit and the wavefunction formalism\citep{book:Blum-1981,book:Abragam-1983}.

\section{Quantum Markovian master equation approach to CW magnetic resonance}
\subsection{Derivation of Eq. (52)}\label{subsec:Derivation}
The derivation follows, to some extent, the usual routine involved in such microscopic derivations\citep{book:Breuer-2007}. 
\par Eq. (50) of the text reads:
			\begin{multline}
			\label{eq:50}
			\frac{d}{dt} \varrho^{(0)}(t)  = -i B_1 \sum_r \sum_{n,\omega_o}   \left(e^{-it(\omega_o-\omega_r n)} + e^{-it(\omega_o+\omega_r n)} \right) \left[\xi^x(n, \omega_o) , \varrho^{(0)}(0)\right] \\
			-  \left(\sum_{r,r'} \sum_{n,\omega_o}\sum_{n',\omega'_o} e^{it\left[ (\omega_o - \omega'_o) - (\omega_r n - \omega_{r'} n')\right]}\ \Gamma(\omega'_o - n'\omega_{r'}) \left[\xi^{x\dagger}(n,\omega_o) ,\xi^{x}(n',\omega'_o)  \varrho^{(0)}(t)\right] + h.c. \right)\\
			-  \left(\sum_{r,r'} \sum_{n,\omega_o}\sum_{n',\omega'_o} e^{it\left[ (\omega_o - \omega'_o) + (\omega_r n - \omega_{r'} n')\right]}\ \Gamma(\omega'_o + n'\omega_{r'}) \left[\xi^{x\dagger}(n,\omega_o) ,\xi^{x}(n',\omega'_o)  \varrho^{(0)}(t)\right] + h.c. \right)
			\end{multline}
with $\Gamma(\omega'_o \pm n'\omega_{r'})$ defined in Eq. (51) as:
			\begin{equation}
			\label{eq:51}
			\begin{split}
			\Gamma(\omega'_o \pm \omega_{r'}n')& := B^2_1\int^{+\infty}_0 d\tau \ e^{i\tau(\omega'_o \pm \omega_{r'}n')} 
			 = B^2_1\left[\pi \delta(\omega'_o \pm n'\omega_{r'}) + i \ \mathcal{P}\left( \frac{1}{\omega'_o \pm n'\omega_{r'}} \right)\right] \ .
			\end{split}
			\end{equation}
If we impose the secular approximation on the last two terms of Eq. \eqref{eq:50} by setting $(\omega_o - \omega'_o) - (\omega_r n - \omega_{r'} n')=0$, we see that the easiest way to let this hold is as follows:
			\begin{equation}
			\label{eq:secular_approx}
			\omega_o = \omega'_o \ , \quad n= n' \ ,  \quad  \omega_r = \omega_{r'} \ .
			\end{equation}
and it is also consistent with our initial assumption that the cross-terms involving $\mathscr{V}^{(0)}_+(t)$ and $\mathscr{V}^{(0)}_-(t)$ do not contribute to the equation of motion.
\par With Eq. \eqref{eq:secular_approx}, the sum of the last two terms of Eq. \eqref{eq:50} reduces to:
			\begin{equation}
			-  \sum_{r} \sum_{n,\omega_o} \bigg( \Gamma(\omega_o - n\omega_{r}) + \Gamma(\omega_o + n\omega_{r}) \bigg) \left[\xi^{x\dagger}(n,\omega_o) ,\xi^{x}(n,\omega_o)  \varrho^{(0)}(t)\right] + h.c.  
			\end{equation}
It is convenient at this point to decompose $\Gamma(\omega_o \pm \omega_{r}n)$ into the sum:
			\begin{equation}
			\Gamma(\omega_o \pm \omega_{r}n) = \frac{1}{2}\eta^{xx}(\omega_o \pm \omega_r n) + i \zeta^{xx}(\omega_o \pm \omega_r n)
			\end{equation}
where,
			\begin{subequations}
			\label{eq:eta_xx_zeta_xx}
			\begin{align}
			\eta^{xx}(\omega_o \pm \omega_r n) & : = 2\pi B^2_1 \ \delta(\omega_o \pm n\omega_{r}) \\
			\zeta^{xx}(\omega_o \pm \omega_r n) & := B^2_1 \ \mathcal{P}\left( \frac{1}{\omega_o\pm n\omega_{r}} \right) \ .
			\end{align}
			\end{subequations}
If we now assume a continuous distribution of the frequencies in the applied radiation field, then $\sum_r \mapsto \int d\omega' \rho_f(\omega')$ and $\omega_r \mapsto \omega'$ -- where $\rho_f(\omega')$ is the probability density function for the radiation field's frequencies (still centered on $\omega$, as specified in text). And it can be verified, after some algebraic manipulations, that:
			\begin{multline}
			\label{eq:diff_varrho_n=0_4}
		 	-  \int d\omega' \rho_f(\omega') \sum_{n,\omega_o} \ \bigg( \Gamma(\omega_o-n\omega') + \Gamma(\omega_o+n\omega')\bigg) \left[\xi^{x\dagger}(n,\omega_o) ,\xi^{x}(n,\omega_o)  \varrho^{(0)}(t)\right] + h.c. \\
		 	= - i \left[H_{LS+} + H_{LS-}, \varrho^{(0)}(t) \right] + \mathcal{D}_+\left[\varrho^{(0)}(t) \right] + \mathcal{D}_-\left[\varrho^{(0)}(t) \right]
			\end{multline}
where,
			\begin{equation}
			H_{LS\pm} \equiv \sum_{n,\omega_o} \int^{+\infty}_{-\infty} d\omega' \rho_f (\omega') \zeta^{xx}(\omega_o \mp \omega' n)\xi^{x\dagger}(n,\omega_o) \xi^{x}(n,\omega_o)\label{eq:Lamb-shift}
			\end{equation}
			\begin{multline}	
			\mathcal{D}_\pm \left[\varrho^{(0)}(t) \right]  \equiv \sum_{n,\omega_o} \int^{+\infty}_{-\infty} d\omega' \ \rho_f(\omega') \eta^{xx}(\omega_o \mp \omega' n) \left[\xi^{x}(n,\omega_o)\varrho^{(0)}(t)\xi^{x\dagger}(n,\omega_o)  - \frac{1}{2}\left\lbrace \xi^{x\dagger}(n,\omega_o)\xi^{x}(n,\omega_o) , \varrho^{(0)}(t) \right\rbrace  \right]\label{eq:dissipator} \ .
			\end{multline}
The sign subscripts for $H_{LS}$ and $\mathcal{D}$ are such chosen to indicate their origins: the subscript `$+$' means the object originates from the second term of Eq. (41), while `$-$' indicates its origin is the third term of Eq. (41). 
Furthermore, in Eqs. \eqref{eq:Lamb-shift} and \eqref{eq:dissipator}, we have extended the lower limit of the integral over $\omega'$ from zero to ($-\infty$). This causes no appreciable error in subsequent calculations since the frequency $\omega$ at which $\rho_f(\omega')$ is centered on is in the order of MHz, and is usually quite sharply peaked. Naturally, the normalization condition $\int^{+\infty}_{-\infty}d\omega' \ \rho_f(\omega')=1$ holds. 
\par A closer look at the integrals in Eqs. \eqref{eq:Lamb-shift} and \eqref{eq:dissipator} shows that:
			\begin{equation}
			\label{eq:integrals_Lamb_shift}
			\int^{+\infty}_{-\infty} d\omega' \rho_f (\omega') \zeta^{xx}(\omega_o \mp \omega' n)= \begin{cases}
			\pm \frac{\pi B^2_1}{n}  \ \rho^\succ_f(\pm \omega_o/n), & (n \neq 0)\\
			B^2_1 \ \mathcal{P}\left(\frac{1}{\omega_o} \right), & (n=0)
			\end{cases}
			\end{equation}
			
			\begin{equation}
			\label{eq:integrals_dissipator}
			\int^{+\infty}_{-\infty} d\omega' \ \rho_f(\omega') \eta^{xx}(\omega_o \mp \omega' n)=\begin{cases}
			\frac{2\pi B^2_1}{\vert n \vert} \rho_f(\pm \omega_o/n), & (n \neq 0)\\
			2\pi B^2_1 \ \delta(\omega_o), & (n = 0)
			\end{cases}
			\end{equation}								
where $\rho^\succ_f(\pm \omega_o/n)$ is the Hilbert transform\citep{book:King-2009} of $\rho_f$ centered on $\pm \omega_o/n$:
			\begin{equation}
			\rho^\succ_f(\pm \omega_o/n) := \frac{1}{\pi} \int^{+\infty}_{-\infty} d\omega' \ \mathcal{P} \left( \frac{\rho_f(\omega')}{\pm \omega_o/n-\omega'}\right) \ .
			\end{equation}
Note that given a specific $\xi^x(n,\omega_o)$, it follows from Eq. (45) of the text that if we make the transformation $n \mapsto -n$, then $\omega_o \mapsto -\omega_o$ also follows. Making use of this property and the fact that $\xi^x(0,0)=0$, we may conveniently rewrite the dissipator terms $\mathcal{D}_\pm$ as:
			\begin{multline}
			\label{eq:dissipator_2}
			\mathcal{D}_\pm \left[\varrho^{(0)}(t) \right] = \sum_{n>0,\omega_o}  \frac{2\pi B^2_1}{ n } \rho_f(\pm \omega_o/n) \left[ \xi^{x}(n,\omega_o)\varrho^{(0)}(t)\xi^{x\dagger}(n,\omega_o) - \frac{1}{2}\left\lbrace \xi^{x\dagger}(n,\omega_o)\xi^{x}(n,\omega_o) , \varrho^{(0)}(t) \right\rbrace  \right]\\
			+ \sum_{n>0,\omega_o}  \frac{2\pi B^2_1}{ n } \rho_f(\pm \omega_o/n) \left[ \xi^{x\dagger}(n,\omega_o)\varrho^{(0)}(t)\xi^{x}(n,\omega_o) - \frac{1}{2}\left\lbrace \xi^{x}(n,\omega_o)\xi^{x\dagger}(n,\omega_o) , \varrho^{(0)}(t) \right\rbrace  \right]
			\end{multline}
and for the Lamb shift Hamiltonians, $H_{LS\pm}$, Eq. \eqref{eq:Lamb-shift}, we may write:
			\begin{equation}
			H_{LS\pm} = \pm \sum_{n>0,\omega_o} \frac{\pi B_1^2}{n} \rho^\succ_f(\pm \omega_o/n) \left[\xi^{x\dagger}(n,\omega_o), \xi^{x}(n,\omega_o)\right] \label{eq:Lamb-shift_2} \ .
			\end{equation}
Drawing on the fact that the operator $\xi^x$ is the $q=\pm1$ component of a rank $k=1$ spherical tensor, we see that $n=+1$ in Eqs. \eqref{eq:dissipator_2} and \eqref{eq:Lamb-shift_2}. Thus, Eqs. \eqref{eq:dissipator_2} and \eqref{eq:Lamb-shift_2} reduce to Eqs. (60) and (57), respectively.
\par Let us now turn to the first term on the r.h.s. of Eq. \eqref{eq:50} and split it into two:
			\begin{equation}
			\label{eq:LRT_split}
			-i B_1 \sum_r \sum_{n,\omega_o}   e^{-it(\omega_o-\omega_r n)}  \left[\xi^x(n, \omega_o) , \varrho^{(0)}(0)\right]
			-i B_1 \sum_r \sum_{n,\omega_o}    e^{-it(\omega_o+\omega_r n)}  \left[\xi^x(n, \omega_o) , \varrho^{(0)}(0)\right] \ .
			\end{equation}
We focus now on the first term. Assuming here a continuous distribution of the frequencies of the applied radiation field, we get:
			\begin{equation}
			-i  B_1 \sum_r \sum_{n,\omega_o}  e^{-it(\omega_o-\omega_r n)} \left[\xi^x(n, \omega_o)  , \varrho^{(0)}(0)\right] \\ 
			 \mapsto  -i  B_1  \sum_{n,\omega_o} \int^{+\infty}_{-\infty}d\omega' \ \rho_f(\omega') \ e^{it(\omega' n-\omega_o)} \left[\xi^x(n, \omega_o)  , \varrho^{(0)}(0)\right]	\ .
			\end{equation}
But,
			\begin{equation}
			\begin{split}
			-i  B_1  \sum_{n,\omega_o} \int^{+\infty}_{-\infty}d\omega' \ \rho_f(\omega') \ & e^{it(\omega' n-\omega_o)} \left[\xi^x(n, \omega_o)  , \varrho^{(0)}(0)\right]\\
			& = -i  B_1  \sum_{\omega_o} e^{-it \omega_o} \int^{+\infty}_{-\infty}d\omega' \ \rho_f(\omega') \ e^{it \omega' } \left[\xi^x(+1, \omega_o)  , \varrho^{(0)}(0)\right] + h.c.\\
			& = -i  B_1  \sum_{\omega_o} e^{-it \omega_o} \varphi_{f}(t) \left[\xi^x(+1, \omega_o)  , \varrho^{(0)}(0)\right] + h.c.\\
			& = -i \left[H_{LR+}(t),  \varrho^{(0)}(0) \right]
			\end{split}
			\end{equation}
where we can easily recognize $\varphi_{f}(t)$ in the above expression as the characteristic function of $\rho_f(\omega')$:
			\begin{equation}
			\varphi_{f}(t) \equiv  \int^{+\infty}_{-\infty}d\omega' \ \rho_f(\omega') \ e^{it \omega' } . 
			\end{equation}
and
			\begin{equation}
			H_{LR+}(t) \equiv B_1 \varphi_{f}(t) \sum_{\omega_o} e^{-it \omega_o}  \xi^x(+1, \omega_o)  + h.c. \ .
			\end{equation}
Analogously, for the second term of Eq. \eqref{eq:LRT_split}, we have
			\begin{equation}
			-i B_1 \sum_r \sum_{n,\omega_o}    e^{-it(\omega_o+\omega_r n)}  \left[\xi^x(n, \omega_o) , \varrho^{(0)}(0)\right] = -i \left[H_{LR-}(t),  \varrho^{(0)}(0) \right]
			\end{equation}
where
			\begin{equation}
			H_{LR-}(t) \equiv B_1 \varphi^*_{f}(t) \sum_{\omega_o} e^{-it \omega_o}  \xi^x(+1, \omega_o)  + h.c.
			\end{equation}
($\varphi^*_{f}(t)$ is the complex conjugate of $\varphi_{f}(t)$.) Putting the two results together, we get Eqs. (53) and (54).
\subsection{Derivation of the equation for $\Lambda(t)$, Eq. (64)}\label{subsec:Derivation_Lambda}
We want to find the map $\Lambda(t)$ associated with the zeroth-order master equation in Eq. (52) of the paper, which reads
			\begin{equation}
			\label{eq:varrho_eq_motion}
			\frac{d}{dt}\varrho^{(0)}(t) = \mathcal{A}(t)\varrho^{(0)}(0) + \mathcal{L} \varrho^{(0)}(t) \ .
			\end{equation}	
Following [\onlinecite{art:Giovannetti-2013}], let
			\begin{equation}
			\label{eq:varrho_Lambda_t}
			\varrho^{(0)}(t) = \Lambda(t) \varrho^{(0)}(0) \ .
			\end{equation}
Then, from Eq. \eqref{eq:varrho_eq_motion} we have
			\begin{equation}
			\label{eq:Lambda_eq_motion}
			\frac{d}{dt}\Lambda(t) = \mathcal{A}(t) + \mathcal{L} \ \Lambda(t)
			\end{equation}
which after performing the Laplace transform (indicated by the operational symbol $\mathsf{L}$) becomes
			\begin{equation}
			\label{eq:Lambda_eq_motion_LT}
			s\widetilde{\Lambda}(s) - \mathbb{I} = \widetilde{\mathcal{A}}(s) + \mathcal{L} \ \widetilde{\Lambda}(s)
			\end{equation}
where $\mathsf{L}\left[ \Lambda(t)\right]=\widetilde{\Lambda}(s)$ and $\mathsf{L}\left[ \mathcal{A}(t)\right]=\widetilde{\mathcal{A}}(s)$. Upon a rearrangement of the terms in Eq. \eqref{eq:Lambda_eq_motion_LT}, we end up with
			\begin{equation}
			\label{eq:Lambda_eq_motion_LT_2}
			\left( s\mathbb{I} - \mathcal{L} \right) \widetilde{\Lambda}(s) = \widetilde{\mathcal{A}}(s) + \mathbb{I} \ .
			\end{equation}
That is,
			\begin{equation}
			\label{eq:Lambda_eq_motion_LT_2}
			\widetilde{\Lambda}(s) = \left( s\mathbb{I} - \mathcal{L} \right)^{-1} \left(\widetilde{\mathcal{A}}(s) + \mathbb{I}\right) \ .
			\end{equation}
If we now expand the resolvent $\left( s\mathbb{I} - \mathcal{L} \right)^{-1}$ in powers of the generator $\mathcal{L}$, we get
			\begin{equation}
			\left( s\mathbb{I} - \mathcal{L} \right)^{-1} = \sum^{\infty}_{n=0} \frac{\mathcal{L}^n}{s^{n+1}}
			\end{equation}	
with $\mathcal{L}^n = \mathbb{I}$ for $n=0$. Thus, Eq. \eqref{eq:Lambda_eq_motion_LT_2} becomes
			\begin{equation}
			\widetilde{\Lambda}(s) = \sum^{\infty}_{n=0} \mathcal{L}^n  \left(\widetilde{\mathcal{A}}(s)\frac{1}{s^{n+1}} + \mathbb{I}\frac{1}{s^{n+1}}\right) \ .			
			\end{equation}
The inverse Laplace transform of this last equation is
			\begin{equation}
			\begin{split}
			\Lambda(t) & = \sum^{\infty}_{n=0} \mathcal{L}^n  \left(\mathsf{L}^{-1}\left[\widetilde{\mathcal{A}}(s)\frac{1}{s^{n+1}}\right] + \mathbb{I} \cdot \mathsf{L}^{-1}\left[\frac{1}{s^{n+1}}\right]\right)
			 = \int^t_0 d\tau \left(\sum^{\infty}_{n=0} \mathcal{L}^n \frac{\tau^n}{n!} \right) \mathcal{A}(t-\tau)  + \mathbb{I} \cdot \left(\sum^{\infty}_{n=0} \mathcal{L}^n\frac{t^n}{n!} \right)\\
			& = \int^t_0 d\tau \ e^{\mathcal{L}\tau}  \mathcal{A}(t-\tau)  + e^{\mathcal{L}t}\ = \int^t_0 d\tau \ e^{\mathcal{L}(t-\tau)}  \mathcal{A}(\tau)  + e^{\mathcal{L}t} \ . 
			\end{split}		
			\end{equation}
\subsection{$\Lambda(t)$ is not CP (Completely Positive)}\label{subsec:L1}
Certainly, $\Lambda(t)$ preserves trace for any input state $\varrho^{(0)}(0)$. For $\Lambda(t)$ to qualify as a CP map, its output state must always be positive definite, irrespective of the input $\varrho^{(0)}(0) \neq 0$. It can be easily shown that this is not the case for $\Lambda(t)$. Again, it has to do with the presence of the superoperator $\mathcal{A}(t)$ in the expression for $\Lambda(t)$ Eq. (64). If we take a look at Eq. (63),
			\begin{equation}
			\label{eq:Lambda_t_pre_Thompson}
			\varrho^{(0)}(t) = e^{\mathcal{L} t}\varrho^{(0)}(0) + \int^t_0 dt'\  e^{\mathcal{L}(t-t')} \mathcal{A}(t')\varrho^{(0)}(0)
			\end{equation}
we see that while the first term on the right clearly is certainly thus CP, the second term involves the composition of two superoperators acting on $\varrho^{(0)}(0)$. In the Kraus operator sum representation, Eq. \eqref{eq:Lambda_t_pre_Thompson} may be rewritten as
			\begin{equation}
			\label{eq:Lambda_t_Thompson_Kraus}
			\varrho^{(0)}(t) = \sum_\alpha \mathscr{K}_\alpha(t)\varrho^{(0)}(0)\mathscr{K}^\dagger_\alpha(t) + \sum_\alpha \int^t_0 d\tau\   \mathscr{K}_\alpha(t-\tau)\left[\mathcal{A}(\tau)\varrho^{(0)}(0)\right]\mathscr{K}^\dagger_\alpha(t-\tau) \ . 
			\end{equation}
where the Kraus operators $\{\mathscr{K}_\alpha(t)\}$ obey the usual completeness relation $\sum_\alpha \mathscr{K}^\dagger_\alpha(t)\mathscr{K}_\alpha(t)= \mathbb{I}$. Or, more explicitly, 
			\begin{multline}
			\label{eq:Lambda_t_Thompson_Kraus_2}
			\varrho^{(0)}(t) = \sum_\alpha \mathscr{K}_\alpha(t)\varrho^{(0)}(0)\mathscr{K}^\dagger_\alpha(t) + \sum_\alpha \int^t_0 d\tau\   \mathscr{K}_\alpha(t-\tau)\mathscr{M}(\tau)\varrho^{(0)}(0)\mathscr{M}^\dagger(\tau)\mathscr{K}^\dagger_\alpha(t-\tau) \\
			- \sum_\alpha \int^t_0 d\tau\   \mathscr{K}_\alpha(t-\tau)\mathscr{M}^\dagger(\tau)\varrho^{(0)}(0)\mathscr{M}(\tau)\mathscr{K}^\dagger_\alpha(t-\tau)  
			\end{multline}
with\citep{art:Thompson-1958, art:Gaines-1966}
			\begin{equation}
			\mathscr{M}(t) \equiv \frac{1}{\sqrt{2}} \big[\mathbb{I} - i H_{LR}(t) \big] \ .
			\end{equation}
where $H_{LR}(t)$ is the linear response Hamiltonian, Eq. (54). Furthermore, we note that
			\begin{multline}
			\mathbb{I}= \sum_\alpha \mathscr{K}^\dagger_\alpha(t) \mathscr{K}_\alpha(t) + \sum_\alpha \int^t_0 d\tau\   \mathscr{M}^\dagger(\tau)\mathscr{K}^\dagger_\alpha(t-\tau) \mathscr{K}_\alpha(t-\tau)\mathscr{M}(\tau) \\
			- \sum_\alpha \int^t_0 d\tau\   \mathscr{M}(\tau)\mathscr{K}^\dagger_\alpha(t-\tau)\mathscr{K}_\alpha(t-\tau)\mathscr{M}^\dagger(\tau) 
			\end{multline}
-- which confirms again that $\Lambda(t)$ is trace-preserving. 
However, from Eq. \eqref{eq:Lambda_t_Thompson_Kraus_2}, we see that $\varrho^{(0)}(t)$ cannot be guaranteed to be always positive for an arbitrary $\varrho^{(0)}(0)$ -- given that it is the difference between two positive operators. In fact, $\Lambda(t)$ can be seen as the difference between two CP maps: 
				\begin{equation}
				\Lambda(t) = \Phi_{1,t} - \Phi_{2,t}
				\end{equation}
				\begin{subequations}
				\begin{align}
				\Phi_{1,t}[\varrho^{(0)}(0)] & \equiv \sum_\alpha \mathscr{K}_\alpha(t)\varrho^{(0)}(0)\mathscr{K}^\dagger_\alpha(t) + \sum_\alpha \int^t_0 d\tau\   \mathscr{K}_\alpha(t-\tau)\mathscr{M}(\tau)\varrho^{(0)}(0)\mathscr{M}^\dagger(\tau)\mathscr{K}^\dagger_\alpha(t-\tau)\\
				\Phi_{2,t}[\varrho^{(0)}(0)] & \equiv \sum_\alpha \int^t_0 d\tau\   \mathscr{K}_\alpha(t-\tau)\mathscr{M}^\dagger(\tau)\varrho^{(0)}(0)\mathscr{M}(\tau)\mathscr{K}^\dagger_\alpha(t-\tau) \ .
				\end{align}
				\end{subequations}
Thus, the map $\Lambda(t)$ is not CP\citep{art:Carteret-2008, art:Yu-2000, inbook:Shabani_co-2014}.
\subsection{Positivity of $\Lambda(t)$ on its specified domain, some indications}\label{subsec:L2}
\par Even though $\Lambda(t)$ is not CP, there are strong indications it is positive on its domain. These indications stem from the parameters involved in the theory and assumptions like  $\frac{B_1}{B_o} \ll 1$. The positivity of $\Lambda(t)$ on its domain, if that be the case, may therefore be attributed to these parameters and the assumptions of the theory (which are to some extent, experimental constraints in many cases), rather than an elegant theorem like Choi's\citep{art:Choi-1975}. We present these arguments in the following. A better analysis may, perhaps, be necessary in the future.
\par Let $\ket{\nu}$ be a generic vector of the spin system's Hilbert space $\mathcal{H}_S$. We show below that there are indications that
			\begin{equation}
			\label{eq:ineq_test}
			\sum_\alpha \matrixel{\nu}{\mathscr{K}_\alpha(t)\varrho^{(0)}(0)\mathscr{K}^\dagger_\alpha(t)}{\nu} > \left\vert \sum_\alpha \int^t_0 d\tau\  \matrixel{\nu}{ \mathscr{K}_\alpha(t-\tau)\left[\mathcal{A}(\tau)\varrho^{(0)}(0)\right]\mathscr{K}^\dagger_\alpha(t-\tau)}{\nu} \right\vert
			\end{equation}
for the Boltzmann input state $\varrho^{(0)}(0)=\frac{e^{-\beta \mathscr{Z}_o}}{\mbox{Tr}\left[e^{-\beta \mathscr{Z}_o} \right]}$ -- making, therefore, the map $\Lambda(t)$ positive for this particular input state. 
\par To begin with, from the expressions for $\mathcal{A}(t)$ and $\varrho^{(0)}(0)$, we have that
			\begin{equation}
			 \mathcal{A}(t)\varrho^{(0)}(0) = i 2B_1 \Re[\varphi_f(t)] \sum_{\omega_o} \sum_{\mathbb{n},\mathbb{n}'} \delta_{\omega_o,\epsilon_{\mathbb{n}'}- \epsilon_{\mathbb{n}}} \delta_{+1,M_{\mathbb{n}'} - M_{\mathbb{n}}} \left( P^{(0)}_{\mathbb{n}'} - P^{(0)}_{\mathbb{n}} \right)
				\bigg(e^{i\omega_o t}  \ket{\mathbb{n}'}\!\matrixel{\mathbb{n}'}{\xi^x}{\mathbb{n}}\bra{\mathbb{n}} - h.c. \bigg) 
			\end{equation}
-- where $P^{(0)}_\mathbb{n} =  \frac{e^{-\beta \epsilon_{\mathbb{n}}}}{\mbox{Tr}\left[e^{-\beta \mathscr{Z}_o} \right]}$, Eq. (128) --, which means
{\small
			\begin{multline}
			\left\vert \sum_\alpha \int^t_0 d\tau\  \matrixel{\nu}{ \mathscr{K}_\alpha(t-\tau)\left[\mathcal{A}(\tau)\varrho^{(0)}(0)\right]\mathscr{K}^\dagger_\alpha(t-\tau)}{\nu} \right\vert \\
			= \bigg\vert 4 B_1   \sum_{\omega_o} \sum_{\mathbb{n},\mathbb{n}'} \delta_{\omega_o,\epsilon_{\mathbb{n}'}- \epsilon_{\mathbb{n}}} \delta_{+1,M_{\mathbb{n}'} - M_{\mathbb{n}}} \left( P^{(0)}_{\mathbb{n}'} - P^{(0)}_{\mathbb{n}} \right)\matrixel{\mathbb{n}'}{\xi^x}{\mathbb{n}}
				 \Im \left( \int^t_0 d\tau \ e^{i\omega_o \tau} \Re[\varphi_f(\tau)]   \sum_\alpha \matrixel{\nu}{\mathscr{K}_\alpha(t-\tau)}{\mathbb{n}'}\matrixel{\mathbb{n}}{\mathscr{K}^\dagger_\alpha(t-\tau)}{\nu}  \right) \bigg\vert \\
				\leq  4 B_1   \sum_{\omega_o} \sum_{\mathbb{n},\mathbb{n}'} \delta_{\omega_o,\epsilon_{\mathbb{n}'}- \epsilon_{\mathbb{n}}} \delta_{+1,M_{\mathbb{n}'} - M_{\mathbb{n}}} \left\vert P^{(0)}_{\mathbb{n}'} - P^{(0)}_{\mathbb{n}} \right\vert \ \left\vert\! \matrixel{\mathbb{n}'}{\xi^x}{\mathbb{n}}\right\vert 
				\times \bigg\vert \Im \left( \int^t_0 d\tau \ e^{i\omega_o \tau} \Re[\varphi_f(\tau)]   \sum_\alpha \matrixel{\nu}{\mathscr{K}_\alpha(t-\tau)}{\mathbb{n}'}\matrixel{\mathbb{n}}{\mathscr{K}^\dagger_\alpha(t-\tau)}{\nu}  \right) \bigg\vert \\
				\leq 4 B_1   \sum_{\omega_o} \sum_{\mathbb{n},\mathbb{n}'} \delta_{\omega_o,\epsilon_{\mathbb{n}'}- \epsilon_{\mathbb{n}}} \delta_{+1,M_{\mathbb{n}'} - M_{\mathbb{n}}} \left\vert P^{(0)}_{\mathbb{n}'} - P^{(0)}_{\mathbb{n}} \right\vert \ \left\vert\! \matrixel{\mathbb{n}'}{\xi^x}{\mathbb{n}}\right\vert 
				 \times \bigg\vert \Im \left( \int^t_0 d\tau \ e^{i\omega_o \tau} \Re[\varphi_f(\tau)] \right)  \bigg\vert \times  \max_{\ket{\mathbb{n}''},t} \left( \sum_\alpha \left\vert \matrixel{\nu}{\mathscr{K}_\alpha(t)}{\mathbb{n}''}\right\vert^2 \right) \ .
			\end{multline}
}		
Furthermore, since $\varphi_f(t)$ is the characteristic function of a symmetric distribution function, we know from Pòlya's theorem\citep{inproceed:Polya-1949} that $\left\vert \Re[\varphi_f(\tau)] \right\vert \leq 1$, so
			\begin{multline}
			\left\vert \sum_\alpha \int^t_0 d\tau\  \matrixel{\nu}{ \mathscr{K}_\alpha(t-\tau)\left[\mathcal{A}(\tau)\varrho^{(0)}(0)\right]\mathscr{K}^\dagger_\alpha(t-\tau)}{\nu} \right\vert \\
				\leq 4 B_1   \sum_{\omega_o} \sum_{\mathbb{n},\mathbb{n}'} \delta_{\omega_o,\epsilon_{\mathbb{n}'}- \epsilon_{\mathbb{n}}} \delta_{+1,M_{\mathbb{n}'} - M_{\mathbb{n}}} \left\vert P^{(0)}_{\mathbb{n}'} - P^{(0)}_{\mathbb{n}} \right\vert \ \left\vert\! \matrixel{\mathbb{n}'}{\xi^x}{\mathbb{n}}\right\vert 
				\times \bigg\vert \Im \left( \int^t_0 d\tau \ e^{i\omega_o \tau} \right)  \bigg\vert \times  \max_{\ket{\mathbb{n}''},t} \left( \sum_\alpha \left\vert \matrixel{\nu}{\mathscr{K}_\alpha(t)}{\mathbb{n}''}\right\vert^2 \right)\\
				= 4 B_1   \sum_{\omega_o} \sum_{\mathbb{n},\mathbb{n}'} \delta_{\omega_o,\epsilon_{\mathbb{n}'}- \epsilon_{\mathbb{n}}} \delta_{+1,M_{\mathbb{n}'} - M_{\mathbb{n}}} \left\vert P^{(0)}_{\mathbb{n}'} - P^{(0)}_{\mathbb{n}} \right\vert \ \left\vert\! \matrixel{\mathbb{n}'}{\xi^x}{\mathbb{n}}\right\vert 
				\times \bigg\vert \frac{\cos(\omega_o t)-1}{\omega_o}  \bigg\vert \times  \max_{\ket{\mathbb{n}''},t} \left( \sum_\alpha \left\vert \matrixel{\nu}{\mathscr{K}_\alpha(t)}{\mathbb{n}''}\right\vert^2 \right) \ .
			\end{multline}	
Given that $\xi^x=-\sum_i \gamma_i S^x_i$, Eq. (3), we thus have
			\begin{multline}
			\label{eq:vert_inequality_e}
			\left\vert \sum_\alpha \int^t_0 d\tau\  \matrixel{\nu}{ \mathscr{K}_\alpha(t-\tau)\left[\mathcal{A}(\tau)\varrho^{(0)}(0)\right]\mathscr{K}^\dagger_\alpha(t-\tau)}{\nu} \right\vert \\
				\leq  4    \sum_{\omega_o} \sum_{\mathbb{n},\mathbb{n}'} \delta_{\omega_o,\epsilon_{\mathbb{n}'}- \epsilon_{\mathbb{n}}} \delta_{+1,M_{\mathbb{n}'} - M_{\mathbb{n}}} \left\vert P^{(0)}_{\mathbb{n}'} - P^{(0)}_{\mathbb{n}} \right\vert \  \left(\sum_i \frac{\left\vert \omega_1(i) \right\vert}{\left\vert \omega_o \right\vert}  \left\vert\! \matrixel{\mathbb{n}'}{S^x_i}{\mathbb{n}}\right\vert \right) 
				\times \left\vert \cos(\omega_o t)-1\right\vert \times  \max_{\ket{\mathbb{n}''},t} \left( \sum_\alpha \left\vert \matrixel{\nu}{\mathscr{K}_\alpha(t)}{\mathbb{n}''}\right\vert^2 \right)
			\end{multline}
where $\left\vert \omega_1(i)\right\vert \equiv \left\vert \gamma_i \right\vert B_1$. If the spin configurations in the states $\ket{\mathbb{n}}$ and $\ket{\mathbb{n}'}$ are such the $i-$th spin is the only spin which alters its spin state so that: 1) $\matrixel{\mathbb{n}'}{S^x_i}{\mathbb{n}} \neq 0$, 2) $\omega_o=\epsilon_{\mathbb{n}'}- \epsilon_{\mathbb{n}}$ and 3) $1=M_{\mathbb{n}'} - M_{\mathbb{n}}$, then it follows from Eq. (137) that $\left\vert \omega_o \right\vert \approx \left\vert \gamma_i \right\vert B_o$. This implies that for all the nonzero terms on the r.h.s. of Eq. \eqref{eq:vert_inequality_e}, $\left\vert \frac{\omega_1(i)}{\omega_o}\right\vert \approx \frac{B_1}{B_o} \ll 1$. In high-resolution NMR\citep{book:Corio-1966}, for example, the ratio $\frac{B_1}{B_o}$ is in the order of $10^{-6}$. Note also that all the factors multiplying the sum $\left(4 \sum_i \frac{\left\vert \omega_1(i) \right\vert}{\left\vert \omega_o \right\vert}  \left\vert\! \matrixel{\mathbb{n}'}{S^x_i}{\mathbb{n}}\right\vert \right)$ are all positive numbers which are at most equal to unity, and almost all of them can be compared to a counterpart on the l.h.s. of Eq. \eqref{eq:ineq_test}. Indeed, it may be argued that the validity of the inequality in Eq. \eqref{eq:ineq_test} pivots on the ratio  $\frac{B_1}{B_o}$. In fact, 
				\begin{equation}
				\sum_\alpha \matrixel{\nu}{\mathscr{K}_\alpha(t)\varrho^{(0)}(0)\mathscr{K}^\dagger_\alpha(t)}{\nu} = \sum_\mathbb{n} P^{(0)}_\mathbb{n} \left(\sum_\alpha \left\vert\matrixel{\nu}{\mathscr{K}_\alpha(t)}{\mathbb{n}} \right\vert^2 \right)
				\end{equation}
so,
				\begin{multline}
				\sum_\mathbb{n} P^{(0)}_\mathbb{n} \left(\sum_\alpha \left\vert\matrixel{\nu}{\mathscr{K}_\alpha(t)}{\mathbb{n}} \right\vert^2 \right) 
				\gg  4 \sum_{\mathbb{n},\mathbb{n}'}  \sum_{\omega_o} \delta_{\omega_o,\epsilon_{\mathbb{n}'}- \epsilon_{\mathbb{n}}} \delta_{+1,M_{\mathbb{n}'} - M_{\mathbb{n}}} \left\vert P^{(0)}_{\mathbb{n}'} - P^{(0)}_{\mathbb{n}} \right\vert \  \left(\sum_i \frac{\left\vert \omega_1(i) \right\vert}{\left\vert \omega_o \right\vert}  \left\vert\! \matrixel{\mathbb{n}'}{S^x_i}{\mathbb{n}}\right\vert \right) \\
				\times \left\vert \cos(\omega_o t)-1\right\vert \times  \max_{\ket{\mathbb{n}''},t} \left( \sum_\alpha \left\vert\! \matrixel{\nu}{\mathscr{K}_\alpha(t)}{\mathbb{n}''}\right\vert^2 \right)
				\end{multline}
for the non-null Boltzmann state $\varrho^{(0)}(0)$, due to the fact that $\frac{\left\vert \omega_1(i) \right\vert}{\left\vert \omega_o \right\vert} \ll 1$ (recall $\omega_o \neq 0$).
This confirms the inequality stated in Eq. \eqref{eq:ineq_test}.
\subsection{On restricting $\Lambda(t)$ to its specified domain}\label{subsec:L3}
As discussed in the paper, without the term $\mathcal{A}(t)\varrho^{(0)}(0)$ in Eq. (52) (or Eq. \eqref{eq:varrho_eq_motion}), $\Lambda(t)$ is obviously CP. We have seen in the last subsection an argument suggesting why $\Lambda(t)$ is non-CP for input states 
			\begin{equation}
			\label{eq:varrho_0}
			\varrho^{(0)}(0) = \frac{e^{-\beta \mathscr{Z}_o}}{\mbox{Tr}[e^{-\beta \mathscr{Z}_o}]} 
			\end{equation}
with $\mathscr{Z}_o$ given by Eq. (17a), i.e. 
			\begin{equation}
			\mathscr{Z}_o \equiv -\sum_i \gamma_i B_o S^z_i + \sum_{i>j} T_{ij} S^z_i S^z_j \ .
			\end{equation}
\par It is vital the input states are restricted to Eq. \eqref{eq:varrho_0}, else one risks obtaining unphysical results. We can see this by studying, for example, $\Lambda(t)$ in the limit $\mathcal{L} \to 0$ (this is close to taking the adiabatic process limit). Then, $\Lambda(t)$ reduces to
			\begin{equation}
			\label{eq:Lambda_L=0}
			\Lambda(t) = \mathbb{I} + \int^t_0 d\tau \ \mathcal{A}(\tau) = \mathbb{I} +  \mathcal{F}(t) 
			\end{equation}
where, for an operator $X$,
			\begin{equation}
			\label{eq:F(t)_K(t)}
			\mathcal{F}(t)X = -i \left[ K(t), X \right] \ , \qquad \ K(t) \equiv \int^t_0 d\tau \  H_{LR}(\tau)
			\end{equation}
where $H_{LR}(t)$ is the linear response Hamiltonian, Eq. (54). Consequently, with $\mathcal{L}=0$, 
			\begin{subequations}
			\begin{align}
			\varrho^{(0)}(t) & = \Lambda(t) \varrho^{(0)}(0) \\
			& = \varrho^{(0)}(0) -i \left[ K(t),\varrho^{(0)}(0) \right] \\
			& = \left[\mathbb{I} - i K(t) \right] \varrho^{(0)}(0) \left[\mathbb{I} + i K(t) \right] - K(t)\varrho^{(0)}(0) K(t) \label{eq:difference_positive_ops}\ .
			\end{align}
			\end{subequations}
(This equation can also be derived from Eq. \eqref{eq:Lambda_t_Thompson_Kraus_2} by noting that $\sum_\alpha \mathscr{K}_\alpha(t) X \mathscr{K}_\alpha^\dagger(t) \to X $ as $\mathcal{L}\to 0$.) The resulting $\varrho^{(0)}(t)$ in Eq. \eqref{eq:difference_positive_ops} is still the difference between two positive operators, so it cannot be positive for an arbitrary input state -- unless extra information or assumptions are provided. Nonetheless, it has the Kraus operator sum representation for Hermitian non-CP maps\citep{inbook:Shabani_co-2014}, and satisfies the related trace preserving condition\citep{inbook:Shabani_co-2014} -- namely, 
			\begin{equation}
			\left[\mathbb{I} - i K(t) \right]^\dagger\left[\mathbb{I} - i K(t) \right] - K^\dagger(t) K(t) = \mathbb{I} \ .
			\end{equation}
These last two observations notwithstanding, the action of the map $\Lambda(t)$ needs to be restricted to its specified domain, i.e. Eq. \eqref{eq:varrho_0}. As mentioned above, failing to do so may lead to unphysical results. 
To illustrate this very important point, we shall analyze Eq. \eqref{eq:difference_positive_ops} under two scenarios: 1) without considering Eq. \eqref{eq:varrho_0}, and 2) taking into account Eq. \eqref{eq:varrho_0}.
\subsubsection{Analyzing Eq. \eqref{eq:difference_positive_ops} without taking into account Eq. \eqref{eq:varrho_0}\footnote{This subsection is a paraphrase of an argument originally put forward by Prof. Vittorio Giovannetti in the course of private discussions with the author. All errors are that of the author alone.}} 
\par For example, if $\varrho^{(0)}(0)$ is taken to be a pure state, \emph{i.e.} $\varrho^{(0)}(0)= \ket{0}\!\bra{0}$, and we do not take into account Eq. \eqref{eq:varrho_0}, it can be shown that $\varrho^{(0)}(t)$ in Eq. \eqref{eq:difference_positive_ops} would appear in this case to be nonpositive for $K(t) \neq 0$. Thus, $\Lambda(t)$ would describe an unphysical process. 
\par To prove the proposition, we first observe that with $\varrho^{(0)}(0)= \ket{0}\!\bra{0}$, Eq. \eqref{eq:difference_positive_ops} may be rewritten as
			\begin{equation}
			\label{eq:diff_pure}
			\varrho^{(0)}(t) = \ket{v_1}\!\bra{v_1} - \ket{v_o}\!\bra{v_o}
			\end{equation}
where
			\begin{equation}
			\ket{v_o} \equiv K(t)\ket{0} \ , \qquad \ \ket{v_1} \equiv \left[\mathbb{I} - i K(t) \right]\ket{0} = \ket{0} - i \ket{v_o} 
			\end{equation}
whose norms are
			\begin{equation}
			\left\Vert \ket{v_o} \right\Vert = \sqrt{\matrixel{0}{K(t)^2}{0}} \ , \qquad \ \left\Vert \ket{v_1} \right\Vert = \sqrt{1+\matrixel{0}{K(t)^2}{0}}
			\end{equation}
-- where we have exploited the fact that $K(t)$, Eq. \eqref{eq:F(t)_K(t)}, is Hermitian. For non-null $K(t)$ (which is the case if $H_{LR}(t)$ is nonzero), $\left\Vert \ket{v_o} \right\Vert $ is strictly positive, \emph{i.e.} $\left\Vert \ket{v_o} \right\Vert >0$.
\par Let $x= \sqrt{\matrixel{0}{K(t)^2}{0}}$ and $K(t) \neq 0$. Furthermore, let $\ket{\widetilde{v_o}}$ and $\ket{\widetilde{v_1}}$ be the normalized kets corresponding to $\ket{v_o}$ and $\ket{v_1}$, respectively -- that is, $\ket{\widetilde{v_o}} \equiv \frac{\ket{v_o}}{\left\Vert \ket{v_o} \right\Vert}$ and $\ket{\widetilde{v_1}} \equiv \frac{\ket{v_1}}{\left\Vert \ket{v_1} \right\Vert}$. Then, Eq. \eqref{eq:diff_pure} may be rewritten as
			\begin{equation}
			\label{eq:diff_pure_2}
			\varrho^{(0)}(t)  = (1+x^2) \left[\ket{\widetilde{v_1}}\bra{\widetilde{v_1}} -  \frac{x^2}{1+x^2}\ket{\widetilde{v_o}}\bra{\widetilde{v_o}} \right] \ .
			\end{equation}
We may expand $\ket{\widetilde{v_1}}$ as
			\begin{equation}
			\label{eq:widetilde_v_1}
			\ket{\widetilde{v_1}} = \alpha \ket{\widetilde{v_o}} + \beta \ket{\widetilde{v_o}_{\perp}}
			\end{equation}
where $\ket{\widetilde{v_o}_{\perp}}$ is the normalized ket perpendicular to $\ket{\widetilde{v_o}}$, while $\alpha$ and $\beta$ are complex scalars which also satisfy the condition $\abs{\alpha}^2 + \abs{\beta}^2=1$. Substituting Eq. \eqref{eq:widetilde_v_1} into Eq. \eqref{eq:diff_pure_2} yields
			\begin{equation}
			\label{eq:diff_pure_3}
			\varrho^{(0)}(t)  = (1+x^2) \left[ \left(\abs{\alpha}^2 -  \frac{x^2}{1+x^2}\right)\ket{\widetilde{v_o}}\bra{\widetilde{v_o}} + \alpha \beta^* \ket{\widetilde{v_o}}\bra{\widetilde{v_o}_{\perp}} + \alpha^* \beta \ket{\widetilde{v_o}_{\perp}}\bra{\widetilde{v_o}} + \abs{\beta}^2 \ket{\widetilde{v_o}_{\perp}}\bra{\widetilde{v_o}_{\perp}}\right]
			\end{equation}
-- which in matrix form simply becomes
			\begin{equation}
			\label{eq:diff_pure_4}
			\varrho^{(0)}(t)  = (1+x^2) 
			\begin{bmatrix}
			\left(\abs{\alpha}^2 -  \frac{x^2}{1+x^2}\right) & \alpha \beta^* \\
			\alpha^* \beta & \abs{\beta}^2 
			\end{bmatrix}\ .
			\end{equation}
From this last equation, it follows that the determinant of $\varrho^{(0)}(t)$ is
			\begin{equation}
			\det [\varrho^{(0)}(t)] = - x^2 (1+x^2) \abs{\beta}^2  \ .
			\end{equation}
Thus, $\det [\varrho^{(0)}(t)] < 0$ since $x >0$ for non-null $K(t)$. So, for $\mathcal{L}=0$ and $\mathcal{A}(t) \neq 0$, the map $\Lambda(t)$ would blatantly describe an unphysical process for a pure input state.
\par Note, however, that -- according to Eq. \eqref{eq:Lambda_L=0} -- for $K(t)=0$, while $\mathcal{L}=0$, $\Lambda(t)$ simply becomes the identity operator $\mathbb{I}$. As we shall see in the next subsection, it turns out that this is actually the case if $\varrho^{(0)}(0)$ is pure and we take into account Eq. \eqref{eq:varrho_0}.
\subsubsection{Analyzing Eq. \eqref{eq:difference_positive_ops} taking into account Eq. \eqref{eq:varrho_0}}
\par On a closer examination, if we restrict the input states to $\Lambda(t)$'s domain, i.e. Eq. \eqref{eq:varrho_0}, we realize that $\varrho^{(0)}(0)$ cannot be a pure state unless \emph{all} the particles composing the chemical species have zero spin quantum number  (examples are \ce{^{16}O} and \ce{^{12}C} nuclei). For such a system, it can be shown that $K(t)$ becomes identically zero, therefore, making $\mathcal{A}(t)=0$ and $\Lambda(t)=\mathbb{I}$.
\par To prove the last proposition, let us recall that $H_{LR}(t)$, Eq. (56), is defined as
			\begin{equation}
			H_{LR}(t) = 2B_1 \Re[\varphi_f(t)] \sum_{\omega_o} e^{-i\omega_o t} \xi^x(+1,\omega_o) + h.c.
			\end{equation}
where, according to Eq. (45),
			\begin{equation}
			\label{eq:xi_x}
			\xi^x(+1,\omega_o) = \sum_{\mathbb{n},\mathbb{n}'} \ket{\mathbb{n}}\matrixel{\mathbb{n}}{\xi^x}{\mathbb{n}'}\bra{\mathbb{n}'} \delta_{\omega_o, \epsilon_{\mathbb{n}'}-\epsilon_\mathbb{n}}\delta_{+1, M_{\mathbb{n}'}-M_{\mathbb{n}}} \ .
			\end{equation}
In addition, we know from Eq. (3) that
			\begin{equation}
			\xi^x = -\sum_i \gamma_i S^x_i \ .
			\end{equation}
The operator $\xi^x$ is therefore a zero-trace operator. 
\par Now, if we set $\varrho(0)=\ket{0}\!\bra{0}$, then we are somehow admitting that the spin system in question is a collection of spin zero particles, and $\ket{0}$ is the (spin) state vector of the collection (up to a phase-factor). In this case, we easily deduce from Eq. \eqref{eq:xi_x} that 
			\begin{equation}
			\xi^x(+1,\omega_o) =  \delta_{\omega_o, 0}\delta_{+1,0} \mbox{Tr}[\xi^x] \ket{0}\!\bra{0} 
			\end{equation}
which is identically zero. Consequently, $H_{LR}(t)$ and $K(t)$ also become identically zero. And as a result, $\Lambda(t) \to \mathbb{I}$.
\par We may also note that for a non-entirely-spin-zero system, $\varrho^{(0)}(0)$ \textit{approaches} a pure spin state when $T (\text{temperature}) \to 0$ and/or $B_o \to \infty$. In any case, even under these extreme conditions, it is mathematically impossible for $\varrho^{(0)}(0)$ of a non-entirely-zero-spin system -- as defined in Eq. \eqref{eq:varrho_0} -- to be of rank $1$ (thus, a pure spin state).
\subsection{On higher-order terms}\label{subsec:higher-orders}
\par The equation of motion for higher-order corrections to $\varrho^{(0)}(t)$ easily follows from Eq. (32), which states that
				\begin{equation}
				\label{eq:differential_varrho_n}
				\frac{d}{dt} \varrho^{(n)}(t) = -i \sum^n_{k=0} \left[ \mathscr{V}^{(n)}(t) , \varrho^{(n-k)}(0)\right] 
				- \sum^n_{k=0} \sum^k_{k'=0} \int^{+\infty}_0 d\tau \left[\mathscr{V}^{(n-k)}(t) ,\left[\mathscr{V}^{(k-k')}(t-\tau) , \varrho^{(k')}(t)\right]\right] \ .
				\end{equation}
For any given order $n$ in Eq. \eqref{eq:differential_varrho_n}, the approximations and arguments (see \S \ref{subsec:Derivation} for details)  we laid out for the zeroth-order must be replicated. This leads to an equation of motion of the form given in Eq. (148). Specifically, the secular approximation must be applied to the second term in Eq. \eqref{eq:differential_varrho_n}. But before that, if one had assumed at the zeroth-order that  the cross-terms involving $\mathscr{V}^{(n)}_\pm(t)$ may be neglected (as we did), then the same assumption must be applied to the second term in Eq. \eqref{eq:differential_varrho_n}. Finally, one must also assume a continuous distribution of the frequencies in the oscillating field for both terms in Eq. \eqref{eq:differential_varrho_n}.
\par If we consider the first-order correction to $\varrho(t)$, $\varrho^{(1)}(t)$, for example, it follows from Eq. \eqref{eq:differential_varrho_n} that
				\begin{multline}
				\label{eq:diff_varrho_n=1}
				\frac{d}{dt} \varrho^{(1)}(t)  = -i  \left[ \mathscr{V}^{(0)}(t)  , \varrho^{(1)}(0)\right] -i \left[ \mathscr{V}^{(1)}(t)  , \varrho^{(0)}(0)\right]  \\
				 -  \int^{+\infty}_0 d\tau \left[\mathscr{V}^{(1)}(t) ,\left[\mathscr{V}^{(0)}(t-\tau) , \varrho^{(0)}(t)\right]\right] \\
				 \hspace{3cm} -  \int^{+\infty}_0 d\tau \left[\mathscr{V}^{(0)}(t) ,\left[\mathscr{V}^{(1)}(t-\tau) , \varrho^{(0)}(t)\right]\right] \\
				 -  \int^{+\infty}_0 d\tau \left[\mathscr{V}^{(0)}(t) ,\left[\mathscr{V}^{(0)}(t-\tau) , \varrho^{(1)}(t)\right]\right] \ .
				\end{multline}
We note that the first and last terms in Eq. \eqref{eq:diff_varrho_n=1} are actually copies of  the r.h.s. of Eq. (38), except that $\varrho^{(0)}(0)$ and $\varrho^{(0)}(t)$ are now substituted with  $\varrho^{(1)}(0)$ and $\varrho^{(1)}(t)$, respectively. So, after applying the above mentioned approximations and assumptions, these two terms transform as follows:
				\begin{subequations}
				\begin{align}
				-i  \left[ \mathscr{V}^{(0)}(t)  , \varrho^{(1)}(0)\right] & \mapsto \mathcal{A}(t)\varrho^{(1)}(0)\\
				-  \int^{+\infty}_0 d\tau \left[\mathscr{V}^{(0)}(t) ,\left[\mathscr{V}^{(0)}(t-\tau) , \varrho^{(1)}(t)\right]\right] & \mapsto \mathcal{L} \varrho^{(1)}(t)
				\end{align}
				\end{subequations}		
where $\mathcal{A}(t)$ and $\mathcal{L}$ are still given by Eqs. (53) and (55), respectively.	And the expression for $\varrho^{(1)}(0)$ is derived from Eq. (26).
\par Similarly, for the second term of Eq. \eqref{eq:diff_varrho_n=1}, we may write
				\begin{equation}
				-i \left[ \mathscr{V}^{(1)}(t)  , \varrho^{(0)}(0)\right] \mapsto \mathcal{A}^{(1)}(t)\varrho^{(0)}(0)
				\end{equation}
where $\mathcal{A}^{(1)}(t)$ will be linear in $\mathscr{X}$ and bear some similarities with $\mathcal{A}(t)$. The third and fourth terms of Eq. \eqref{eq:diff_varrho_n=1} may be put together after applying the said approximations and assumptions, and we may write
				\begin{multline}
				-  \int^{+\infty}_0 d\tau \left[\mathscr{V}^{(1)}(t) ,\left[\mathscr{V}^{(0)}(t-\tau) , \varrho^{(0)}(t)\right]\right] 
				-  \int^{+\infty}_0 d\tau \left[\mathscr{V}^{(0)}(t) ,\left[\mathscr{V}^{(1)}(t-\tau) , \varrho^{(0)}(t)\right]\right] 
				 \mapsto \mathcal{L}^{(1)}\varrho^{(0)}(t) \ .
				\end{multline}	
Here too, the superoperator $\mathcal{L}^{(1)}$ will linearly depend on $\mathscr{X}$.
\par Thus, after implementing the said approximations and assumptions, Eq. \eqref{eq:diff_varrho_n=1} simply becomes
				\begin{equation}
				\label{eq:diff_varrho_n=1_s1}
				\frac{d}{dt} \varrho^{(1)}(t) = \mathcal{A}(t)\varrho^{(1)}(0) + \mathcal{L} \varrho^{(1)}(t) + \mathcal{G}^{(1)}(t)
				\end{equation}	
where
				\begin{equation}
				\label{eq:diff_varrho_n=1_s2}
				\mathcal{G}^{(1)}(t) = \mathcal{A}^{(1)}(t)\varrho^{(0)}(0) + \mathcal{L}^{(1)}\varrho^{(0)}(t) \ .
				\end{equation}	
Note that Eqs. \eqref{eq:diff_varrho_n=1_s1} and \eqref{eq:diff_varrho_n=1_s2} are in agreement with Eqs. (148) and (149), respectively. In fact, following the same line of reasoning as we just did for the first-order correction, one can show by induction that Eqs. (148) and (149) hold for any given order $n$. Furthermore, in solving Eq. \eqref{eq:diff_varrho_n=1_s1} for $\varrho^{(1)}(t)$, we must impose the condition $\mbox{Tr}[\varrho^{(1)}(t)]=0$ (this also applies to all $\varrho^{(n)}(t)$ with $n \geq 1$).
\hidefigurestrue
\begin{conditionalfigure}[ht]
  \centering
  \includegraphics[width=5cm]{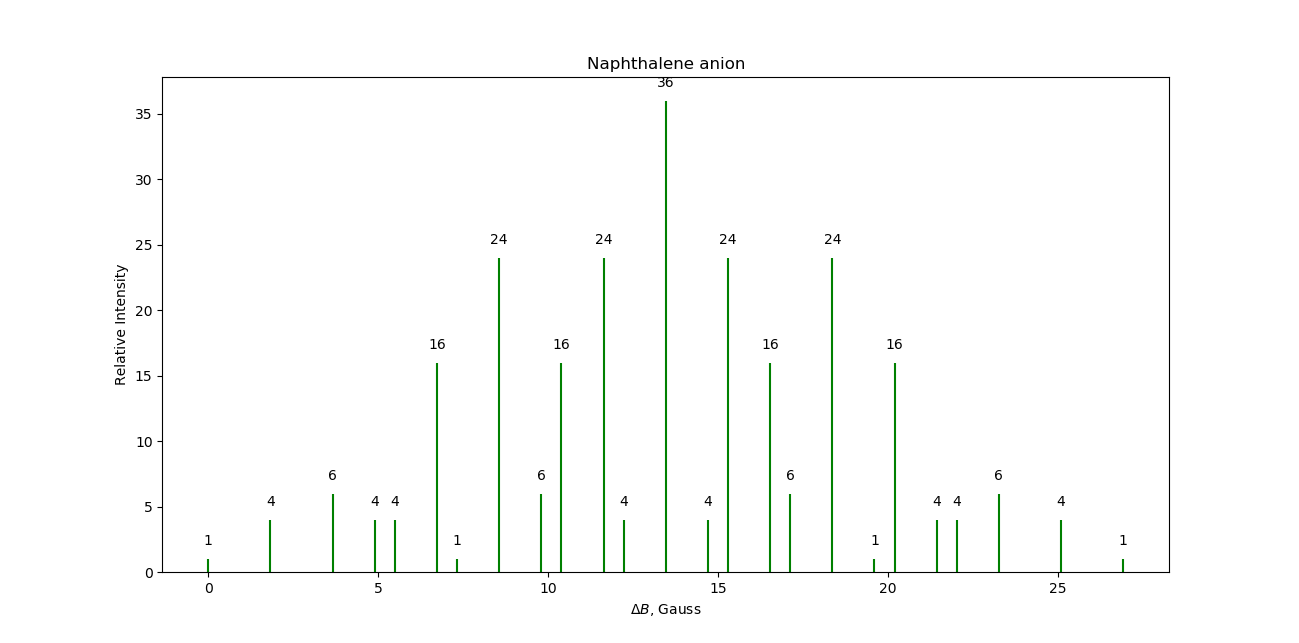}
  \caption{\label{figure1}}
\end{conditionalfigure}															
\bibliography{supplement_biblio}